\documentclass[traditabstract]{aa}
\usepackage{graphicx}
\usepackage{url}
\usepackage{txfonts}
\usepackage{pstricks}
\usepackage{natbib}
\usepackage{subfigure}
\usepackage{rotating}

\bibpunct{(}{)}{;}{a}{}{,}
\begin{document}
\title{COSMOGRAIL: the COSmological MOnitoring of GRAvItational Lenses\thanks{Based on observations made with the NASA/ESA HST Hubble Space Telescope by the CfA-Arizona Space Telescope Lens Survey (CASTLeS) collaboration, obtained from the data archive at the Space Science Institute, which is operated by AURA, the Association of Universities for Research in Astronomy, Inc., under NASA contract NAS-5-26555.}}
\subtitle{X. Modeling based on high-precision astrometry of a sample of 25 lensed quasars: consequences for ellipticity, shear, and astrometric anomalies.}

\titlerunning{Deconvolution and modeling of 25 lensed quasars}

\author{D. Sluse
\inst{1, 2}
\and V. Chantry 
\inst{3}
\and P. Magain 
\inst{3}
\and F. Courbin 
\inst{4}
\and G. Meylan
\inst{4}
}

\institute{Astronomisches Rechen-Institut am Zentrum f\"ur Astronomie der Universitat at Heidelberg,
M\"onchhofstrasse 12-14, 69120 Heidelberg, Germany
\and 
Argelander-Institut f\"ur Astronomie, Auf dem H\"ugel 71, 53121 Bonn, Germany
\email{dsluse@astro.uni-bonn.de}
\and
Institut d'Astrophysique et de G\' eophysique, Universit\' e de Li\`ege, All\'ee du 6 Ao\^ut, 17, 4000 Sart Tilman (Bat. B5C), Li\`ege 1, Belgium
\and 
 Laboratoire d'astrophysique, Ecole Polytechnique F\'ed\'erale de Lausanne (EPFL), Observatoire de Sauverny, 1290 Versoix, Switzerland
}

\date{}

  \abstract
{
Gravitationally lensed quasars can be used as powerful cosmological and astrophysical probes. We can (i) infer the Hubble constant $H_{0}$ based on the so-called time-delay technique,  (ii) unveil substructures along the line-of-sight toward distant galaxies, and (iii) compare the shape and the slope of baryons and dark matter distributions in the inner regions of galaxies. To reach these goals, we need high-accuracy astrometry of the quasar images relative to the lensing galaxy and morphology measurements of the lens. In this work, we first present new astrometry for 11 lenses with measured time delays, namely, JVAS~B0218+357, SBS~0909+532, RX~J0911.4+0551, FBQS~J0951+2635, HE~1104-1805, PG~1115+080, JVAS~B1422+231, SBS~1520+530, CLASS~B1600+434, CLASS~B1608+656, and HE~2149-2745. These measurements proceed from the use of the Magain-Courbin-Sohy (MCS) deconvolution algorithm applied in an iterative way (ISMCS) to near-IR HST images.  We obtain a typical astrometric accuracy of about 1-2.5 mas and an accurate shape measurement of the lens galaxy. Second, we combined these measurements with those of 14 other lensing systems, mostly from the COSMOGRAIL set of targets, to present new mass models of these lenses. The modeling of these 25 gravitational lenses led to the following results: 1) In four double-image quasars (HE0047-1746, J1226-006, SBS~1520+530, and HE~2149-2745), we show that the influence of the lens environment on the time delay can easily be quantified and modeled, hence putting these lenses with high priority for time-delay determination. 2) For quadruple-image quasars, the difficulty often encountered in reproducing the image positions to milli-arcsec accuracy (astrometric anomaly problem) is overcome by explicitly including the nearest visible galaxy/satellite in the lens model. However, one anomalous system  (RXS~J1131-1231) does not show any luminous perturber in its vicinity, and three others (WFI~2026-4536, WFI~2033-4723, and B2045+265) have problematic modeling. These four systems are the best candidates for a pertubation by a dark matter substructure along the line-of-sight. 3) We revisit the correlation between the position angle (PA) and ellipticity of the light and of the mass distribution in lensing galaxies. As in previous studies, we find a significant correlation between the PA of the light and of the mass distributions. However, in contrast with these same studies, we find that the ellipticity of the light and of the mass also correlate well, suggesting that the overall spatial distribution of matter is not very different from the baryon distribution in the inner $\sim$5\,kpc of lensing galaxies. This offers a new test for high-resolution hydrodynamical simulations. }

\maketitle
%
%________________________________________________________________

\section{Introduction}

The measurement of the time delay between the lensed images of a gravitationally lensed quasar offers one of the most elegant ways to measure the \textit{Hubble constant}, $H_{0}$ \citep{Refsdal1964}. This cosmological application of strongly lensed quasars motivated many of the early gravitational lensing studies and time-delay measurement campaigns. Unfortunately,  the precision on $H_{0}$ obtained with the time delay method has been, until now, challenged by systematic errors. Recently, \citet{Suyu2009, Suyu2010} showed that a detailed study of the lensed quasar CLASS~B1608+656, combining different high-resolution data and advanced lens modeling techniques, could lead to an estimate of $H_{0}$  with very small random and systematic uncertainties: $H_{0}= 70.6 \pm 3.1 \ \rm km \ \rm s^{-1} \rm Mpc^{-1}$. The determination of $H_{0}$ based on a large sample of lensed quasars can also reach a precision competitive with the one of standard methods \citep{Oguri2007,Coles2008}.

In order to use the time-delay method as a precision-cosmology tool, it is necessary to derive accurate time-delay measurements and to properly understand the uncertainties associated to the lens environment and to the lens mass distribution. The COSMOGRAIL project, i.e.  the COSmological MOnitoring of GRAvItational Lenses, aims to provide the community with exquisite lightcurves and accurate time-delay measurements for a sample of more than 20 lensed quasars \citep{Cosmograil1, Cosmograil5, Cosmograil7, Courbin2011}. To understand degeneracies affecting lens models, we started to model all systems monitored by COSMOGRAIL as well as those monitored by earlier campaigns in a uniform way. Because the relative astrometry of the lensed images is the main constraint on the lens models, we first worked on deriving an updated astrometry of these systems by applying the \textit{Iterative Strategy coupled with the MCS{\footnote{MCS is an abbreviation build based on the name of the three designers of the method, Magain, Courbin, and Sohy.}} deconvolution algorithm} \citep{MCS98}, i.e. ISMCS \citep{Chantry2007} to HST images. This allows us to obtain accurate relative astrometry and morphological information of the lensing galaxy (i.e. ellipticity and position angle PA of its major axis). We presented the first part of this work in \citet[][ hereafter Paper I]{Chantry2010}, where seven systems currently monitored by COSMOGRAIL were analyzed. We present here the deconvolution of a sample of an additional eleven lensed quasars with published time delays: JVAS~B0218+357, SBS~0909+532, RX~J0911.4+0551, FBQS~J0951+2635, HE~1104-1805, PG~1115+080, JVAS~B1422+231, SBS~1520+530, CLASS~B1600+434, CLASS~B1608+656, and HE~2149-2745. 

High-accuracy relative astrometry of lensed quasars allows one to test the ability of a smooth mass model of the galaxy to reproduce the observed image configuration. The failure of a smooth lens model to do this, which means that we identify an astrometric anomaly, may be caused by a perturbation of the potential by dark matter clump(s) along the line-of-sight, and likely located in the halo of the main lens. During the past five years, the occurrence and amplitude of astrometric anomalies caused by dark matter clumps has been questioned and their use to study the amount of substructures in galaxies has been investigated \citep[][ Paper I and references therein]{Zackrisson2009}. We use the seven quads studied here and in Paper I plus seven other quads with astrometry derived with ISMCS or VLA+HST imaging, to revisit the evidence for astrometric anomalies. We also  quantitatively address the role of the nearby lens environment in producing these anomalies. 

The deconvolution method we used also provides accurate morphological information on the lensing galaxies. With these data, we compared the mass and the light distribution in these systems. This problem has first been approached by \citet{Keeton1998} who compared the mass and the light distribution for 14 multiply-imaged quasars (seven doubles and seven quads). More recently, \citet{Ferreras2008} and \citet{Treu2009} studied a sample of nine (resp. 25) lensing galaxies from the SLACS sample \citep{Bolton2006}. Although these works probe different populations of lenses, the ``SLACS lenses'' lying in typical elliptical galaxy environments while lensed quasars often lie in richer environments \citep{Huterer2005, Oguri2005b, Suyu2009}, they find that the PA of the light and of the mass do agree within 10 degrees. Conversely, no correlation has been found between the ellipticity of the light and of the mass distributions in \citet{Keeton1998} and in \citet{Ferreras2008}. 

The outline of the paper is the following. Section \ref{material11} presents the eleven lensed systems deconvolved with the ISMCS technique. The deconvolution method and the derived astrometry are described in Sect. \ref{dec11}. Section~\ref{sec:model} explains the mass modeling strategy. We discuss the lens models for doubly imaged quasars in Sect.~\ref{sec:doubles}. In Sect.~\ref{sec:quads} we present an enlarged sample of quadruply imaged systems for which we provide additional mass models and a comparison of the shape of the mass and of the light distribution. Finally, our conclusions are presented in in Sect. \ref{Conc11}.

\section{Data}
\label{material11}

We used high-resolution images of eleven gravitationally lensed quasars obtained with NIC2, the Camera 2 of NICMOS, \textit{Near-Infrared Camera and Multi-Object Spectrometer}, on board the \textit{Hubble Space Telescope} (HST). We preferred to use NICMOS data rather than data obtained with  the \textit{Advanced Camera for Surveys} (ACS) for two reasons. First, most of the lensed quasars monitored by COSMOGRAIL have been observed with NICMOS but not always with the ACS. Second, the field distortion and point spread function (PSF) variations are stronger with the ACS than with NICMOS, with damaging consequences for our deconvolution technique, which fails to significantly improve the Tiny Tim PSF for ACS images. The NIC2 data were obtained in the framework of the CASTLES project (Cfa-Arizona Space Telescope LEns Survey\footnote{\url{http://www.cfa.harvard.edu/castles/}}, P.I.: E. Falco). Each data set is composed of four (eight in the case of JVAS~B1422+231) dithered frames acquired through the F160W filter (H-band) in the MULTIACCUM mode, each of them is a combination of 18 to 20 samples. Table \ref{sample_11} summarizes some useful information about the systems: the coordinates in J2000, the type of configuration, the redshift of the lens $z_{l}$, the redshift of the source $z_{s}$, and the angular separation (measured on the NICMOS images, see next section). Table~\ref{sum_dec11} displays information about the image acquisition: the date of observation of every object, the number of frames \emph{NF} and the total exposure time $t_{exp}$.  

Additional details about the lensed quasars under investigation are provided below.

\begin{table*}%[h!]
\begin{center}
\begin{tabular}{lcccccc}
\hline
Object & RA (J2000) & Dec (J2000) & Type & $z_{l}$ & $z_{s}$ & Ang. Sep. (\arcsec)  \\  
\hline
\hline
(a) JVAS~B0218+357 & $02^{\rm h}21^{\rm m}05^{\rm s}.48$ & $+35^{\circ}56'13\farcs78$ & D & $0.6847^{a}$ & $0.944\pm0.002^{b}$ & 0.33     \\
(b) SBS~0909+532 &  $09^{\rm h}13^{\rm m}01^{\rm s}.05$ & $+52^{\circ}59'28\farcs83$ & D & $0.830^{c}$ & $1.377^{d}$ & 1.10     \\
(c) RX~J0911.4+0551 & $09^{\rm h}11^{\rm m}27^{\rm s}.50$ & $+05^{\circ}50'52\farcs00$ & Q & $0.769^{e}$ & $2.8^{f}$ & 3.07     \\
(d) FBQS~J0951+2635 & $09^{\rm h}51^{\rm m}22^{\rm s}.57$ & $+26^{\circ}35'14\farcs10$ & D & $0.260\pm0.002^{g}$ & $1.246\pm0.001^{h}$& 1.10     \\
(e) HE~1104-1805 & $11^{\rm h}06^{\rm m}33^{\rm s}.45$ & $-18^{\circ}21'24\farcs20$ & D & $0.729\pm0.001^{i}$ & $2.319^{j}$ & 3.20     \\
(f) PG~1115+080 & $11^{\rm h}18^{\rm m}17^{\rm s}.00$ & $+07^{\circ}45'57\farcs70$ & Q &  $0.3098\pm0.0002^{k}$ & $1.722^{l}$ & 2.43     \\
(g) JVAS~B1422+231 & $14^{\rm h}24^{\rm m}38^{\rm s}.09$ & $+22^{\circ}56'00\farcs60$ & Q & $0.3366\pm0.0004^{m}$ & $3.62^{n}$ & 1.28     \\
(h) SBS~1520+530 & $15^{\rm h}21^{\rm m}44^{\rm s}.83$ & $+52^{\circ}54'48\farcs60$ & D & $0.761^{o}$ & $1.855\pm0.002^{p}$ & 1.57     \\
(i) CLASS~B1600+434 & $16^{\rm h}01^{\rm m}40^{\rm s}.45$ & $+43^{\circ}16'47\farcs80$ & D & $0.41^{q}$ & $1.59^{r}$ & 1.40     \\
(j) CLASS~B1608+656 & $16^{\rm h}09^{\rm m}13^{\rm s}.96$ & $+65^{\circ}32'29\farcs00$ & Q & $0.6304^{s}$ & $1.394^{t}$ & 2.10    \\
(k) HE 2149-2745 & $21^{\rm h}52^{\rm m}07^{\rm s}.44$ & $-27^{\circ}31'50\farcs20$ & D & $0.603\pm0.001^{u}$ & $2.033^{v}$ & 1.70    \\
\hline
\end{tabular}
\end{center}
{\tiny{Notes: References for the redshifts of the lenses and sources: (a)~\citealt{Browne1993,Stickel1993,Carilli1993} (b)~\citealt{Cohen2003} (c)~\citealt{Lubin2000} (d) \citealt{Kochanek1997,Oscoz1997} (e)~\citealt{Kneib2000} (f)~\citealt{Bade1997} (g)~\citealt{Cosmograil6} (h)~\citealt{Jakobsson2005} (i)~\citealt{Lidman2000} (j)~\citealt{Smette1995} (k)~\citealt{Tonry1998} (l)~\citealt{Weymann1980} (m)~\citealt{Tonry1998} (n)~\citealt{Patnaik1992} (o)~\citealt{Auger2008} (p)~\citealt{Chavushyan1997} (q)~\citealt{Jaunsen1997,Fassnacht1998} (r)~\citealt{Fassnacht1998} (s)~\citealt{Myers1995} (t)~\citealt{Fassnacht1996} (u)~\citealt{Cosmograil6} (v)~\citealt{Wisotzki1996}}}
\vspace{0.2cm}
\caption{General information about the objects of the sample. Type D is for doubly imaged and Q for quadruply imaged system.} 
\label{sample_11}
\end{table*}

\begin{itemize}
 \item \textit{JVAS~B0218+357 (a)}. This system is a doubly imaged blazar (BL Lac type), first considered as such in \citeyear{Patnaik1993} by \citeauthor{Patnaik1993} in the context of the Jodrell Bank/VLA Astrometric Survey (JVAS). It has the smallest image separation found so far in lensed quasar systems.
The lens is a gas-rich late-type galaxy \citep{Keeton1998}. \\

 \item \textit{SBS~0909+532 (b)}. Discovered as a quasar by \citet{Stepanyan1991} in the framework of the Second Byurakan Survey (SBS), the hypothesis of its lensed nature was first proposed by \citet{Kochanek1997} and then supported by \citet{Oscoz1997}. In \citeyear{Engels1998}, \citeauthor{Engels1998} resolved the system into a pair of point sources during the Hamburg-Cfa Bright Quasar Survey. The two lensed images are bright and close to each other, which made it difficult for \citet{Lehar2000} to detect the lensing galaxy even on high-resolution near-IR images: they concluded that the lens is probably an early-type galaxy with a large effective radius and low surface brightness. \\

 \item \textit{RX~J0911.4+0551 (c)}. This lensed quasar was discovered in \citeyear{Bade1997} by \citeauthor{Bade1997} amongst the AGN candidates of the ROSAT All-Sky Survey (RASS). Its lensed nature was confirmed by \citet{Burud1998}. 
The lens is a nearly circular early-type galaxy, which belongs to a massive cluster, centered 38\arcsec~south\,-\,west of the lensed system. It is also known to have a small satellite galaxy in the north\,-\,west direction \citep{Kneib2000}. \\

 \item \textit{FBQS~J0951+2635 (d)}. This system was discovered during the FIRST Bright QSO Survey (FBQS) and \citet{Schechter1998} first identified the two observed objects as lensed images of the same source. The lens is an early-type galaxy \citep[e.g.][]{Cosmograil6}.\\

 \item \textit{HE~1104-1805 (e)}. First considered as a good lens candidate by \citet{Wisotzki1993} in the framework of the Hamburg/ESO Survey (HES) for bright quasars, the nature of this doubly lensed quasar was confirmed by \citet{Courbin1998} and \citet{Remy1998}, who detected the lens, which is an early-type galaxy \citep{Courbin2000b}. \\

 \item \textit{PG~1115+080 (f)}. \citet{Weymann1980} first identified PG~1115+080 as a gravitationally lensed BAL (\emph{Broad Absorption Line}) quasar in the framework of the Palomar-Green survey (PG). At that time they only detected three point sources, while this system is indeed a quad \citep{Hege1981} with two very close-by images. An IR Einstein ring, distorted image of the quasar host galaxy, was detected for the first time by \citet{Impey1998}. According to \citet{Tonry1998}, the lens, which is elliptical \citep{Yoo2005}, is part of a group located at a redshift of $0.311\pm0.001$. The evidence for a dark matter substructure in this lens was studied by \citet{Chiba2005}, while \citet{Yoo2006} investigated its halo morphology. They found an upper limit of the order of 1 mas on the possible offset between the center of the light distribution and of the halo. \\

 \item \textit{JVAS~B1422+231 (g)}. This quadruple system was discovered in \citeyear{Patnaik1992} by \citeauthor{Patnaik1992} during the Jodrell/VLA Astrometric Survey. The lensed nature of JVAS~B1422+231 was first noticed by \citet{Lawrence1992} and then supported by \citet{Remy1993}. According to \citet{Tonry1998}, as first published by \citet{Kundic1997b}, the lens, an early-type galaxy \citep{Impey1996}, is part of a group at a redshift of $\rm z_{g}=0.339\pm0.002$. \\

 \item \textit{SBS~1520+530 (h)}. The Second Byurakan Survey allowed \citet{Chavushyan1997} to discover this doubly imaged BAL quasar. The lensing galaxy was identified by \citet{Crampton1998} and is elliptical \citep[e.g.][]{Auger2008}. \\

 \item \textit{CLASS~B1600+434 (i)}. This system is a two-image lensed quasar first identified as such by \citet{Jackson1995} in the context of the Cosmic Lens All-Sky Survey (CLASS). The primary lensing object is an edge-on spiral galaxy \citep{Jaunsen1997,Fassnacht1998}. \\

 \item \textit{CLASS~B1608+656 (j)}. This lensed quasar, which is the core of a post-starburst radio galaxy \citep{Fassnacht1996}, was discovered by \citet{Myers1995} as part of the Cosmic Lens All-Sky Survey. This object is quadruply lensed by two early-type galaxies located at the same redshift. The main lens belongs to a low-mass group of redshift $\rm z_{g}=0.631$ that is composed of eight confirmed members. Moreover, three other groups are located in the foreground of the primary lens: they contain 10 confirmed members each and are respectively located at $\rm z_{g_{1}}=0.265$, $\rm z_{g_{2}}=0.426$ and $\rm z_{g_{3}}=0.52$ \citep{Fassnacht2006}.\\

 \item \textit{HE 2149-2745 (k)}. This doubly lensed BAL quasar was discovered in \citeyear{Wisotzki1996} by \citeauthor{Wisotzki1996} in the context of the Hamburg/ESO Survey for bright quasars. The main deflector remained undetected until \citet{Lopez1998} and is an elliptical galaxy \citep[e.g.][]{Cosmograil6}. According to \citet{Faure2004} a galaxy group lies along the line of sight to this system.

\end{itemize}

\begin{table}
\centering 
\begin{tabular}{lcccccc}
\hline
\hglue -2mm Object & Obs. date & NF & $t_{exp}$ & NI & $\chi^{2}_{r,f}$ & $\chi^{2}_{r,l}$\\  
\hline
\hline
\hglue -2mm (a) JVAS~B0218+357 & 1997-08-19 & 4 & 43' & 6 & 23.11 & 1.00 \\
\hglue -2mm (b) SBS~0909+532 & 1997-11-07 & 4 & 47' & 3 & 12.70 & 1.43 \\
\hglue -2mm (c) RX~J0911.4+0551 & 1998-10-18 & 4 & 43' & 3 & 9.47 & 1.47 \\
\hglue -2mm (d) FBQS~J0951+2635 & 1998-03-19 & 4 & 43' & 6 & 60.37 & 0.94 \\
% (e) QSO 0957+561 & 98-05-30 & 4 & 47' & 4 & 1.45 \\
\hglue -2mm (e) HE~1104-1805 & 1997-11-22 & 4 & 43' & 4 & 30.26 & 1.61 \\
\hglue -2mm (f) PG~1115+080 & 1997-11-17 & 4 & 43' & 5 & 15.02 & 2.35 \\
\hglue -2mm (g) JVAS~B1422+231 & 1998-02-27 & 8 & 85' & 4 & 118.3 & 1.76 \\
\hglue -2mm (h) SBS~1520+530 & 1998-07-20 & 4 & 47' & 5 & 11.78 & 1.35 \\
\hglue -2mm (i) CLASS~B1600+434 & 1997-10-10 & 4 & 43' & 3 & 5.54 & 1.21 \\
\hglue -2mm (j) CLASS~B1608+656 & 1997-09-29 & 4 & 47' & 4 & 28.50 & 2.09 \\
\hglue -2mm (k) HE 2149-2745 & 1998-09-04 & 4 & 43' & 5 & 18.49 & 1.06 \\

\hline
\end{tabular}
\caption{General information about the HST/NIC2 images and ISMCS.}
\label{sum_dec11}
\end{table}

Three systems with measured time delays but lying in complex or rich environment have been excluded from our analysis: PKS 1830-211 \citep[e.g.][]{Lovell1998}, SDSS~J1004+4112 \citep{Fohlmeister2007,Fohlmeister2008}, QSO~0957+561 \citep[e.g.][]{Oscoz2001,Shalyapin2008,Fadely2010}. Two other systems are not presented here because we already published improved astrometry and time delays: WFI~J2033-4723 \citep{Cosmograil7} and HE~0435-1223 \citep{Courbin2011}.

\section{Iterative deconvolution}
\label{dec11}

\subsection{Method}

The MCS deconvolution algorithm \citep{MCS98} has proved to be very useful for disentangling the contribution of point sources and diffuse elements such as galaxies and arcs (see e.g. \citealt{Burud2000,Letawe2008}). The basic principle of this technique is the following: it does not try to reach an infinite resolution in the resulting deconvolved frame, in order to conform to the sampling theorem. Instead, it aims at obtaining an improved resolution in the final frame in deconvolving not by the total \textit{point spread function} (PSF) but by a partial kernel. Moreover, each frame is decomposed into two contributions: one from the point sources and the other one from the smooth structures in the form of a numerical component. In our case, the resolution of the final deconvolved frame, i.e. the shape of a point source, is chosen as a Gaussian with a full-width-at-half-maximum (FWHM) of two pixels, the final sampling step of which is two times smaller than the original one. The original pixel size with NICMOS depends on the observation date and ranges from 0\farcs075 to 0\farcs076. 

The most delicate point in deconvolution is probably the acquisition of a good PSF. To achieve this, we used a technique described in \citet{PSFsimult}, which allows us to add a numerical component to a first approximation of the PSF so that it fits the actual PSF. But to do this, some PSF stars are needed. In our case, no star is available in the field of NIC2 which, is not large: 19\farcs2$\times$19\farcs2. That is why we used the MCS algorithm in a slightly different way, i.e. coupled with an iterative strategy described in \citet{Chantry2007} and called \emph{ISMCS}. Here is how it works. All available frames for one object are simultaneously deconvolved a first time with PSFs synthesized by the Tiny Tim software \citep{Tinytim} aimed at producing PSFs for every configuration of the HST. These PSFs are satisfactory as a starting point, but not good enough for the level of accuracy we aim to reach \citep[see][ for more details]{Chantry2007}. This first deconvolution produces a first estimate of the diffuse structures which, once reconvolved to the initial resolution, can be subtracted from the original frames. We thus obtain new images with point sources less contaminated by extended components. We are then able to adjust our PSFs on these point sources and obtain a new set of PSFs that we once again use to simultaneously deconvolve our frames: we obtain a new background map (more accurate than the previous one) that we can subtract once more from the original frames. This process has to be repeated until a reasonable reduced chi square, $\chi^{2}_{\rm r}$, is reached on the image area of interest. In practice, we stopped the process when $\Delta \chi^{2}_{\rm r}<0.2$ between two consecutive iterations or when $\chi^{2}_{\rm r} \leq 1.0$. This $\chi^{2}_{\rm r}$ is defined as follows:
\begin{equation}
\chi^{2}_{\rm r} = \frac{1}{N}\Bigg( \frac{\mathcal{M}(\vec{x})-\mathcal{D}(\vec{x})}{\sigma(\vec{x})} \Bigg)^{2}, 
\label{residu_reduit}
\end{equation}
where $N$ is the number of pixels in the image or in the area of interest, $\mathcal{M}(\vec{x})$ is the solution reconvolved by the partial PSF, $\mathcal{D}(\vec{x})$ is the observed signal, and $\sigma(\vec{x})$ is the standard deviation associated to that signal.

\subsection{Results}
\label{Results_11}

The number of iterations \emph{NI} needed to reach convergence with ISMCS and the $\chi^{2}_{\rm r}$ after the first and last iterations (resp. $\chi^{2}_{r,f}$ and $\chi^{2}_{r,l}$) are presented in the three last columns of Table \ref{sum_dec11}. The original frame, i.e. the combination of the original F160W images, as well as the deconvolved frame, are displayed in Fig.~\ref{dec_NICMOS11} for each studied system. The labels of the lensed images are those of previous studies. These results were obtained with a combination of a smoothed numerical background and an analytic galactic profile, which allowed us to obtain complementary constraints on the shape of the light distribution in the lensing galaxies, i.e. PA of the semi-major axis in degrees positive east of north, the ellipticity $e$ ($e=1-b/a$), the effective semi-major axis ($a_{\rm eff}$), the effective semi-minor axis ($b_{\rm eff}$) and the effective radius ($R_{\rm eff}$). The latter is the geometric mean between the effective major and minor axes of the ellipse \citep{Kochanek2002}. These parameters are listed in Table~\ref{gal11} with their $\pm1 \sigma$ internal error bars, i.e. inherent to the deconvolution process (see below how they are obtained). As all the results, they come from the last iteration of ISMCS, but this time performed without the numerical channel to ensure that the maximum amount of light of the galaxy is included in the analytical profile. For early-type galaxies, we fitted a de Vaucouleurs model \citep{1948} and for late-type galaxies, i.e. for JVAS~B0218+357 and B1600+434, we fitted an ellipsoidal exponential luminosity profile \citep{Freeman1970}.

Also displayed in Fig.~\ref{dec_NICMOS11} are the mean residual maps\footnote{The mean residual map is the mean of the differences between the model, i.e. the deconvolved frame reconvolved to the initial resolution, and the individual data frames, in units of the standard deviation.} from the first iteration, i.e. with Tiny Tim PSFs, and last iteration of ISMCS, i.e. with improved PSFs. Their color scale ranges from -5$\sigma$ in black to +5$\sigma$ in white. The black rectangle drawn on each map represents the zone used to estimate the $\chi^{2}_{r}$. The orientation of these rectangles is different for each object as it is the one of the original images. The application of ISMCS removes nearly all artifacts left by the Tiny Tim PSFs. Though there are still some structures in the residual maps from the last iteration, there is no systematic structure underneath the point sources: another iteration of ISMCS would therefore not be able to improve the residuals, which are likely caused by PSF variations with the position on the detector.

\begin{figure*} [pht!]
\centering
\setcounter{subfigure}{0}
  \subfigure{\includegraphics[scale=0.3]{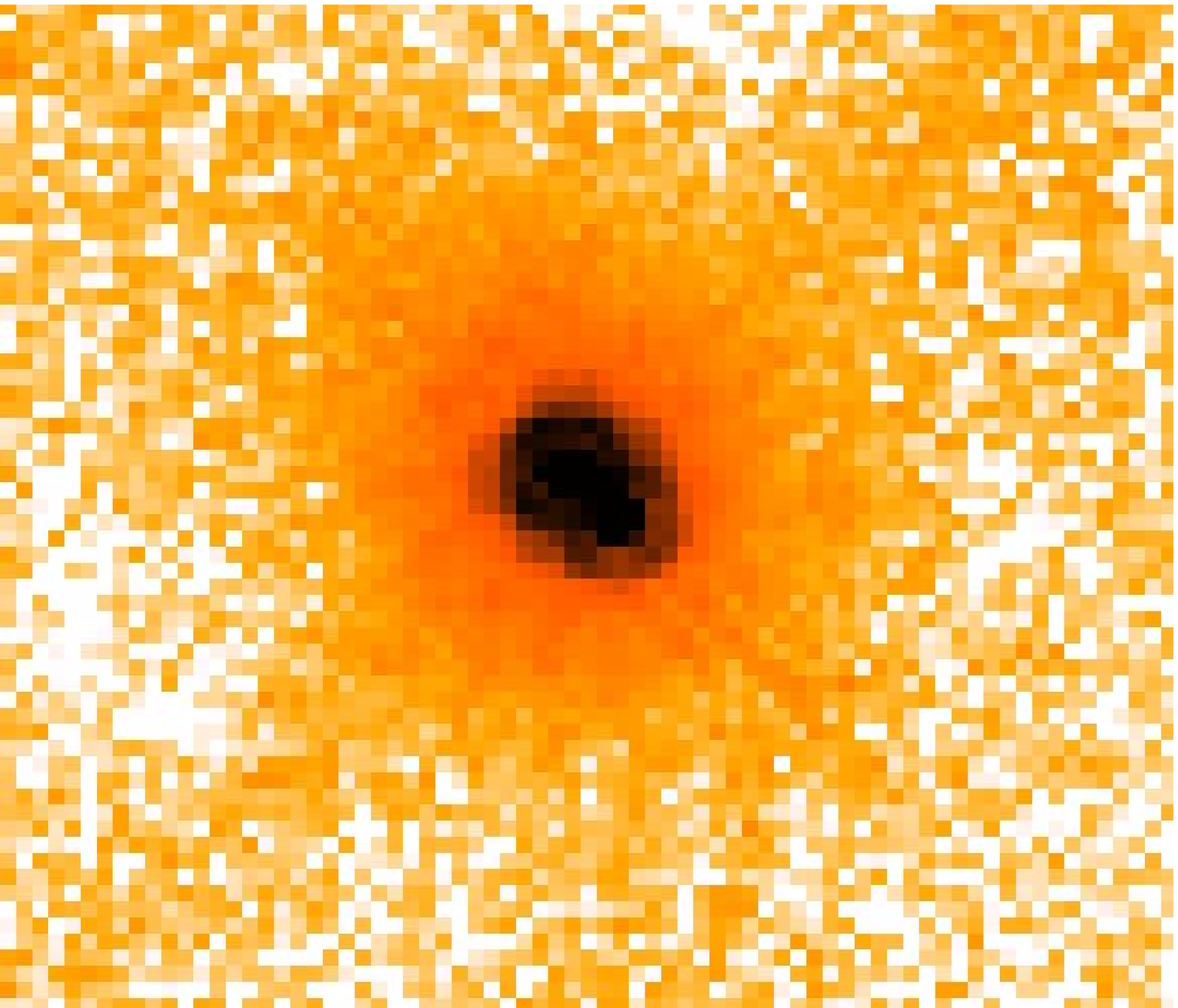}}                
  \subfigure{\includegraphics[scale=0.3]{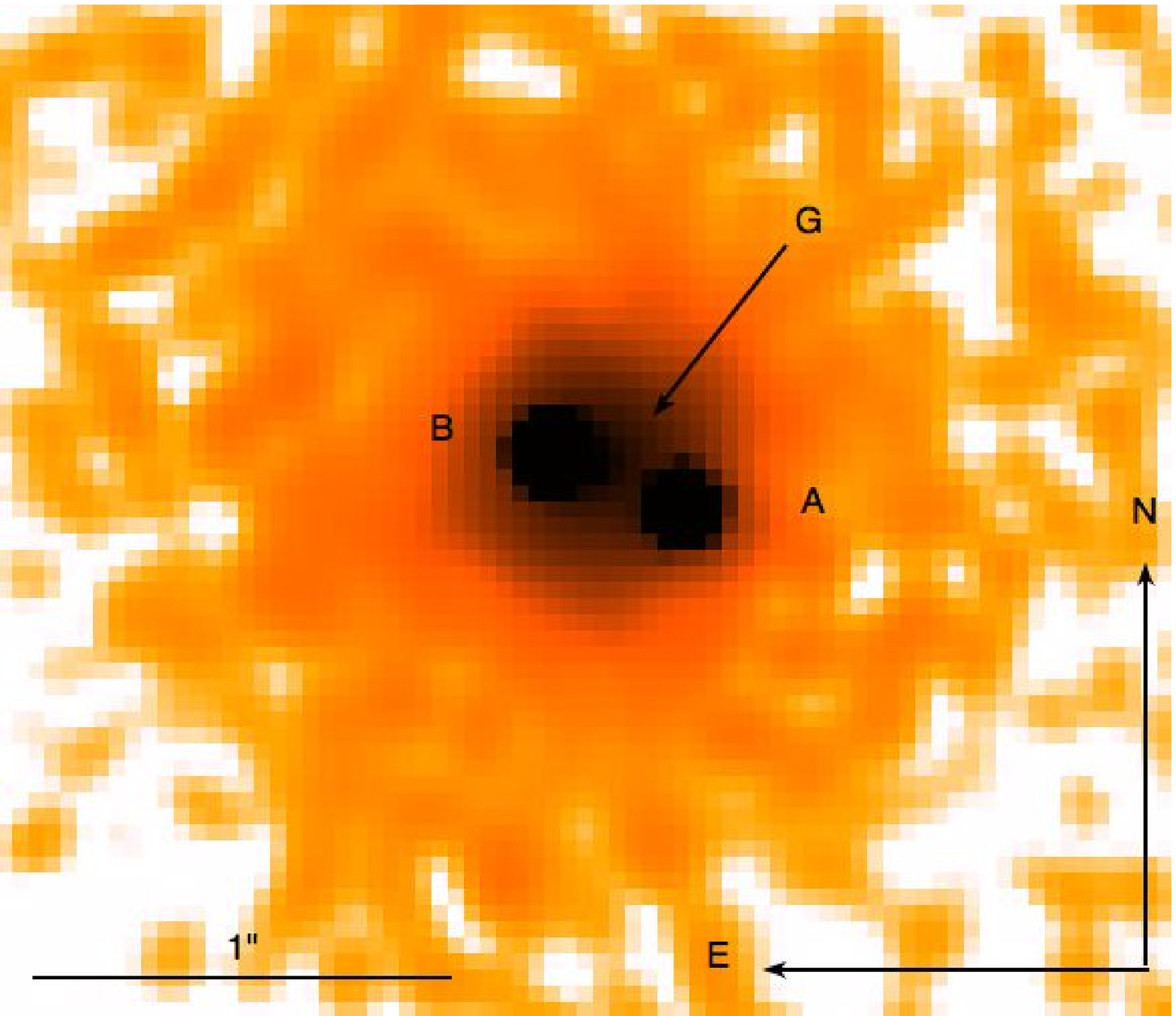}}
\setcounter{subfigure}{0}
  \subfigure[JVAS~B0218+357]{\includegraphics[scale=0.3]{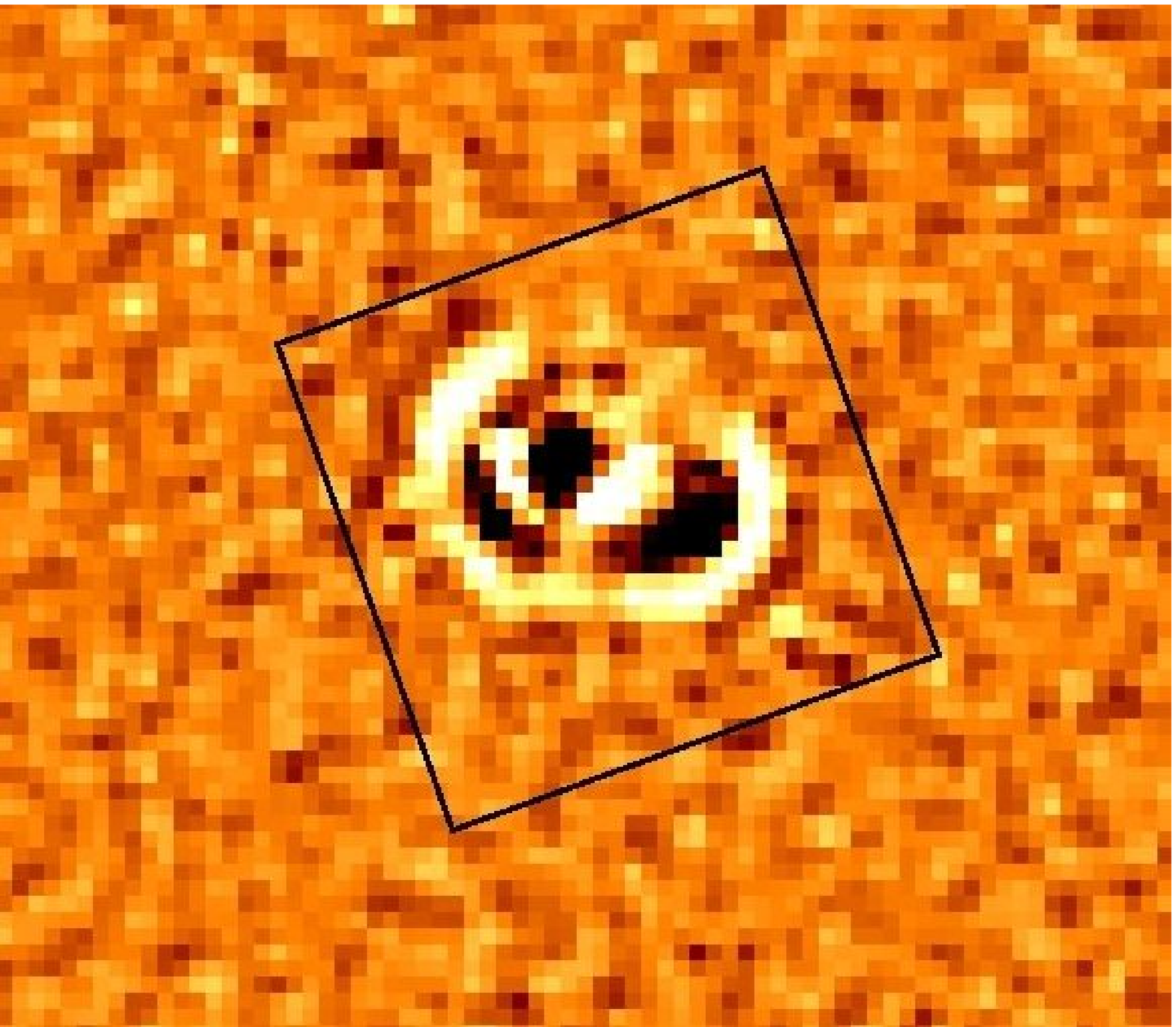}}
  \subfigure{\includegraphics[scale=0.3]{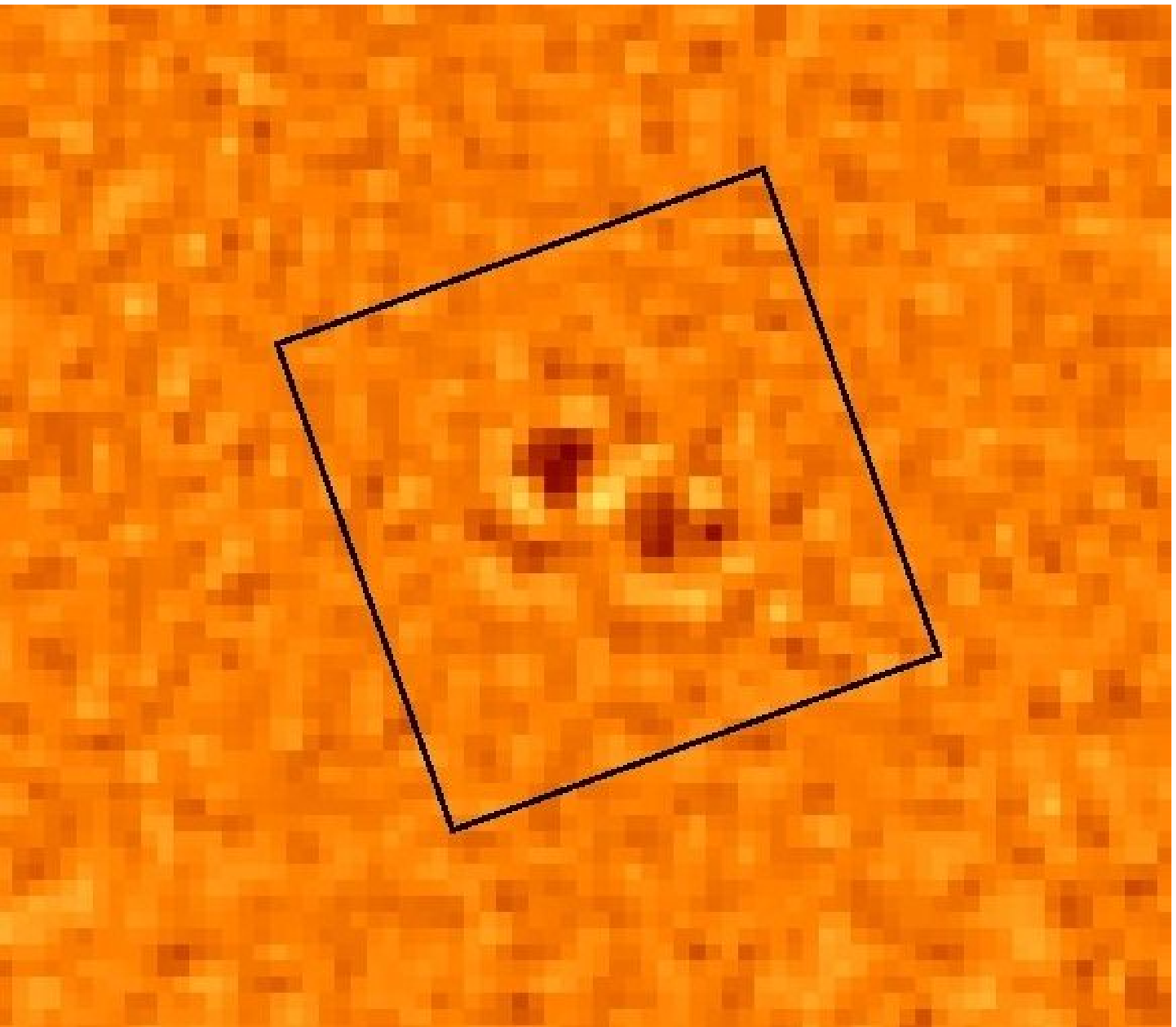}}
  \subfigure{\includegraphics[scale=0.3]{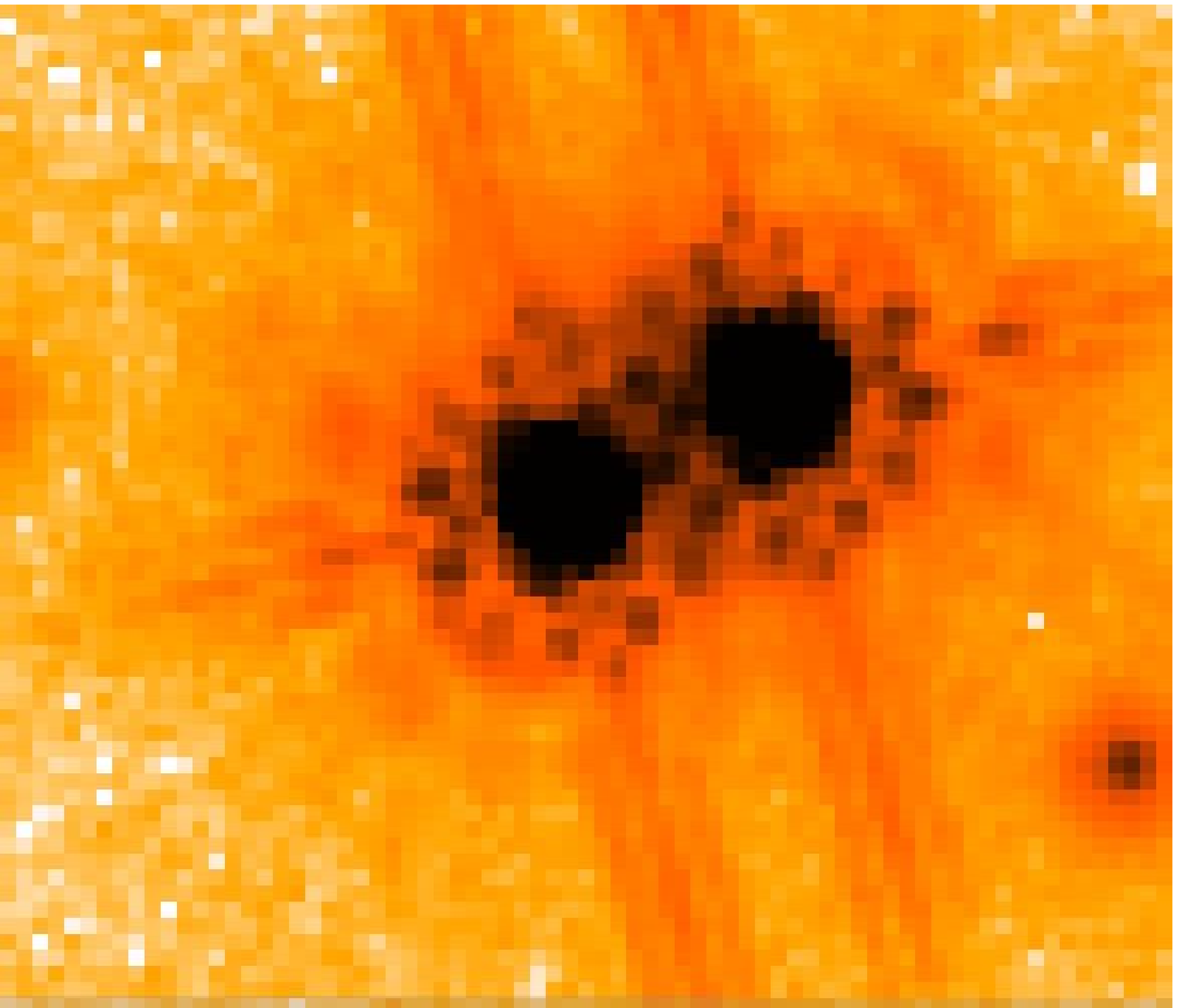}}                
  \subfigure{\includegraphics[scale=0.303]{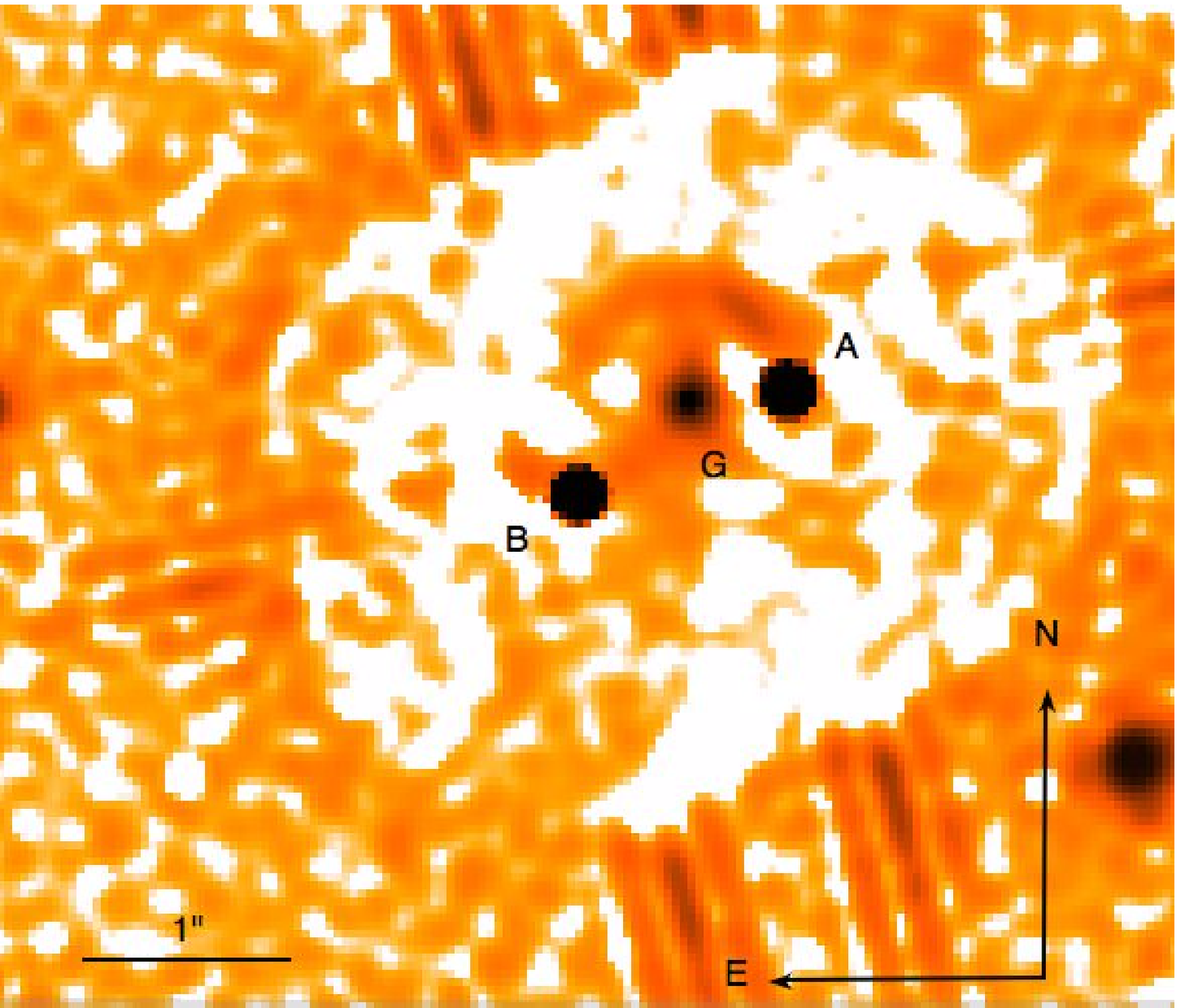}}
\setcounter{subfigure}{1}
  \subfigure[SBS~0909+532]{\includegraphics[scale=0.3]{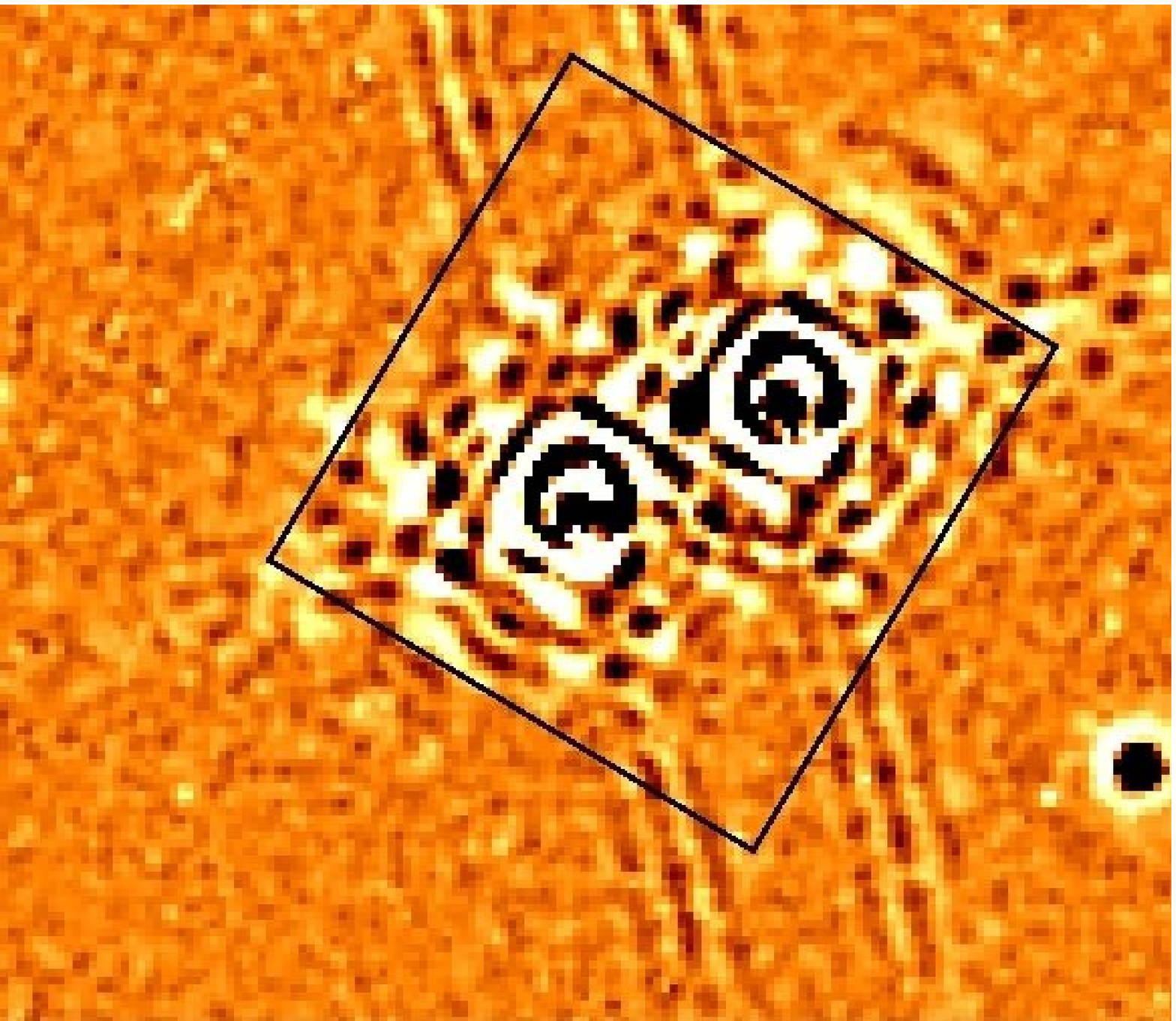}}
  \subfigure{\includegraphics[scale=0.3]{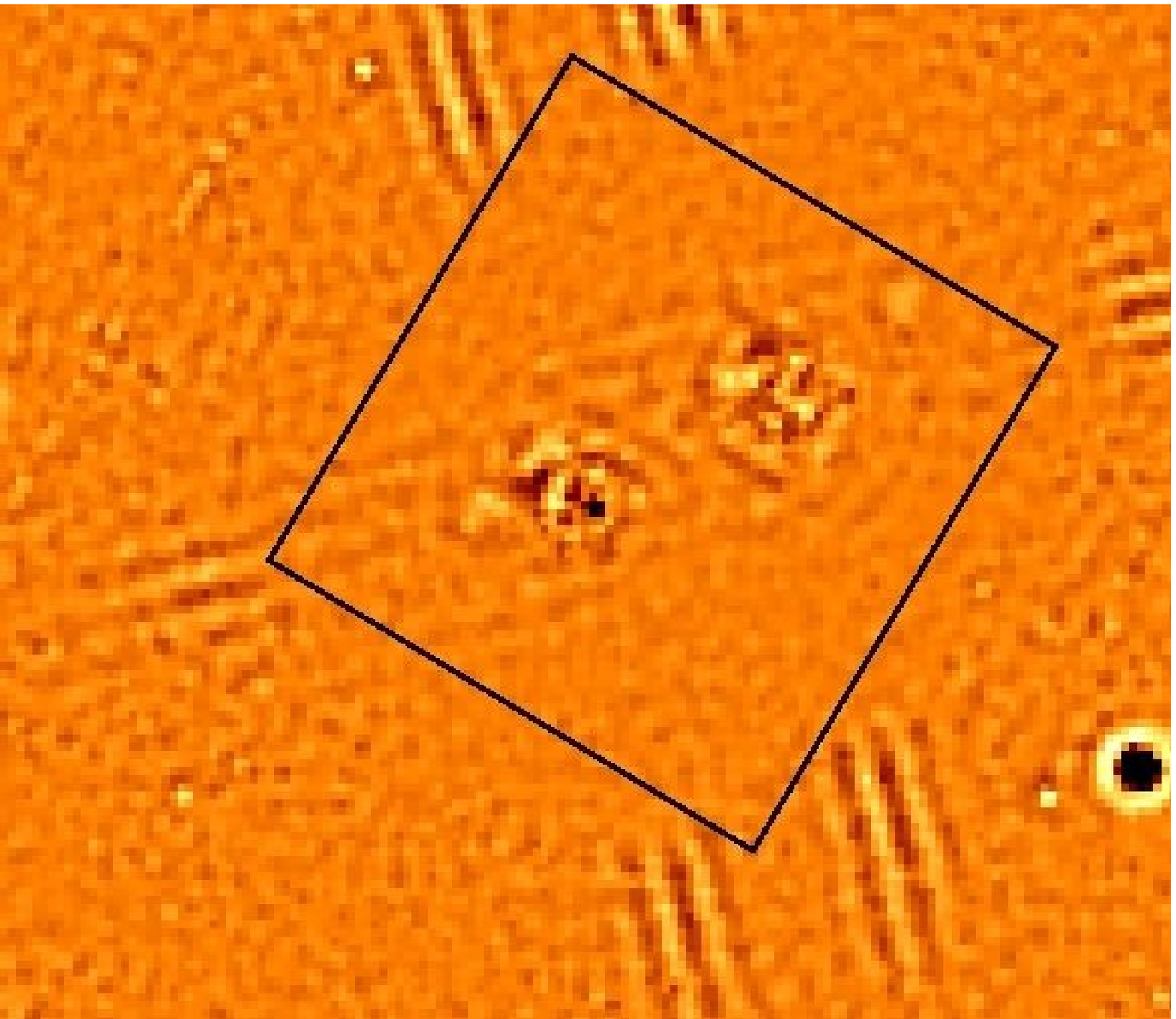}}
\caption{HST/NIC2 original and deconvolved frames in logarithmic flux scale (top left and top right respectively), mean residual maps (linear flux scale, $\pm 5\sigma$ with white corresponding to $> 5\sigma$) from the first and from the last iterations of ISMCS (bottom left and bottom right respectively).}
\label{dec_NICMOS11}
\end{figure*}

\begin{figure*} [pht!]
\centering
  \subfigure{\includegraphics[scale=0.3]{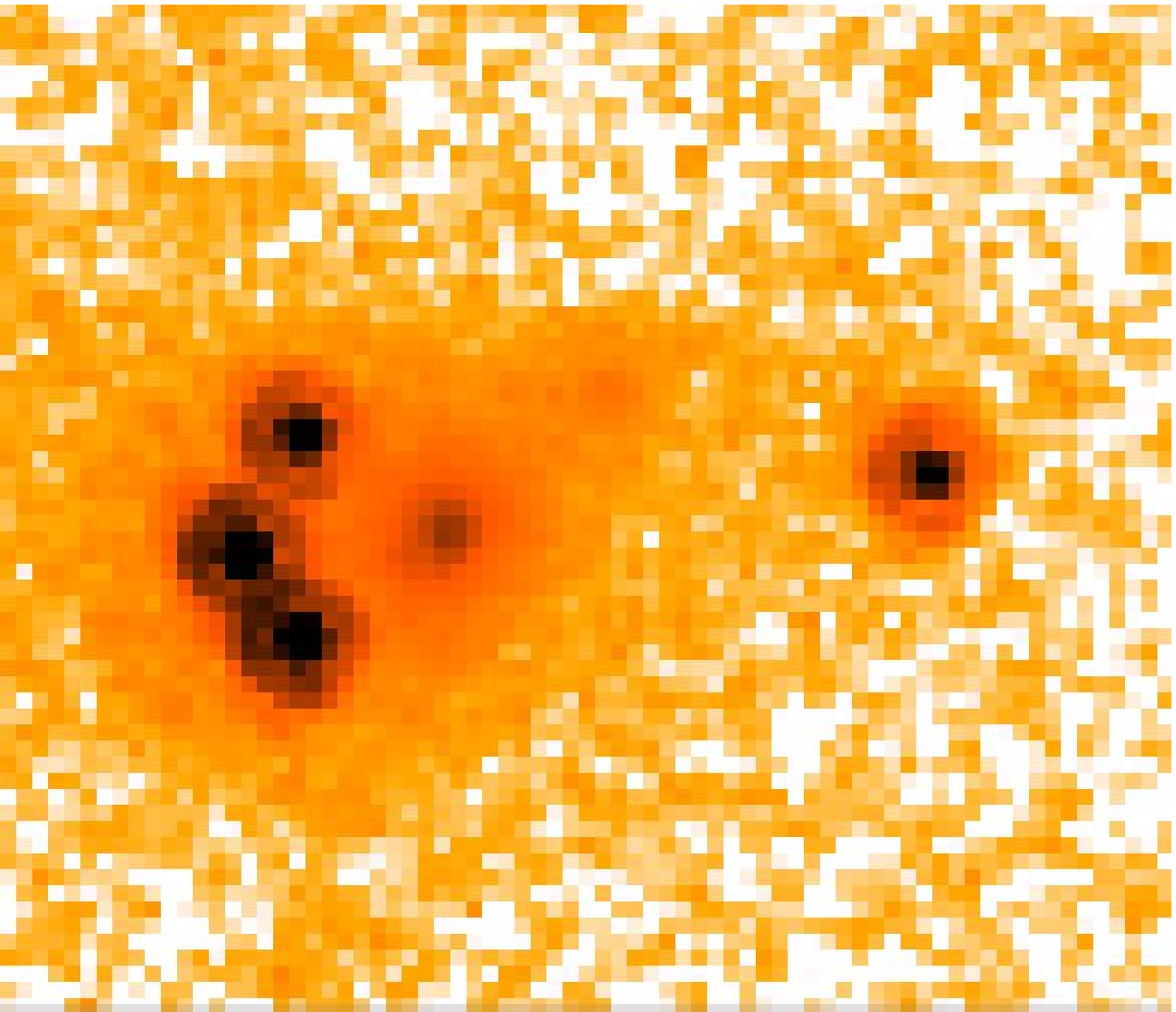}}                
  \subfigure{\includegraphics[scale=0.3]{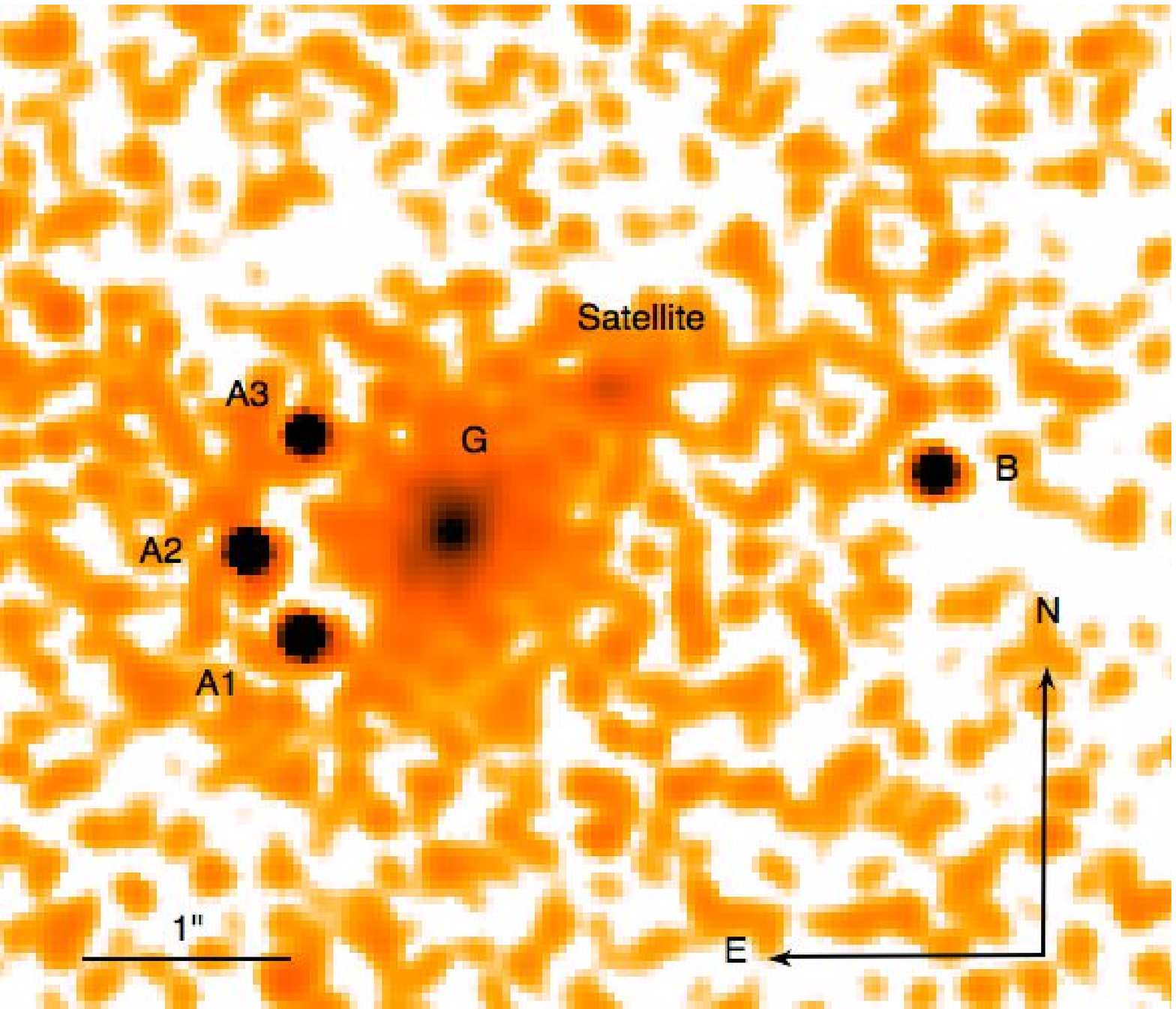}}
\setcounter{subfigure}{2}
  \subfigure[RX~J0911.4+0551]{\includegraphics[scale=0.3]{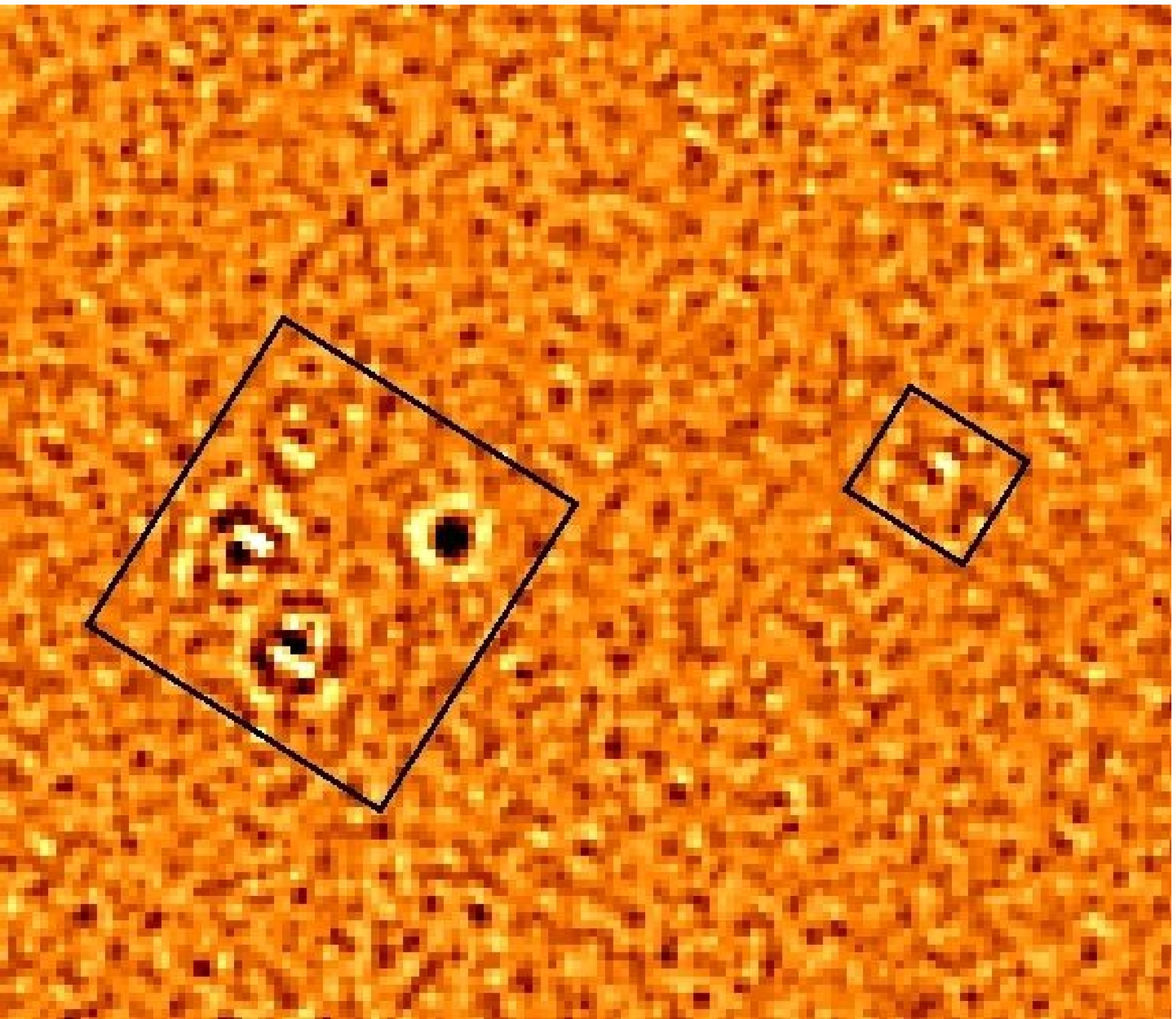}}
  \subfigure{\includegraphics[scale=0.3]{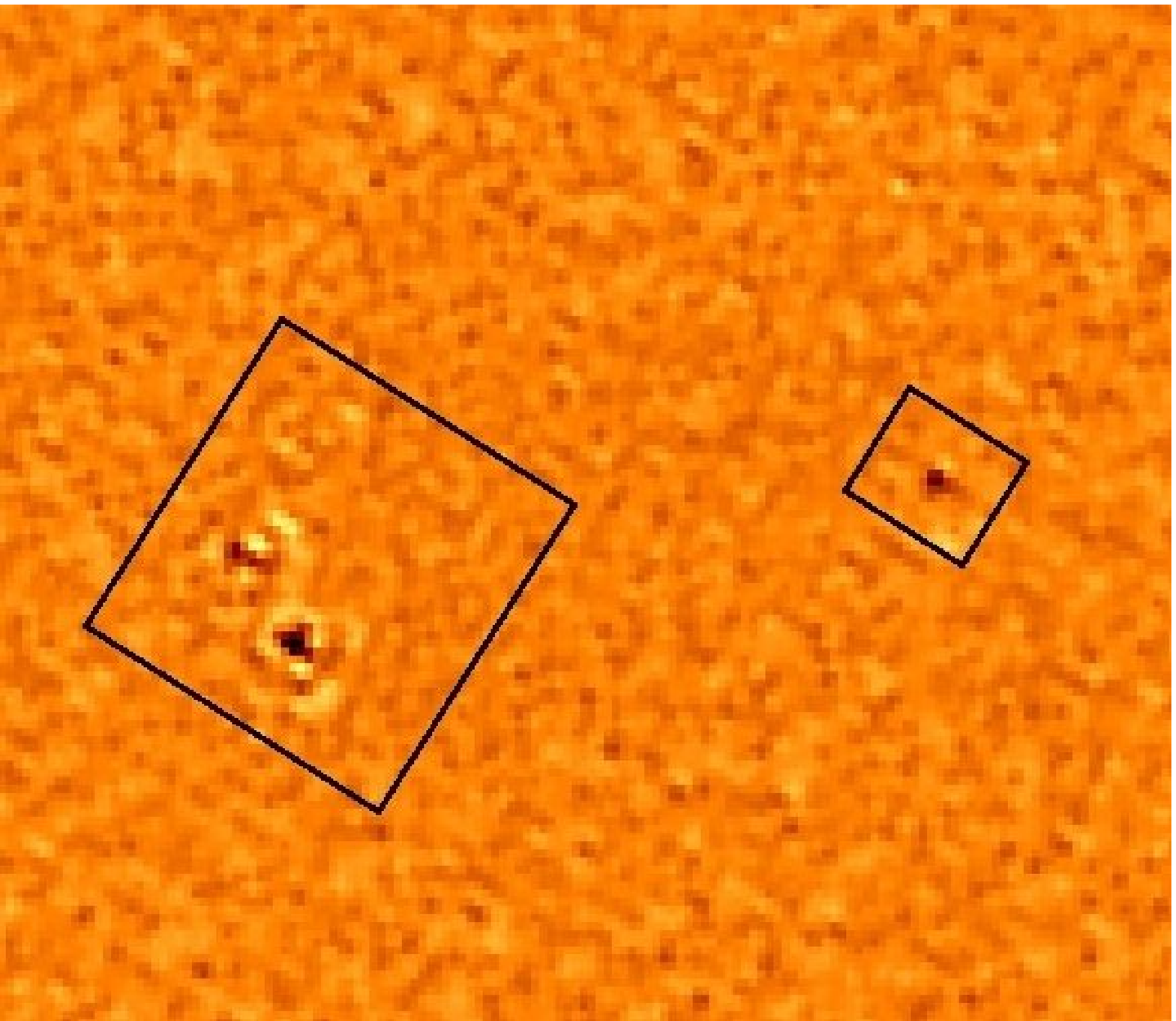}}
  \subfigure{\includegraphics[scale=0.2966]{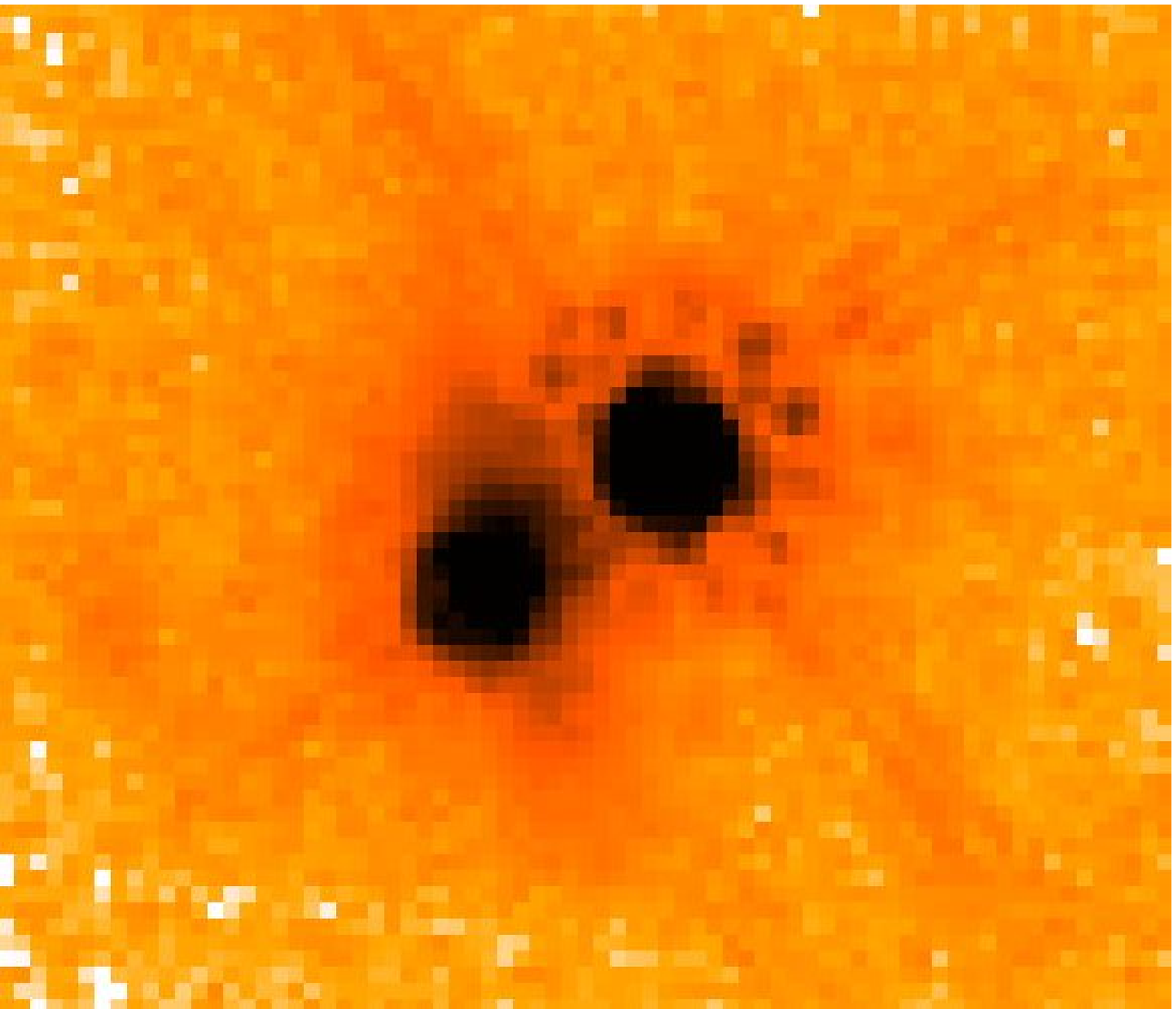}}                
  \subfigure{\includegraphics[scale=0.3]{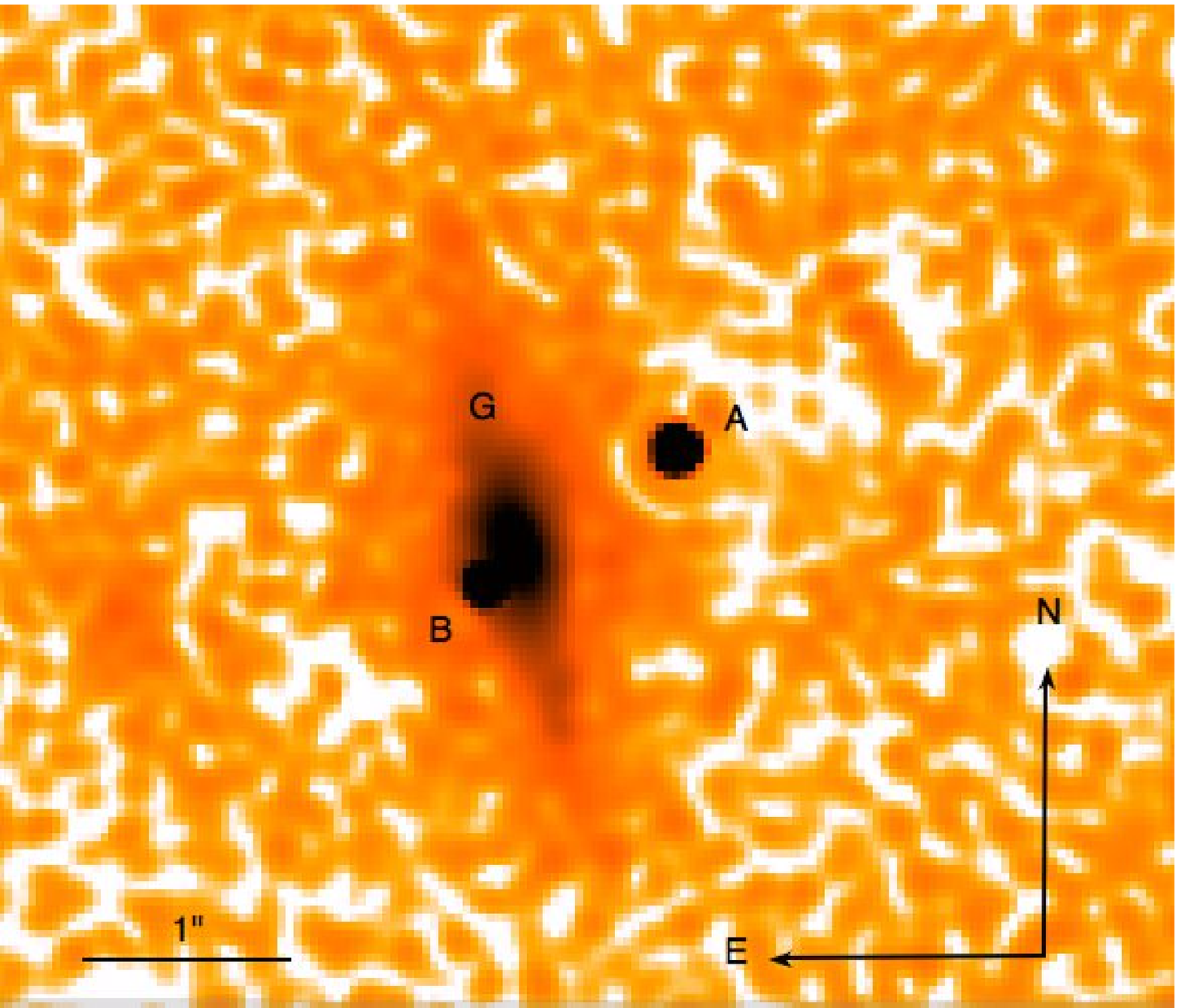}}
\setcounter{subfigure}{3}
  \subfigure[FBQS~J0951+2635]{\includegraphics[scale=0.305]{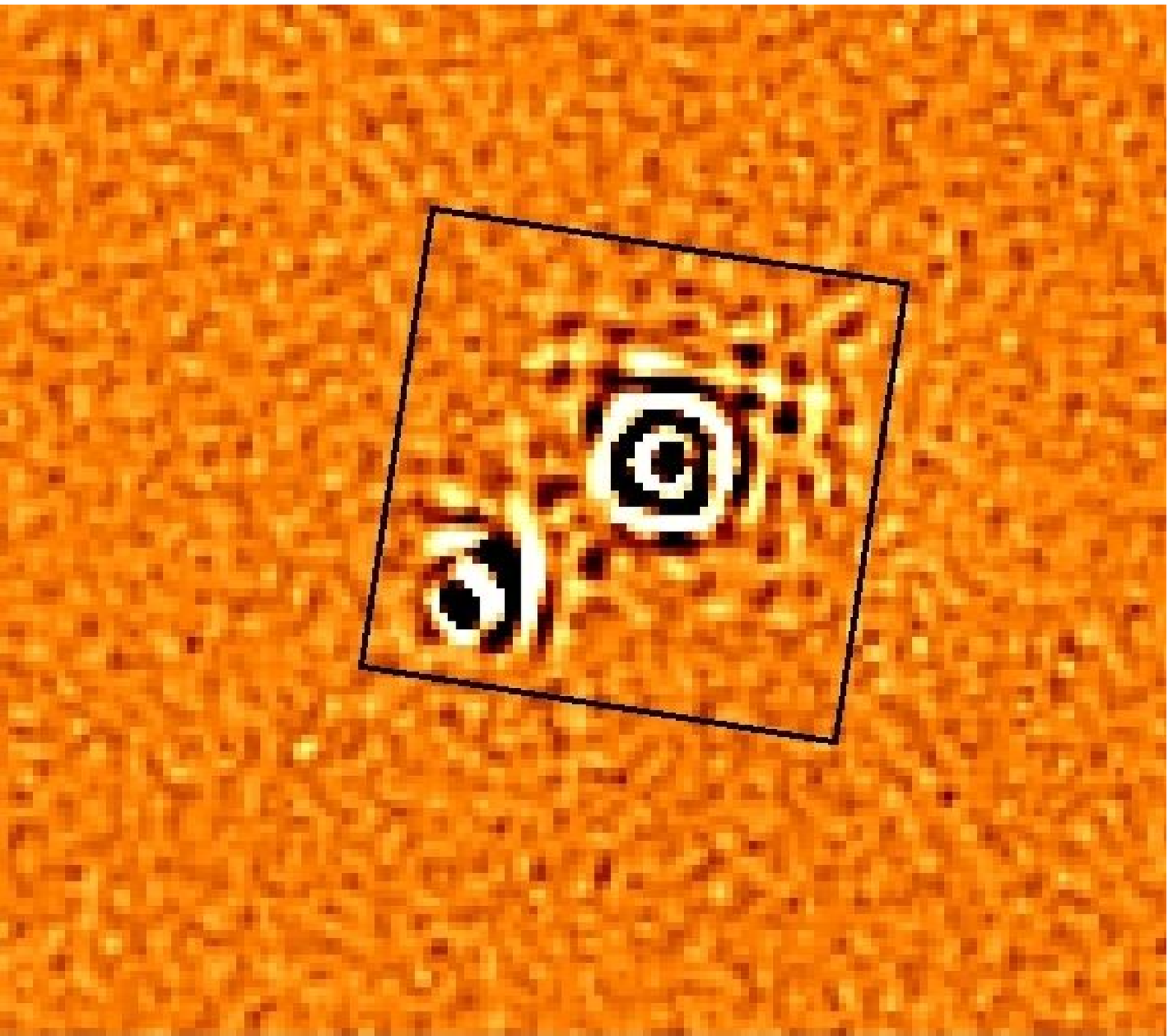}}
  \subfigure{\includegraphics[scale=0.305]{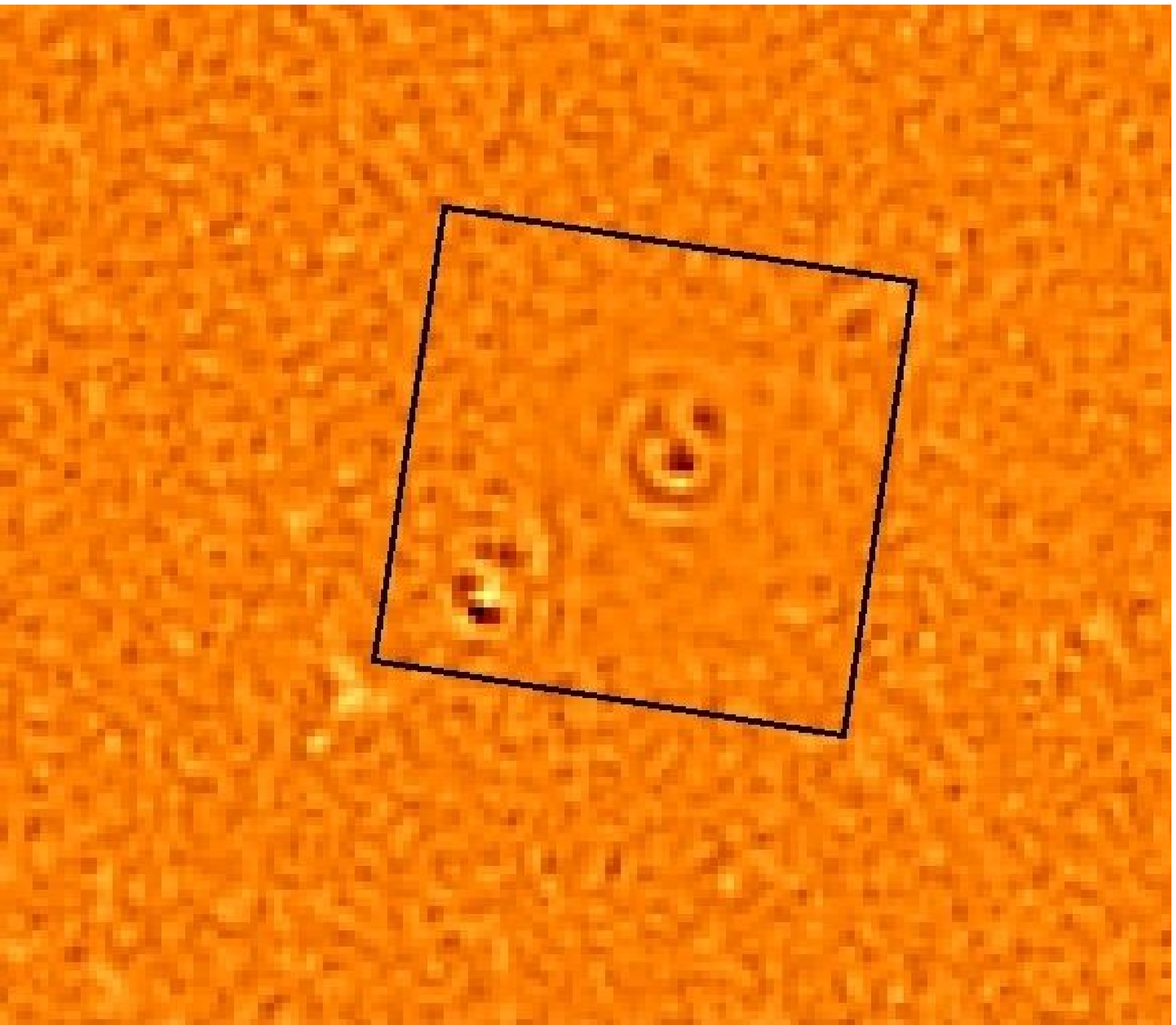}}
\setcounter{figure}{0}
\caption[]{continued.}
\end{figure*}

\begin{figure*} [pht!]
\centering
  \subfigure{\includegraphics[scale=0.3]{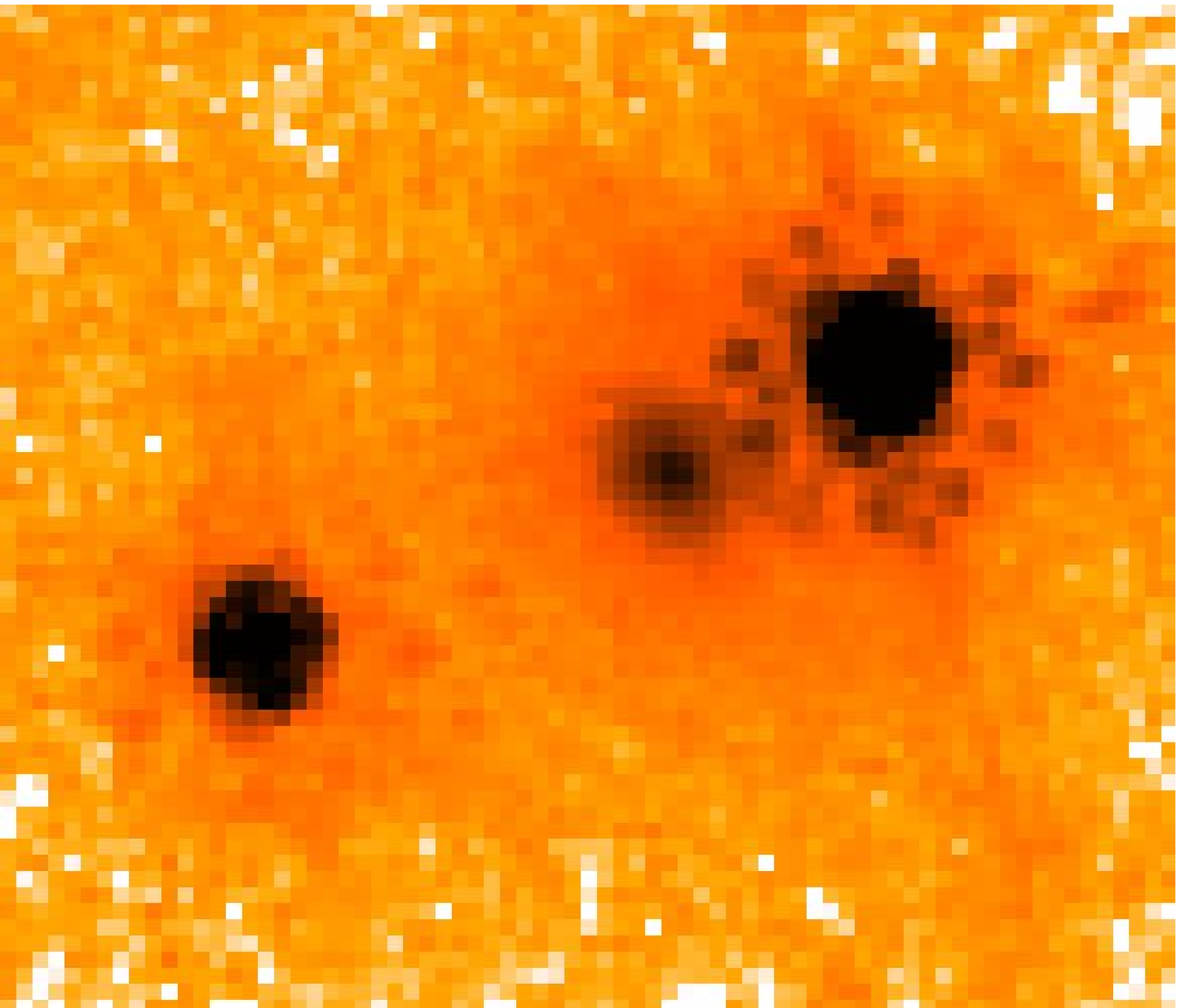}}                
  \subfigure{\includegraphics[scale=0.302]{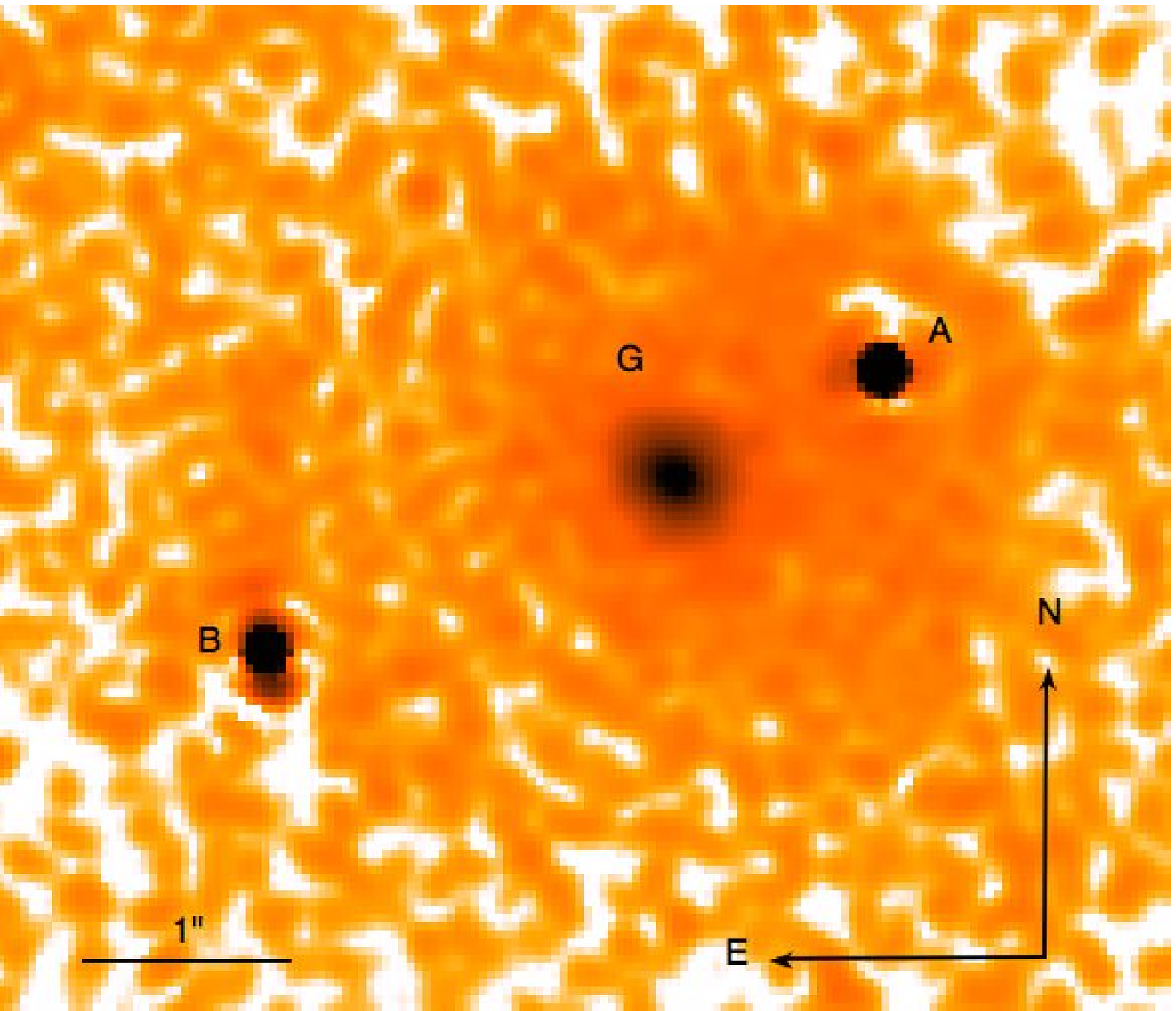}}
\setcounter{subfigure}{4}
  \subfigure[HE~1104-1805]{\includegraphics[scale=0.3]{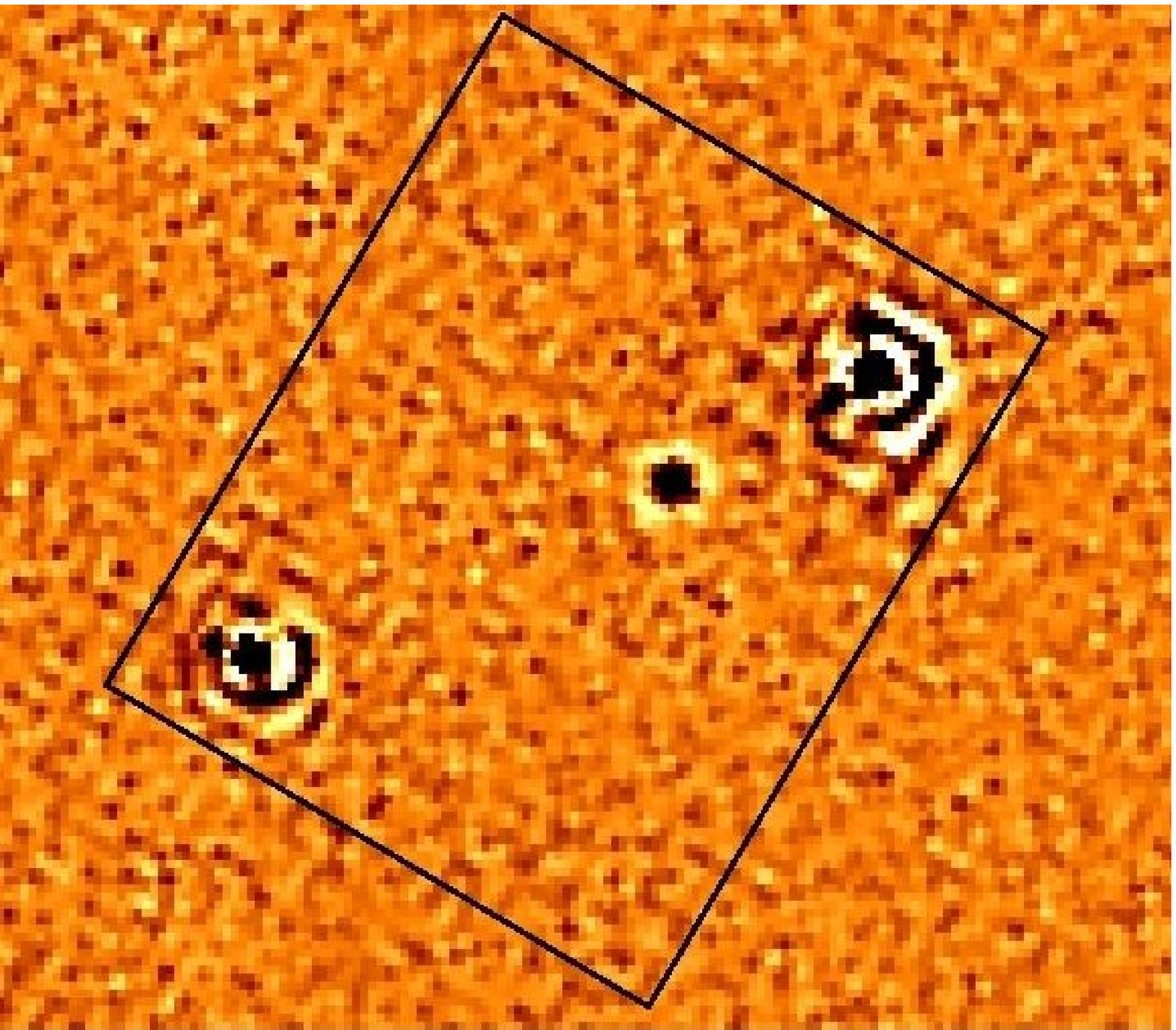}}
  \subfigure{\includegraphics[scale=0.3]{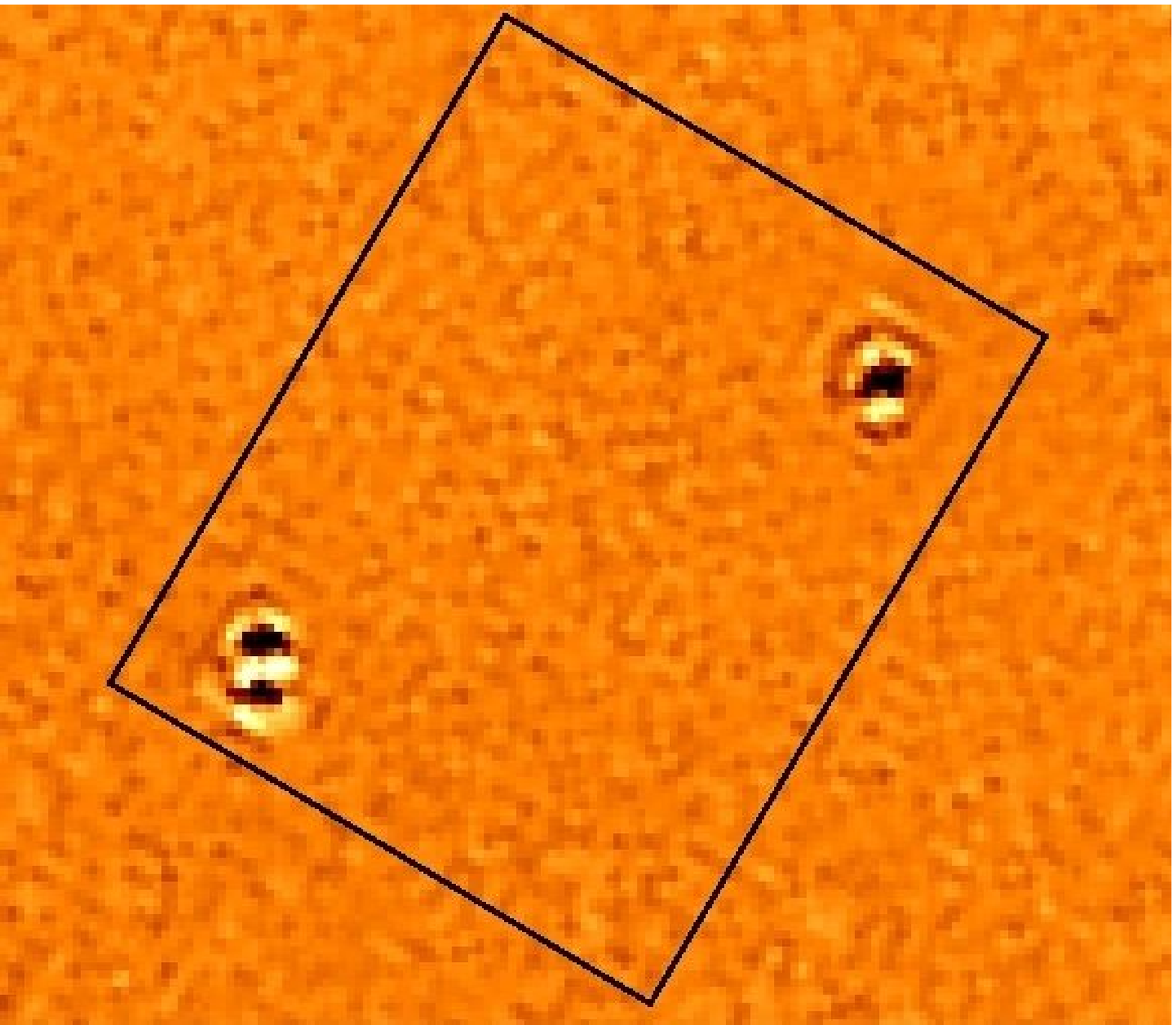}}
  \subfigure{\includegraphics[scale=0.2976]{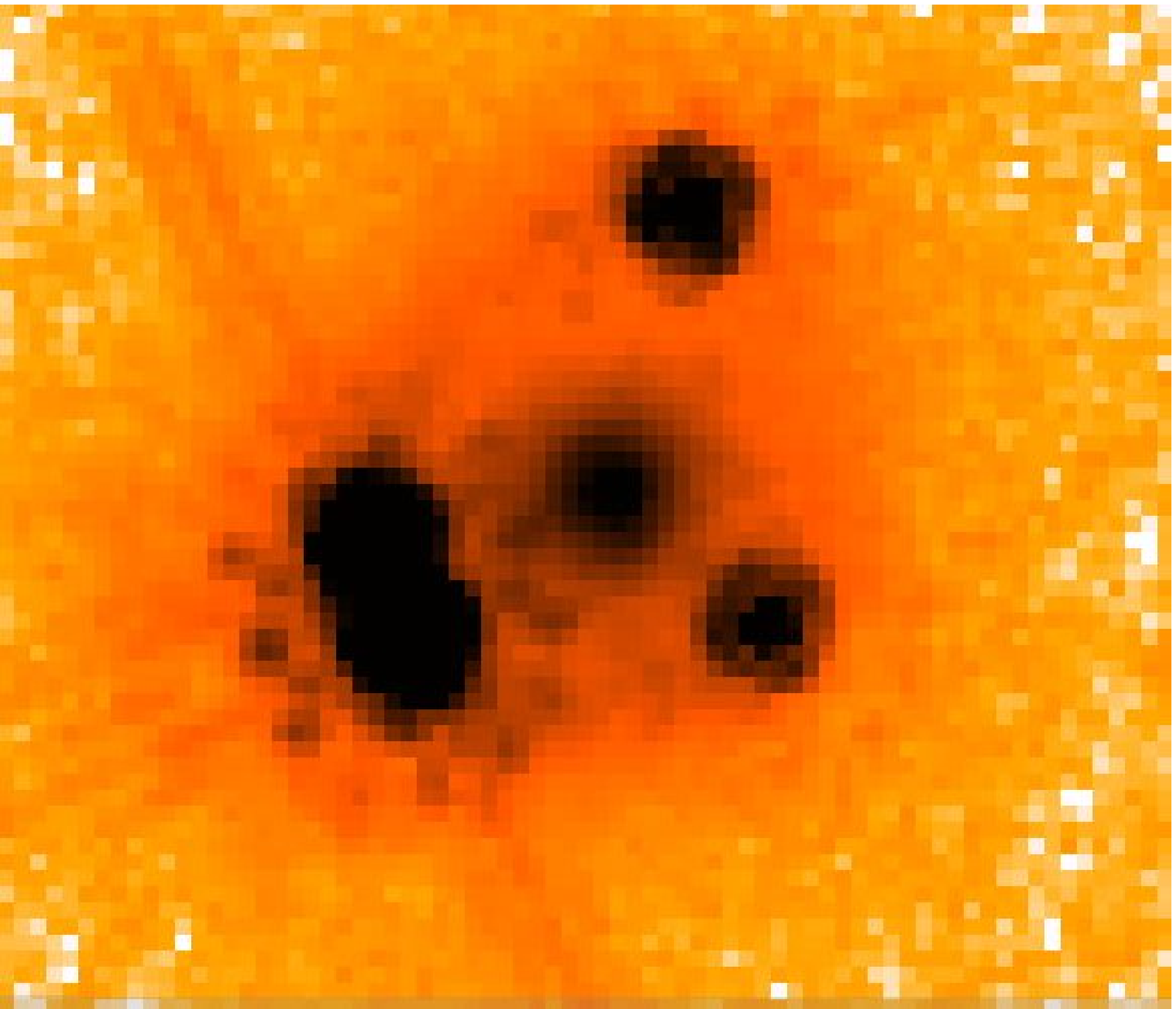}}                
  \subfigure{\includegraphics[scale=0.3]{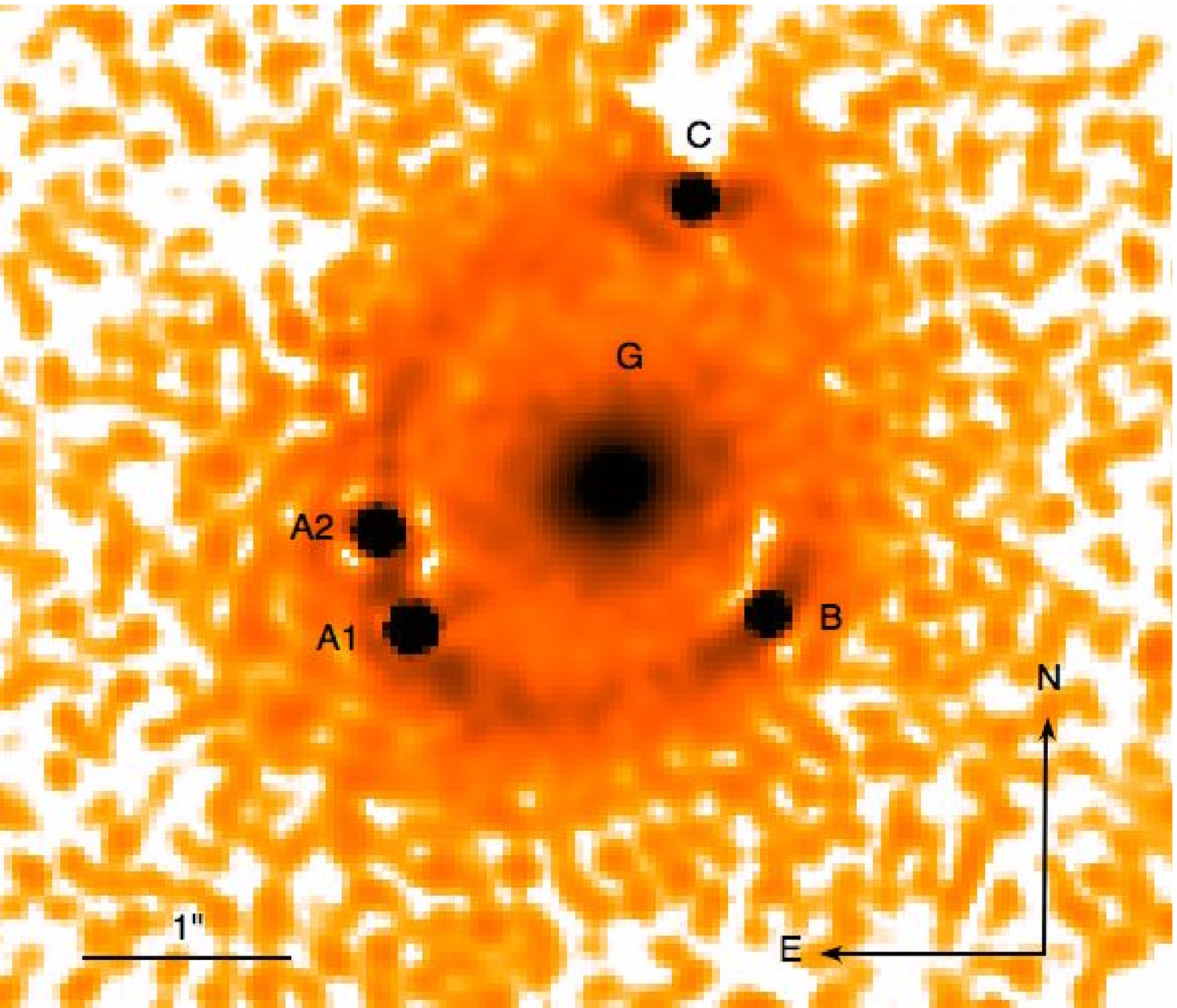}}
\setcounter{subfigure}{5}
  \subfigure[PG~1115+080]{\includegraphics[scale=0.3]{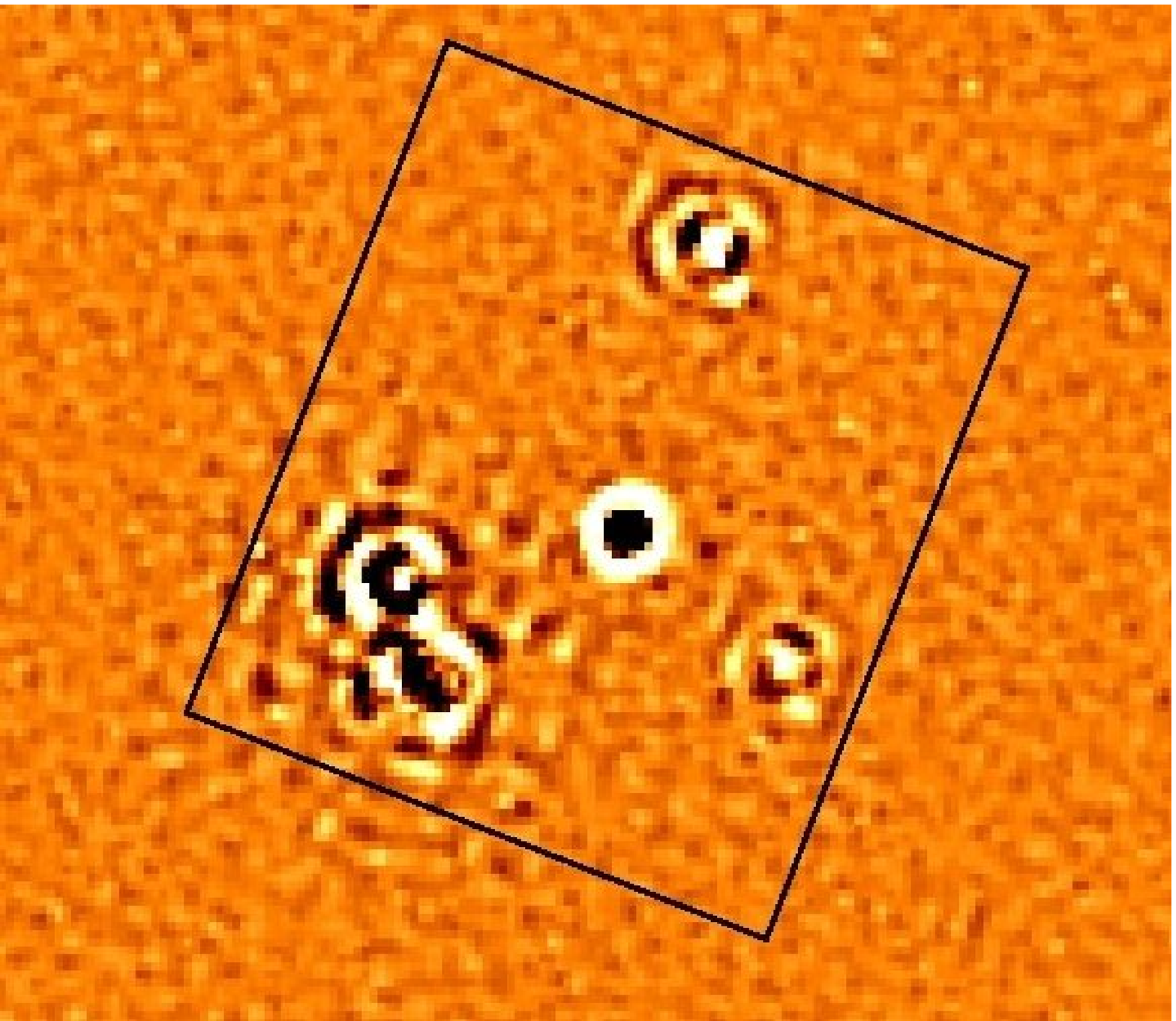}}
  \subfigure{\includegraphics[scale=0.3]{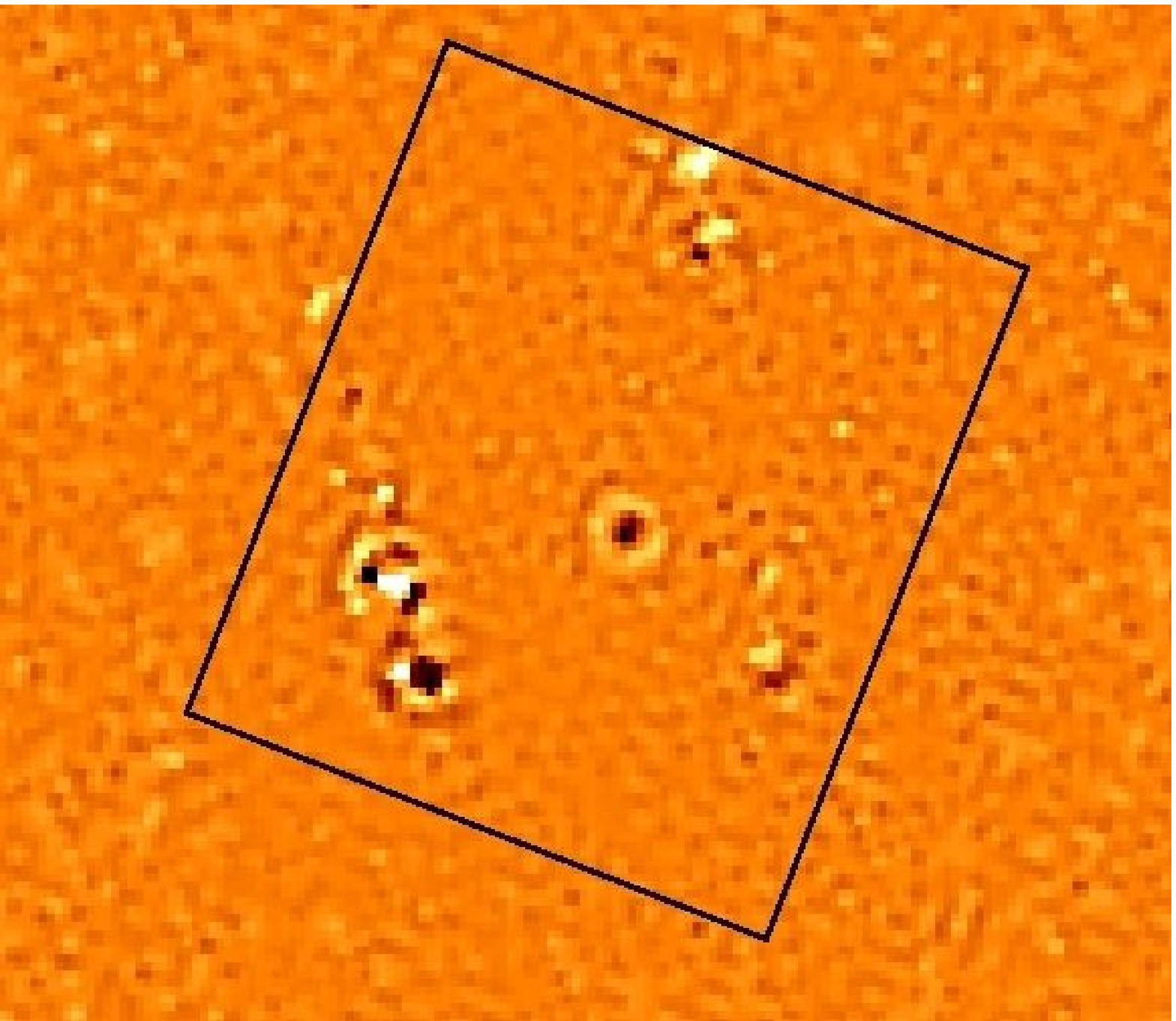}}
\setcounter{figure}{0}
\caption[]{continued.}
\end{figure*}

\begin{figure*} [pht!]
\centering
  \subfigure{\includegraphics[scale=0.302]{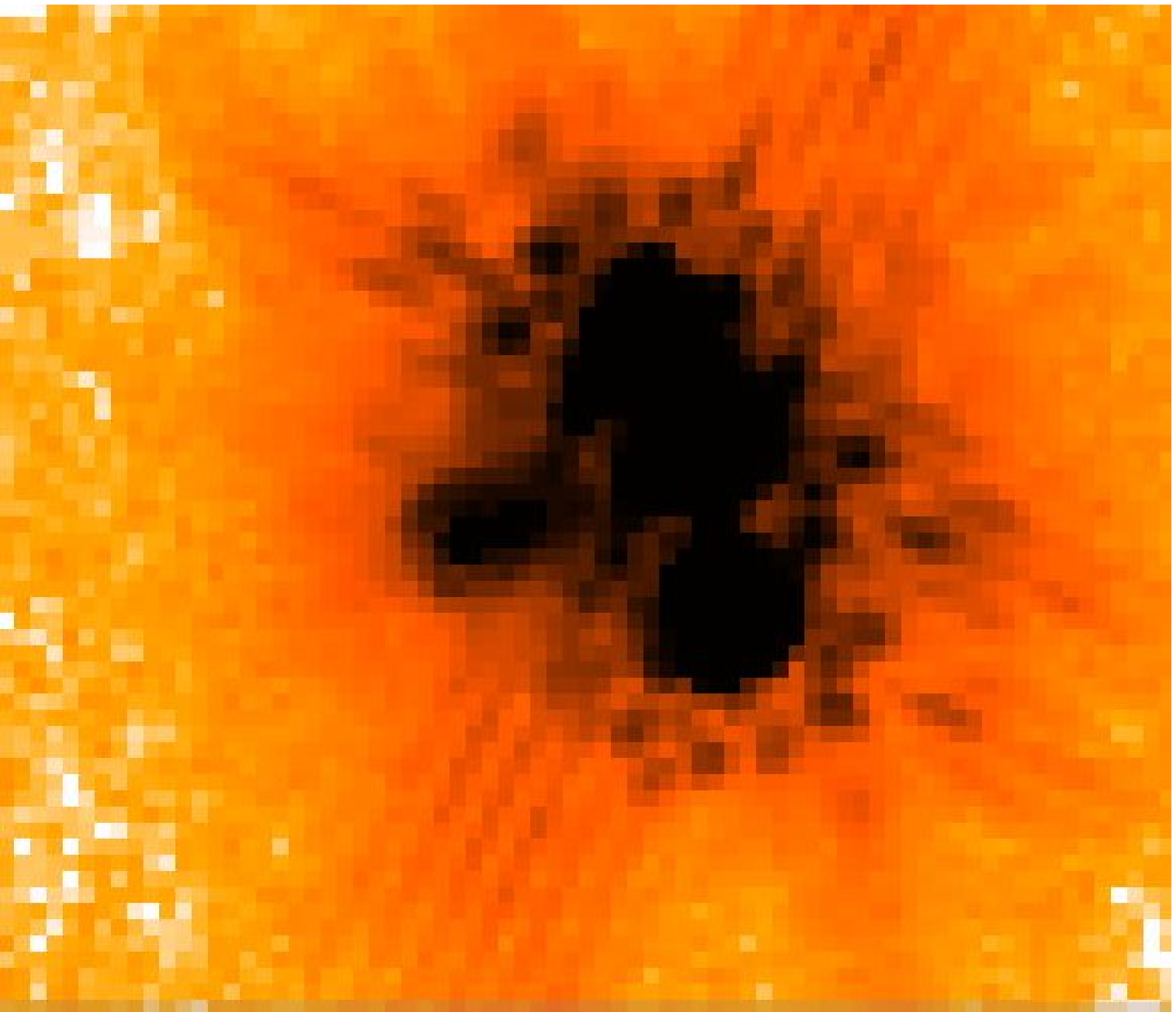}}                
  \subfigure{\includegraphics[scale=0.3]{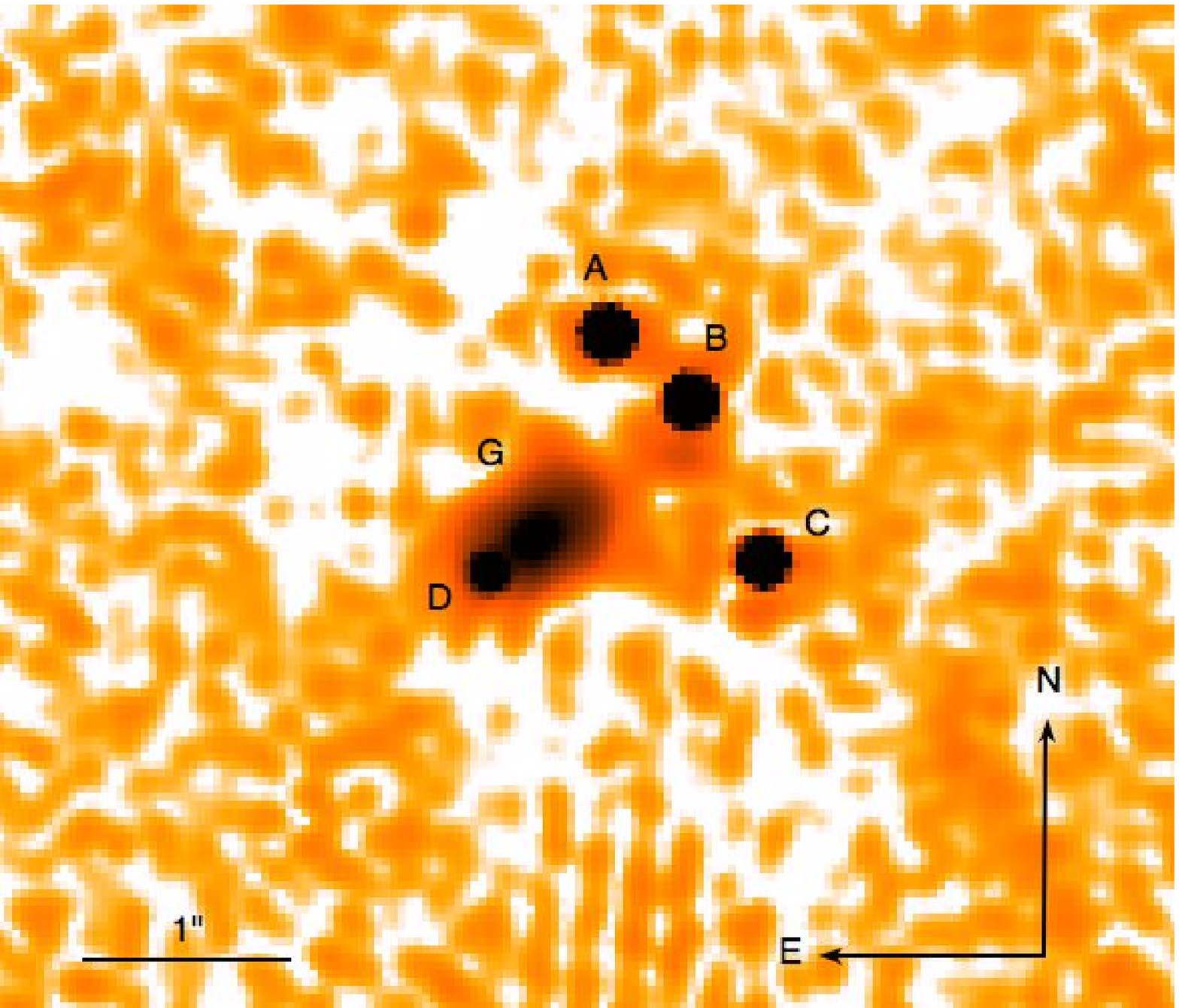}}
\setcounter{subfigure}{6}
  \subfigure[JVAS~B1422+231]{\includegraphics[scale=0.3]{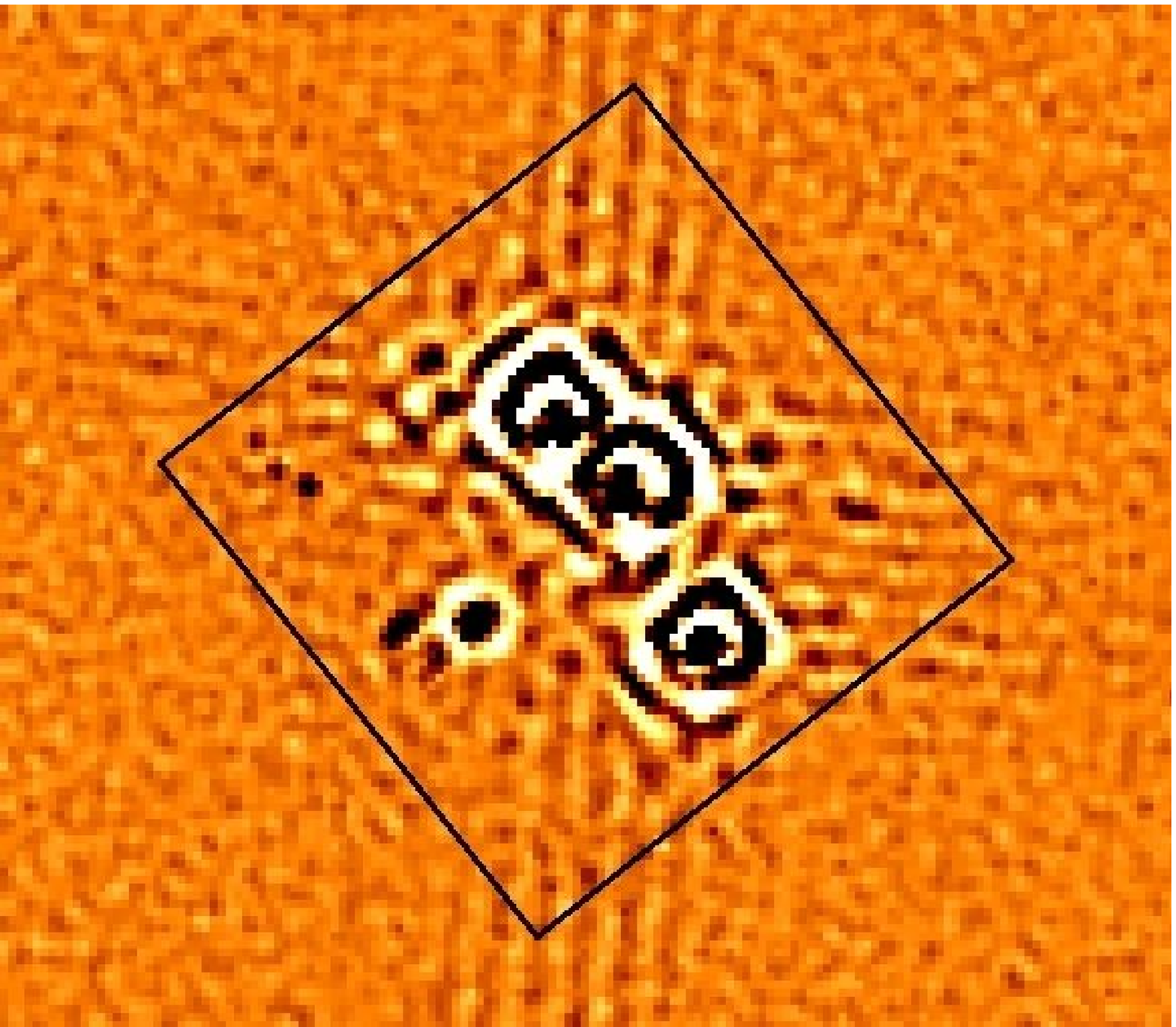}}
  \subfigure{\includegraphics[scale=0.3]{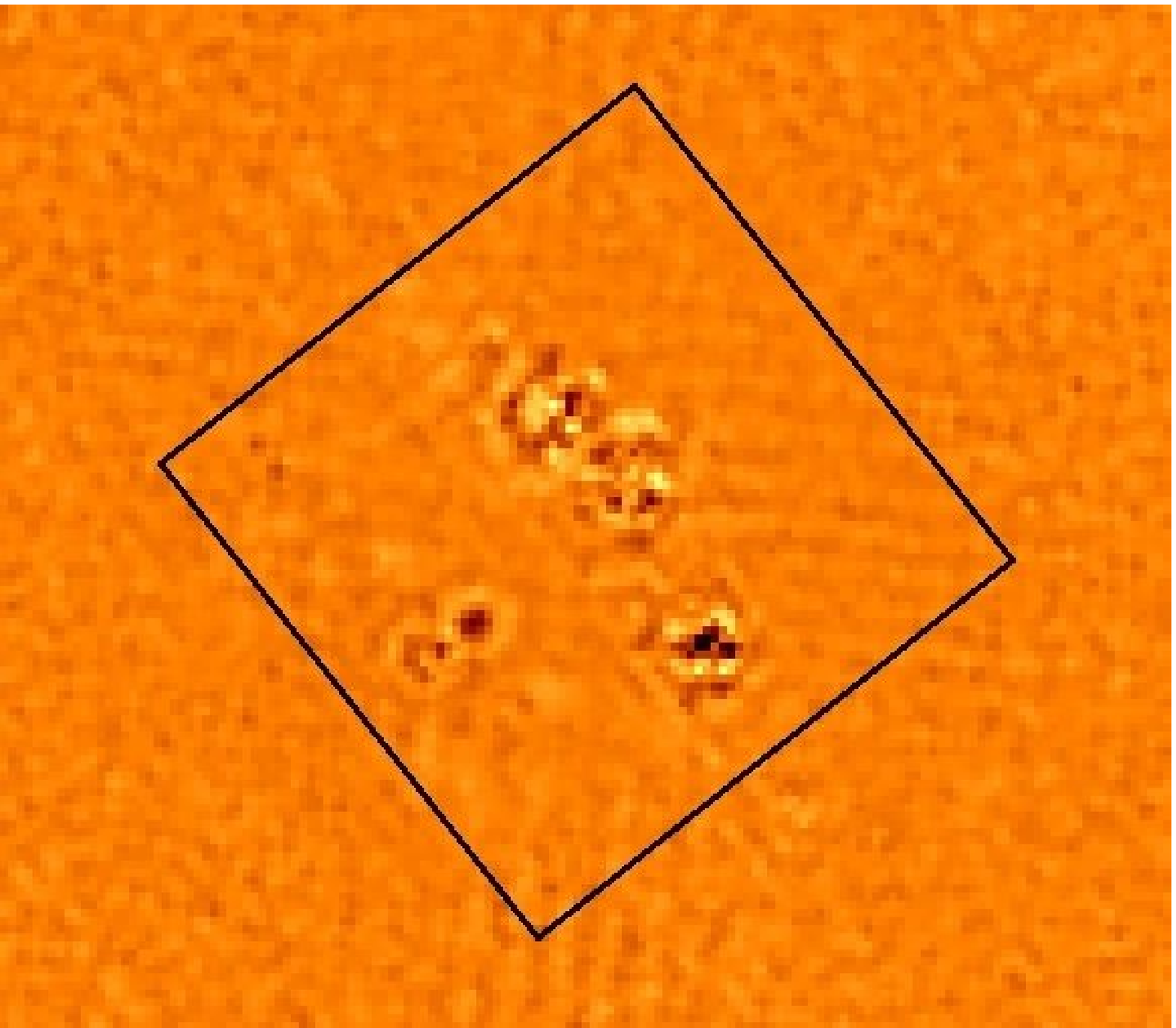}}
  \subfigure{\includegraphics[scale=0.3008]{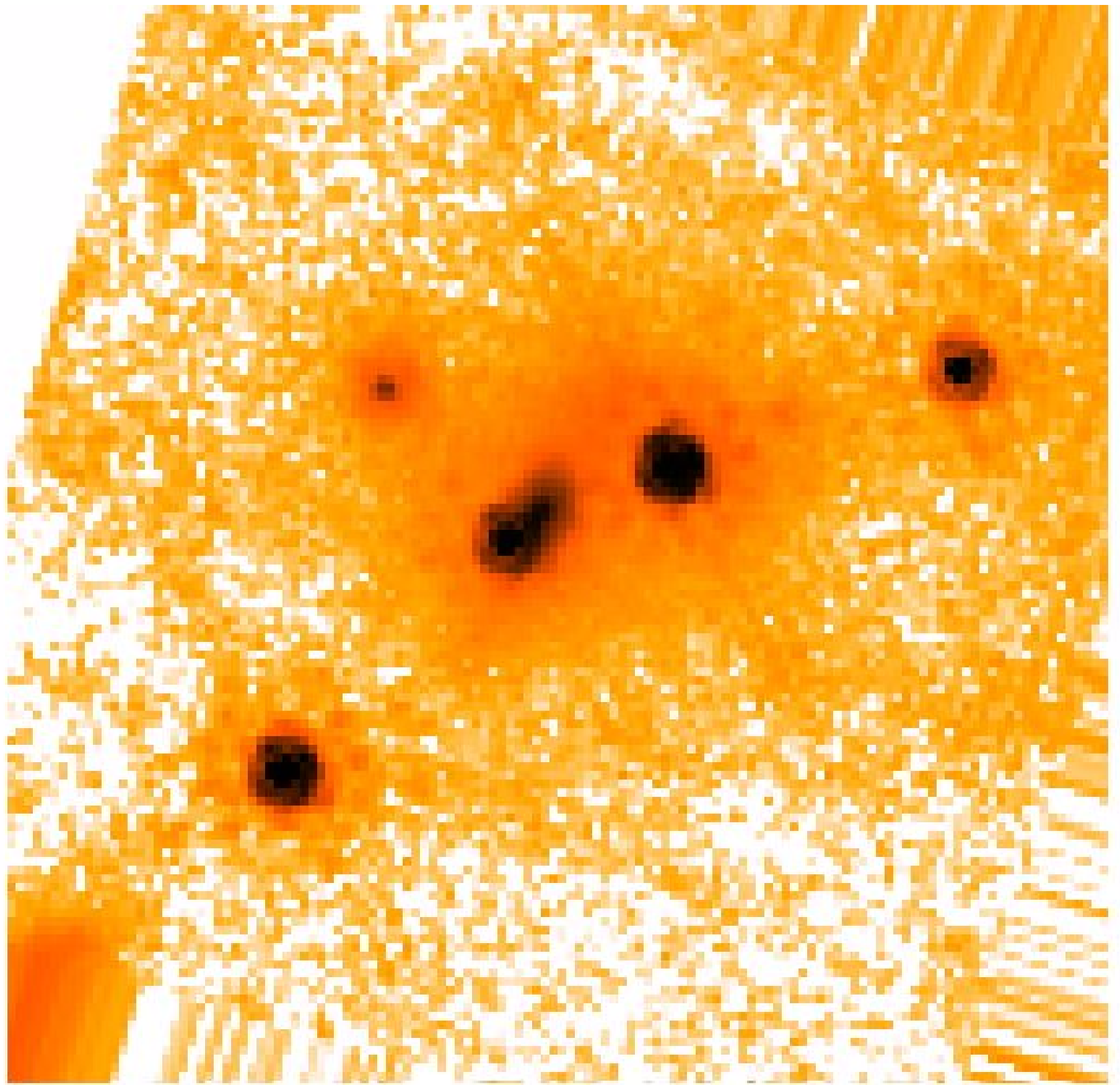}}                
  \subfigure{\includegraphics[scale=0.3]{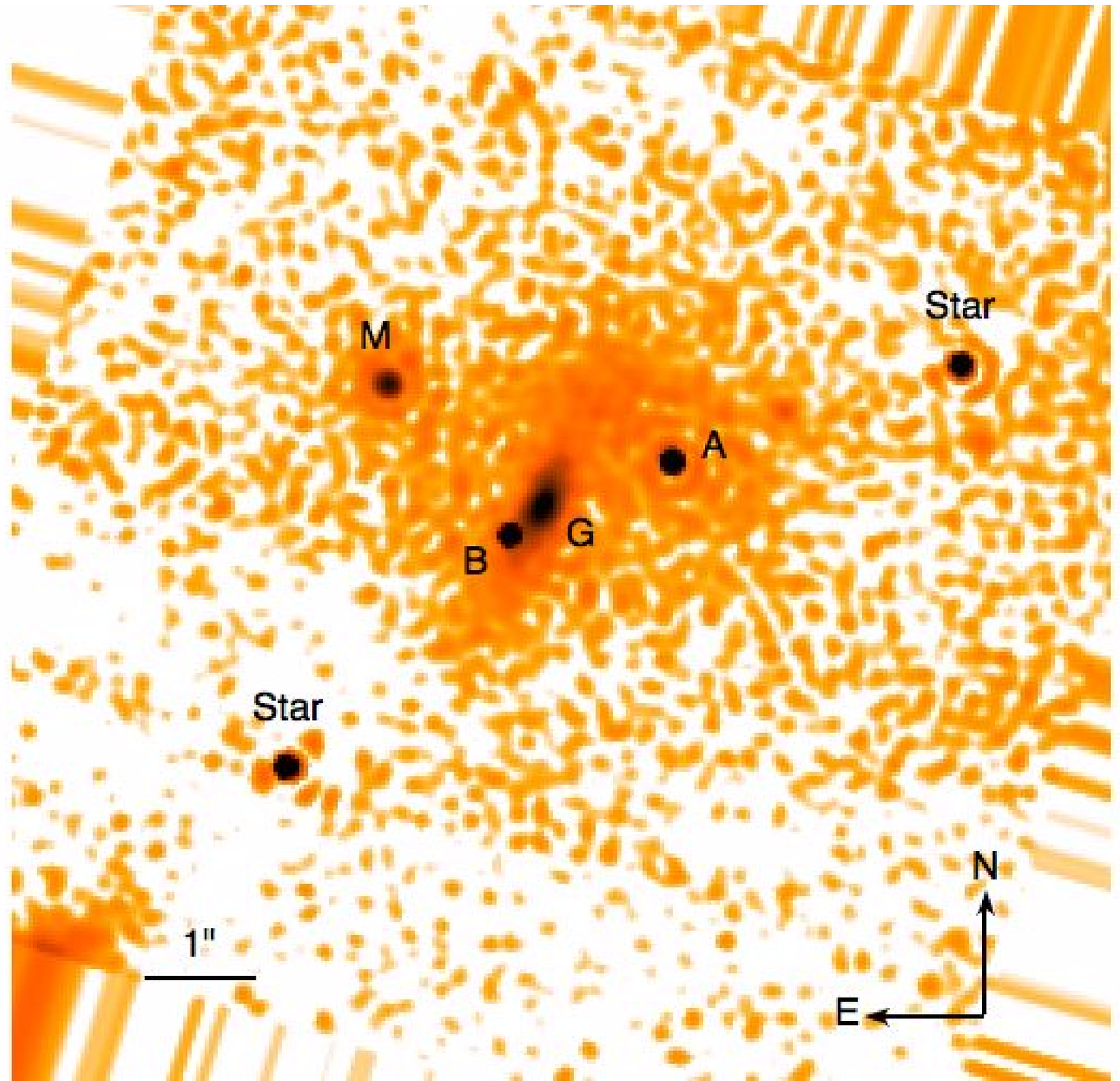}}
\setcounter{subfigure}{7}
  \subfigure[SBS~1520+530]{\includegraphics[scale=0.3]{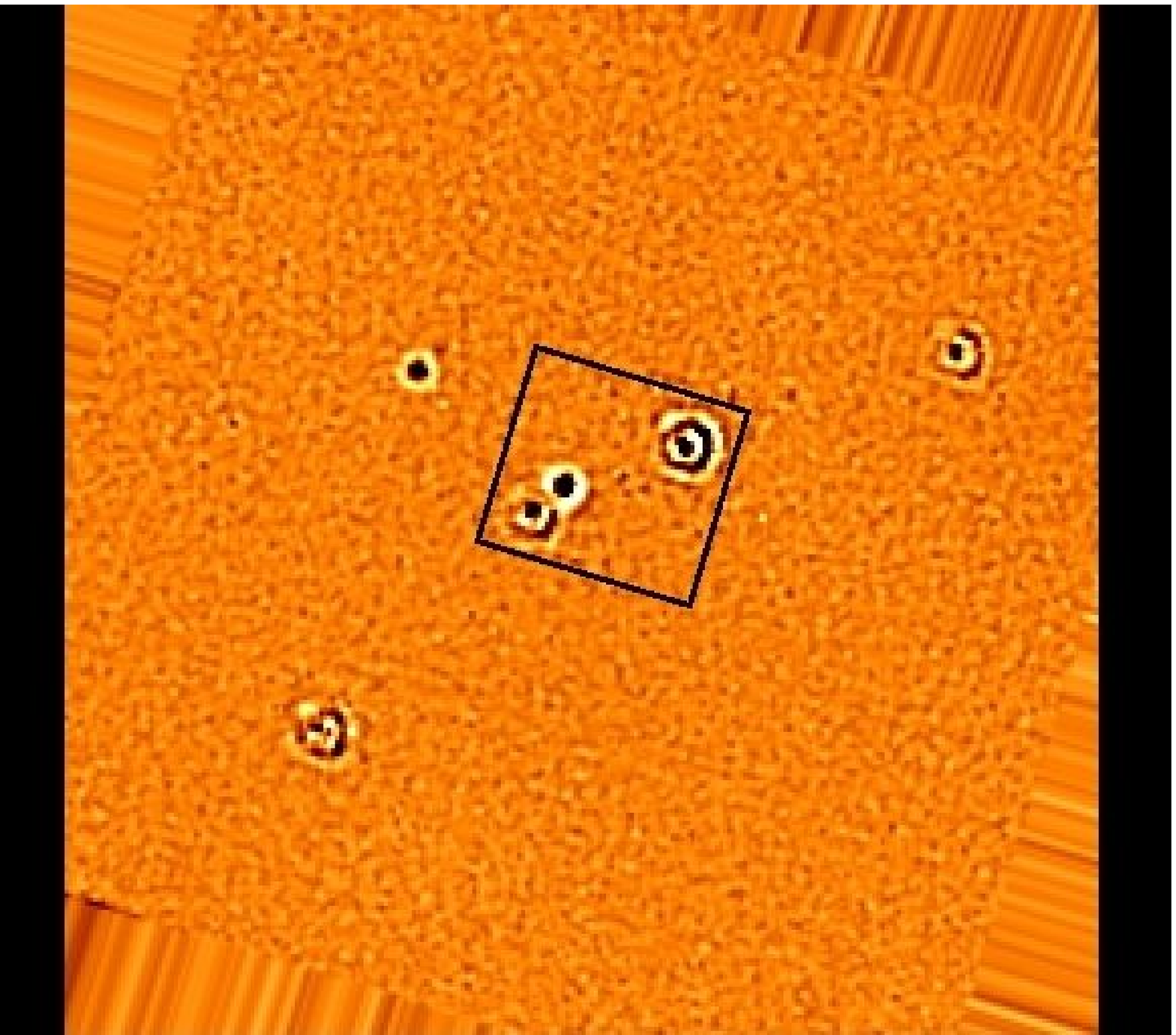}}
  \subfigure{\includegraphics[scale=0.3]{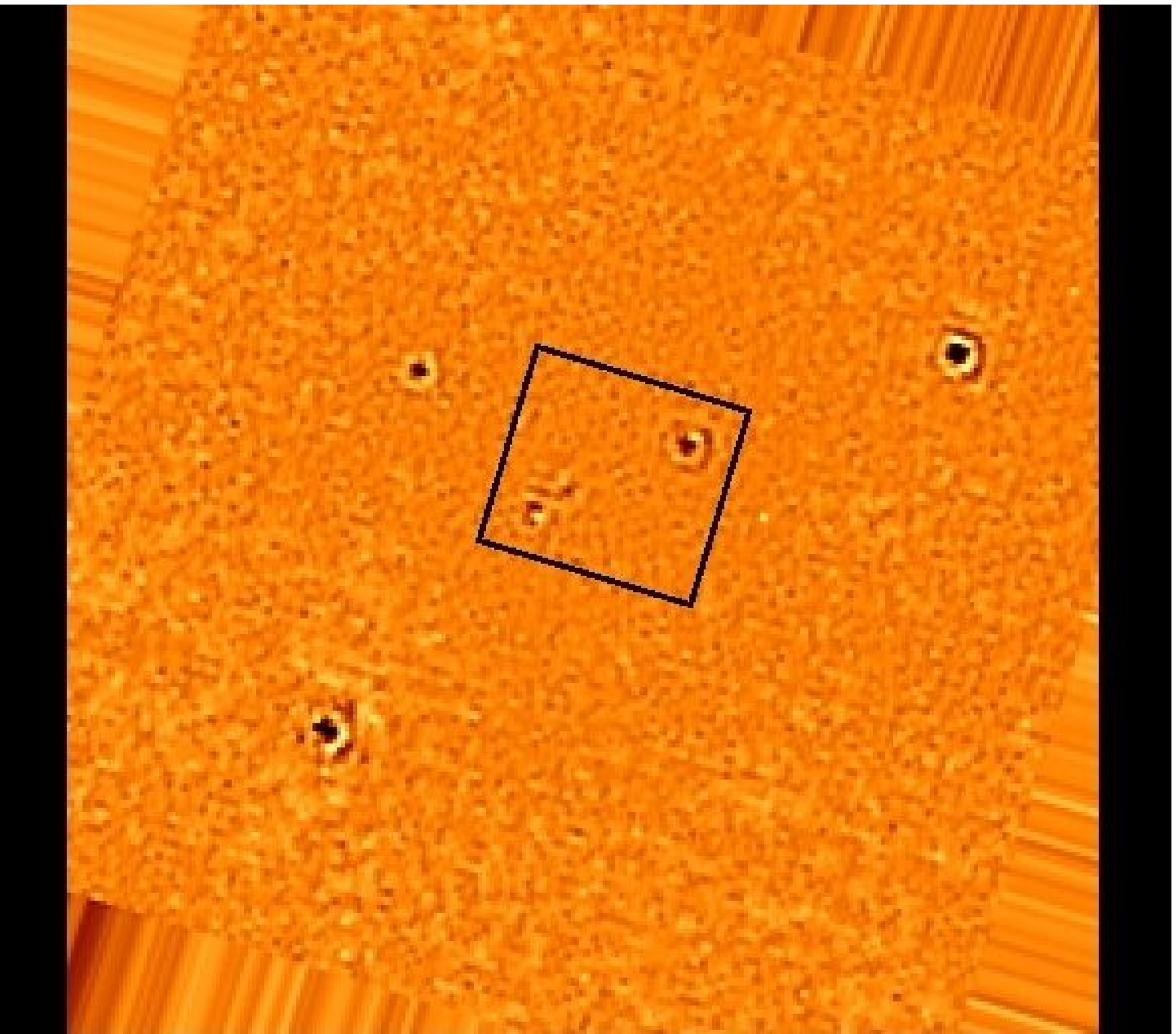}}
\setcounter{figure}{0}
\caption[]{continued.}
\end{figure*}

\begin{figure*} [pht!]
\centering
  \subfigure{\includegraphics[scale=0.3]{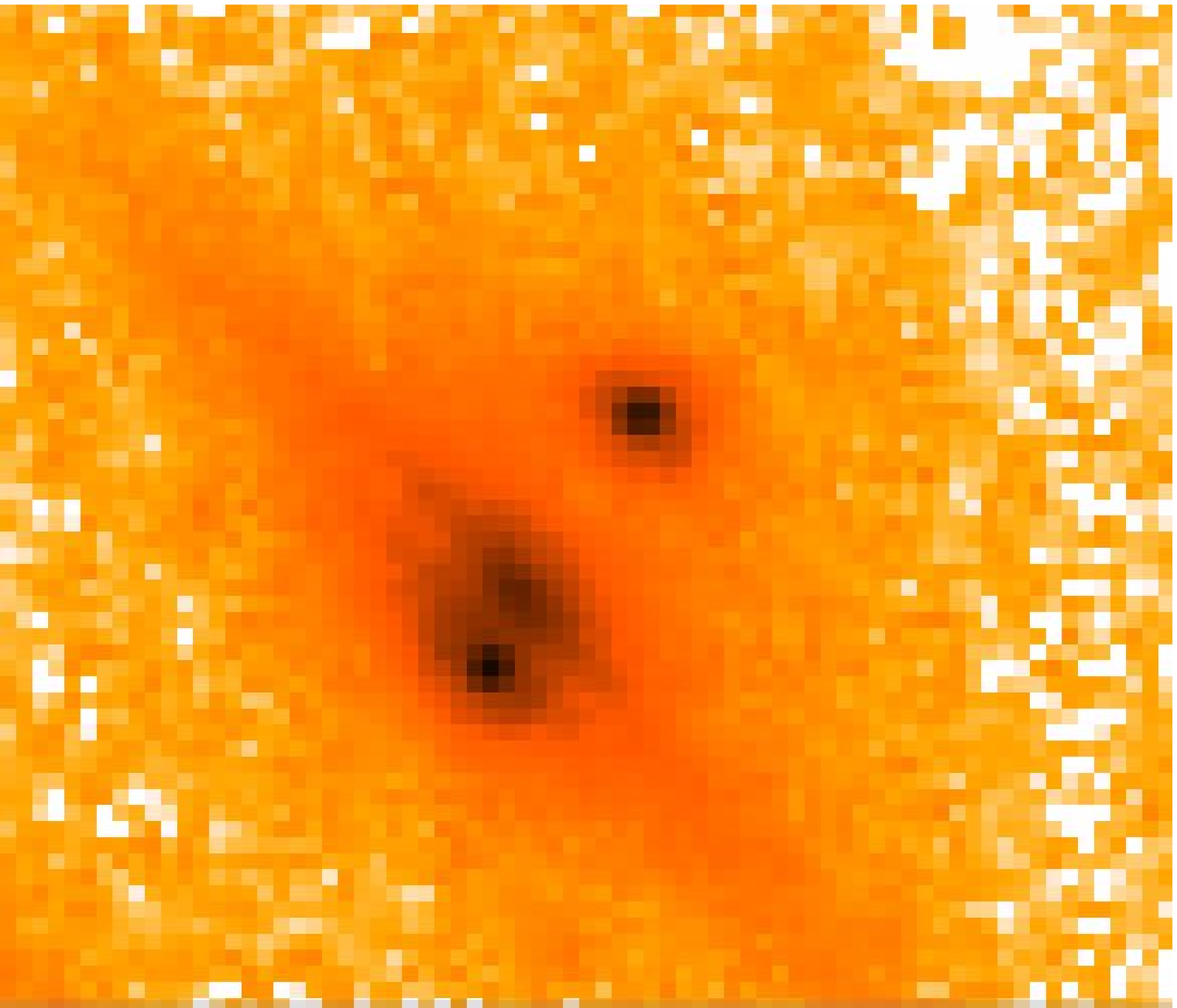}}                
  \subfigure{\includegraphics[scale=0.3]{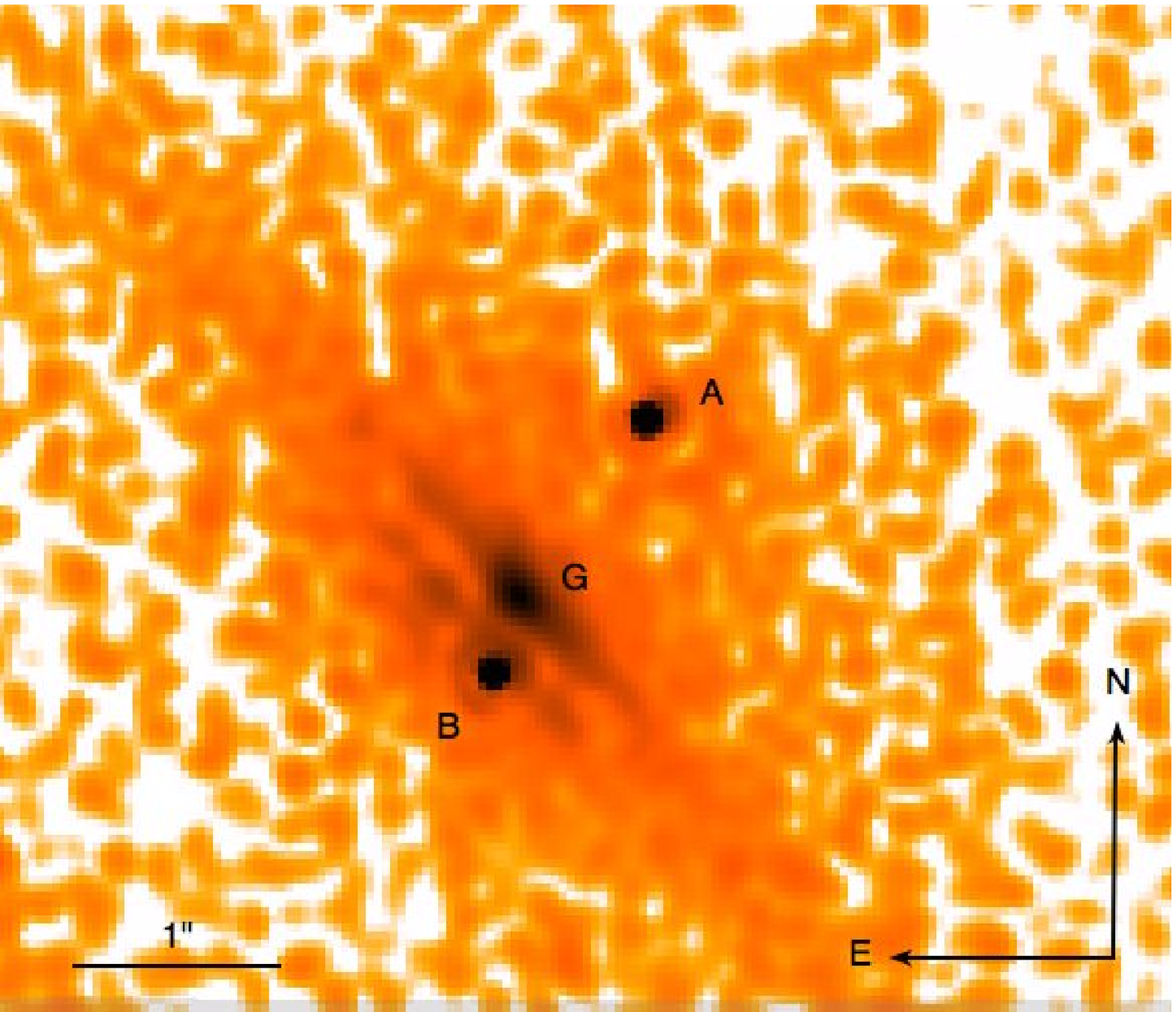}}
\setcounter{subfigure}{8}
  \subfigure[CLASS~B1600+434]{\includegraphics[scale=0.3]{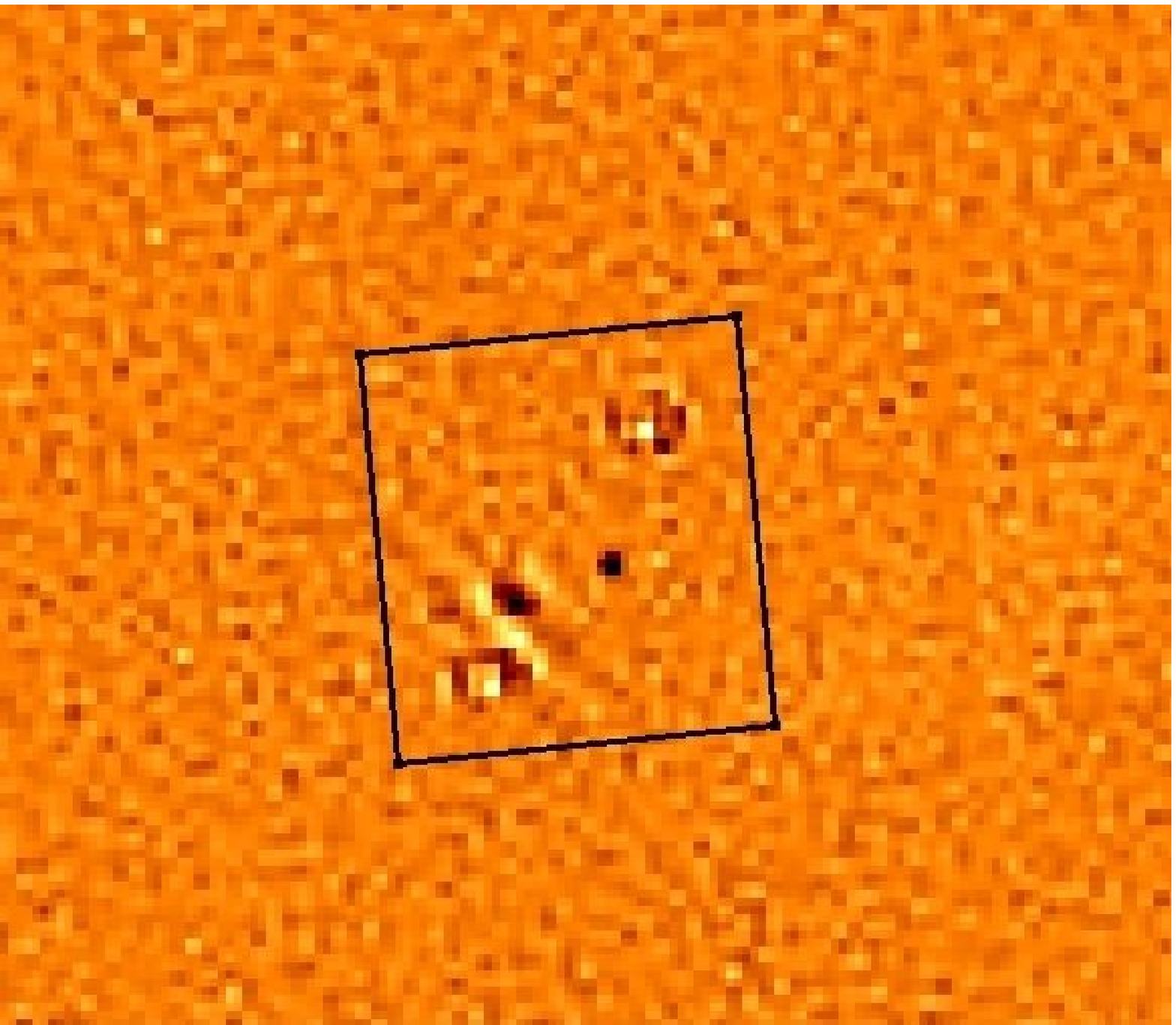}}
  \subfigure{\includegraphics[scale=0.3]{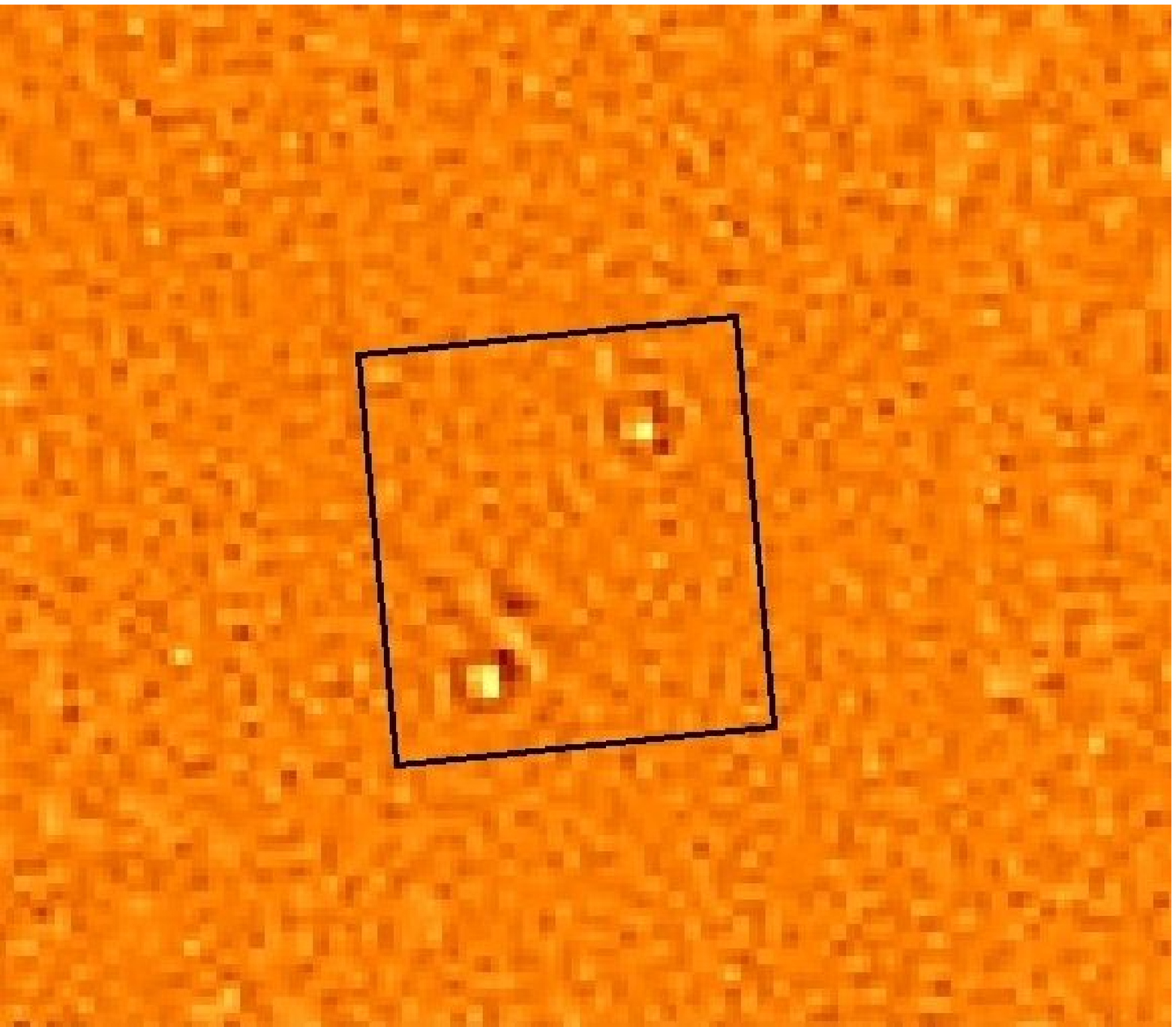}}
  \subfigure{\includegraphics[scale=0.3]{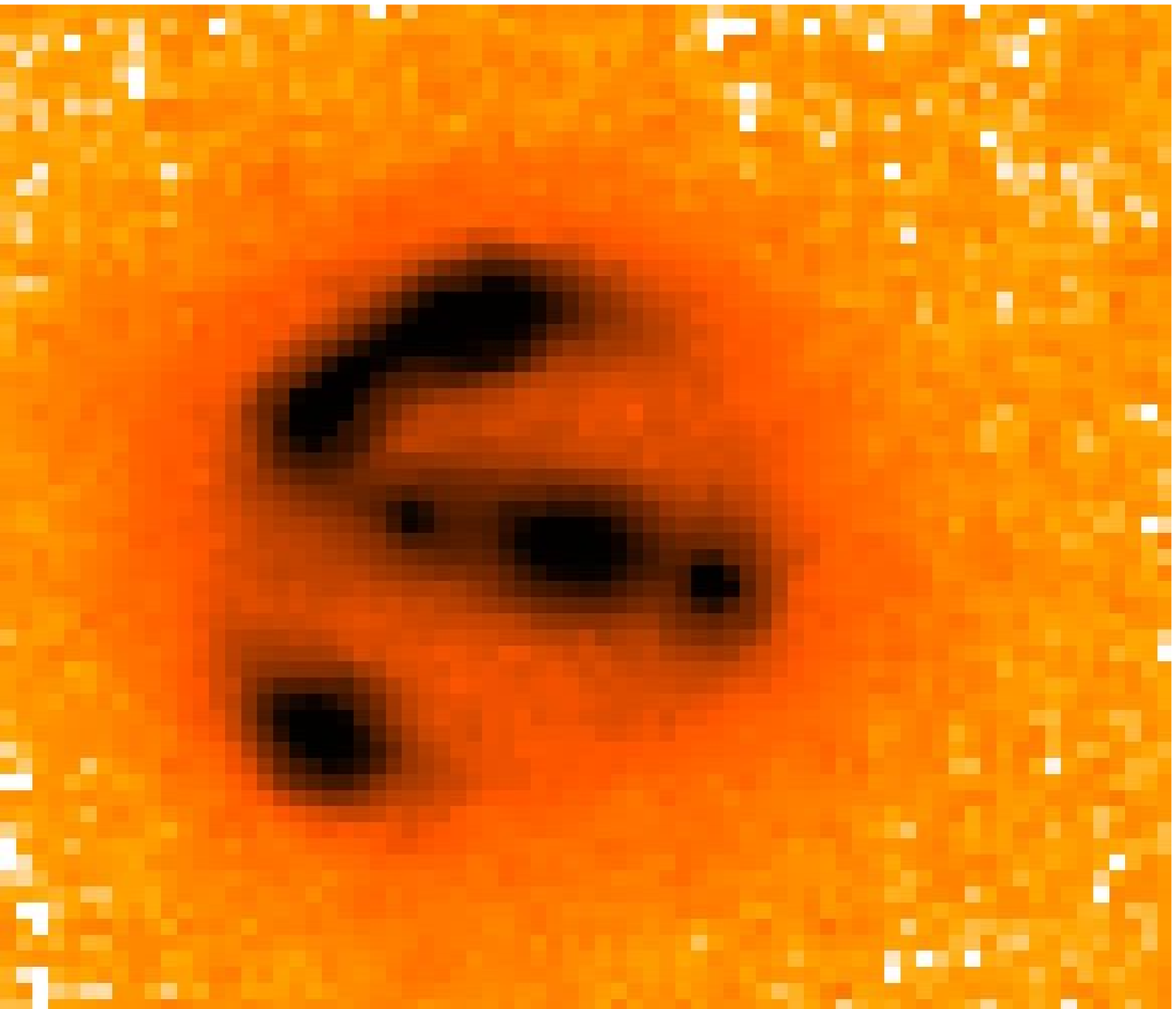}}                
  \subfigure{\includegraphics[scale=0.3013]{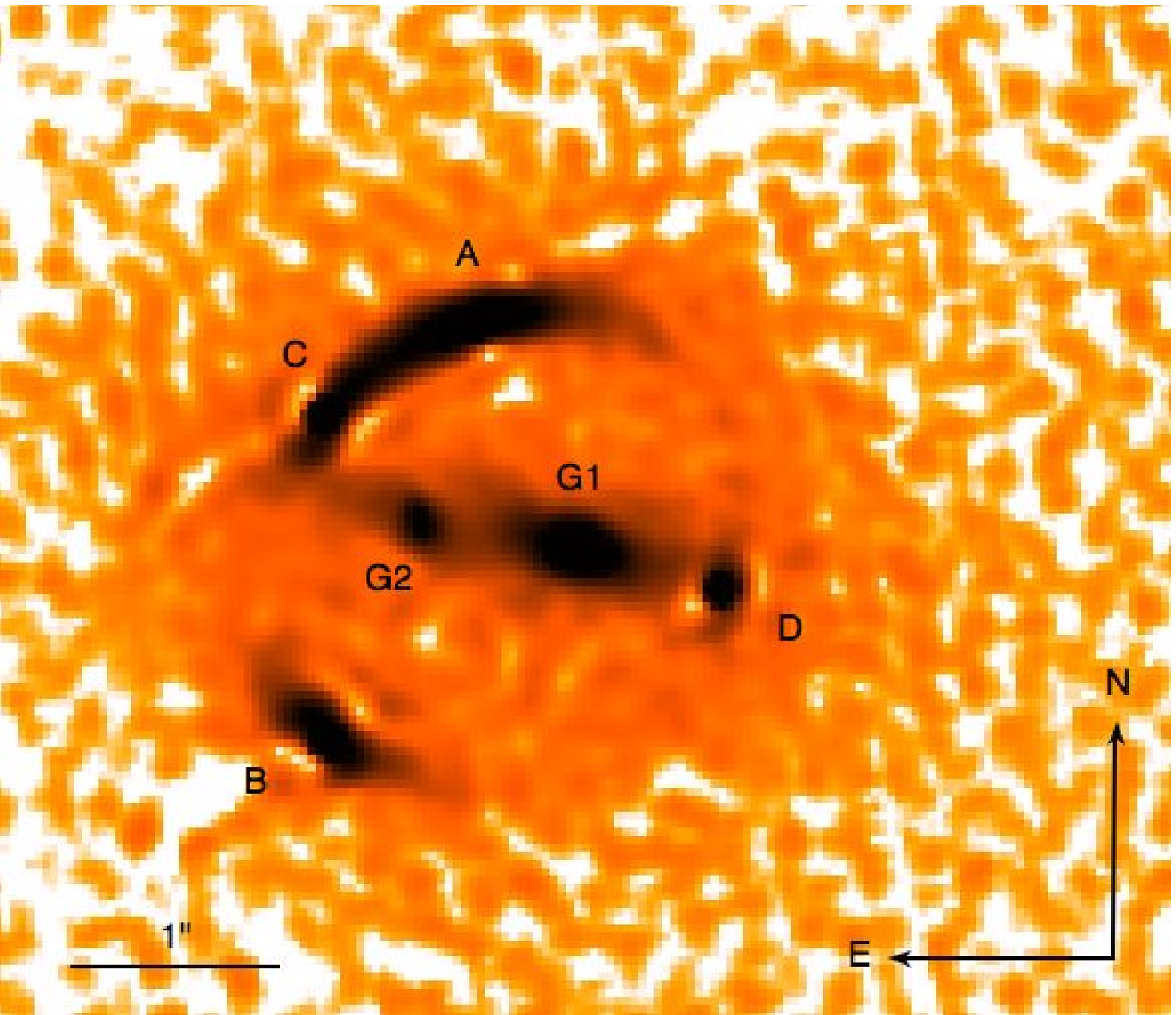}}
\setcounter{subfigure}{9}
  \subfigure[CLASS~B1608+656]{\includegraphics[scale=0.3]{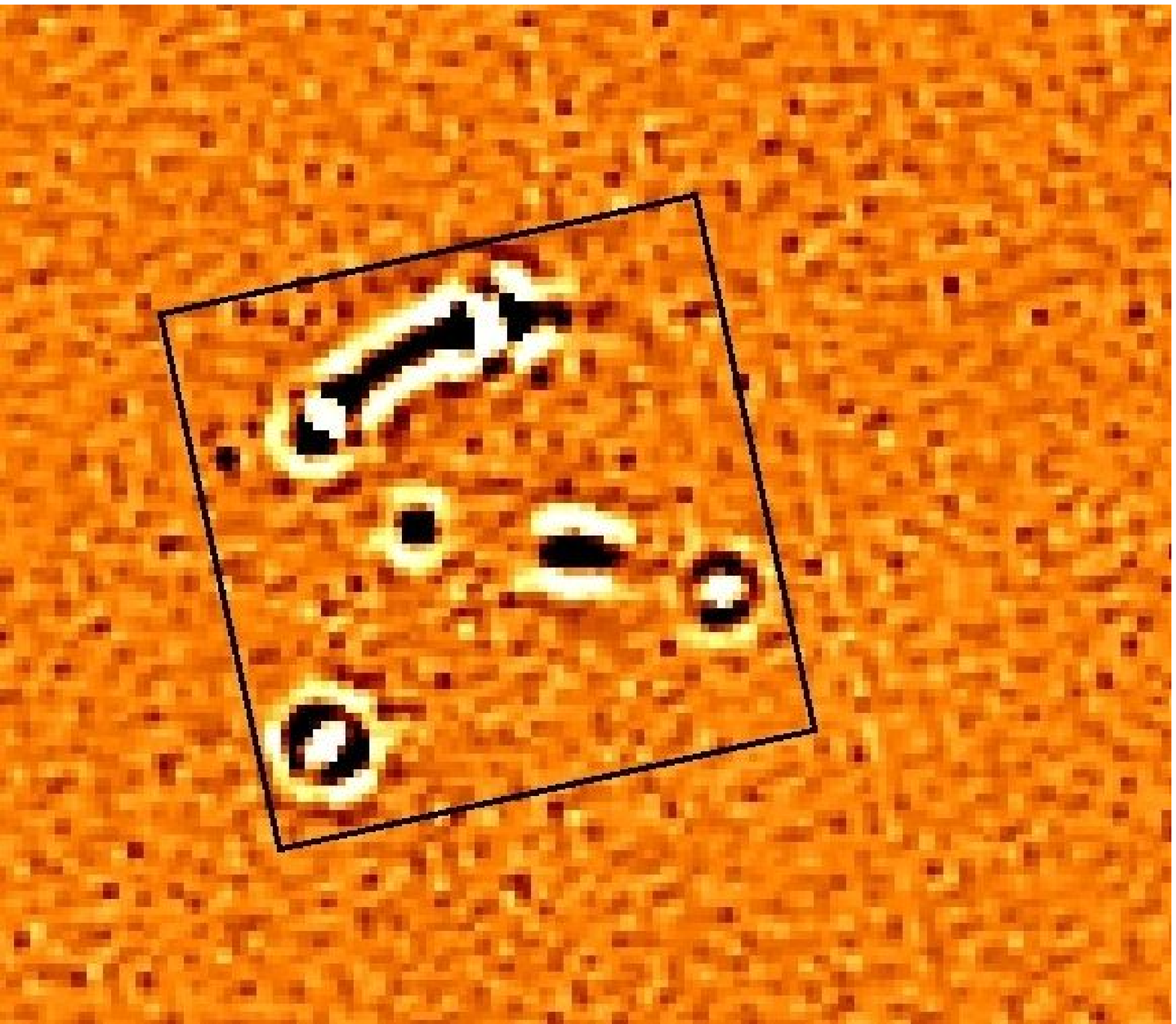}}
  \subfigure{\includegraphics[scale=0.3]{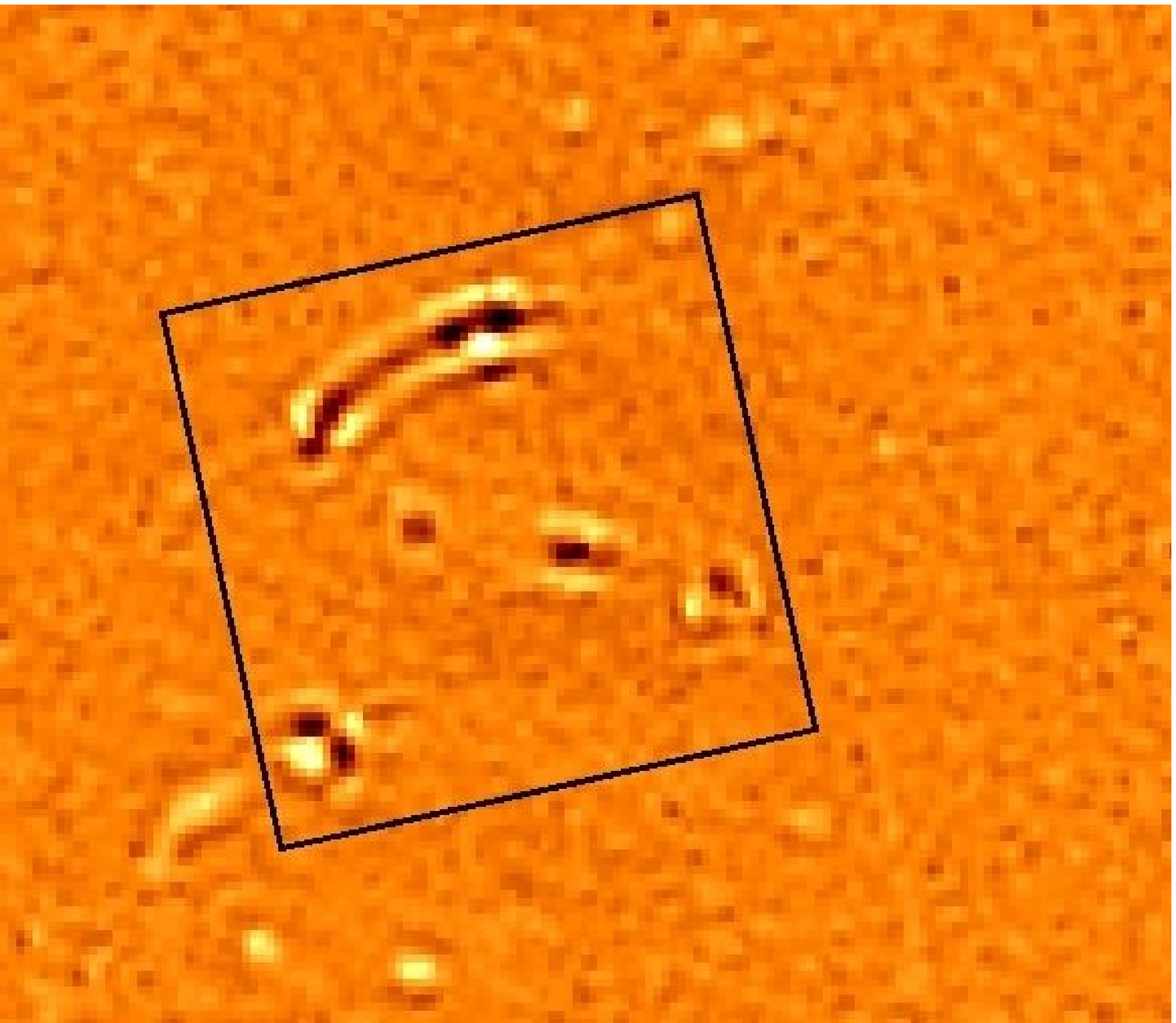}}
\setcounter{figure}{0}
\caption[]{continued.}
\end{figure*}

\begin{figure*} [pht!]
\centering
  \subfigure{\includegraphics[scale=0.3]{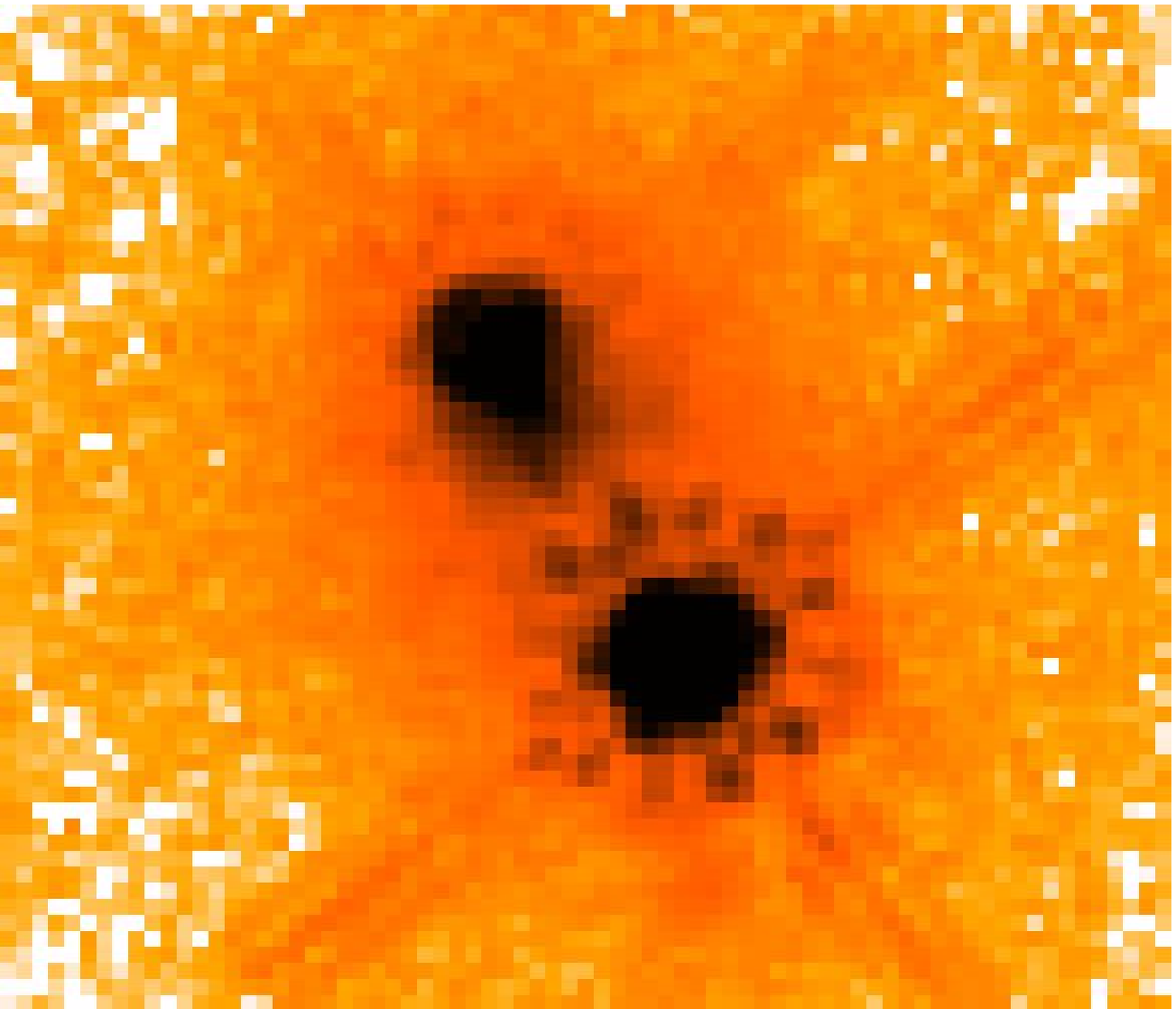}}                
  \subfigure{\includegraphics[scale=0.3013]{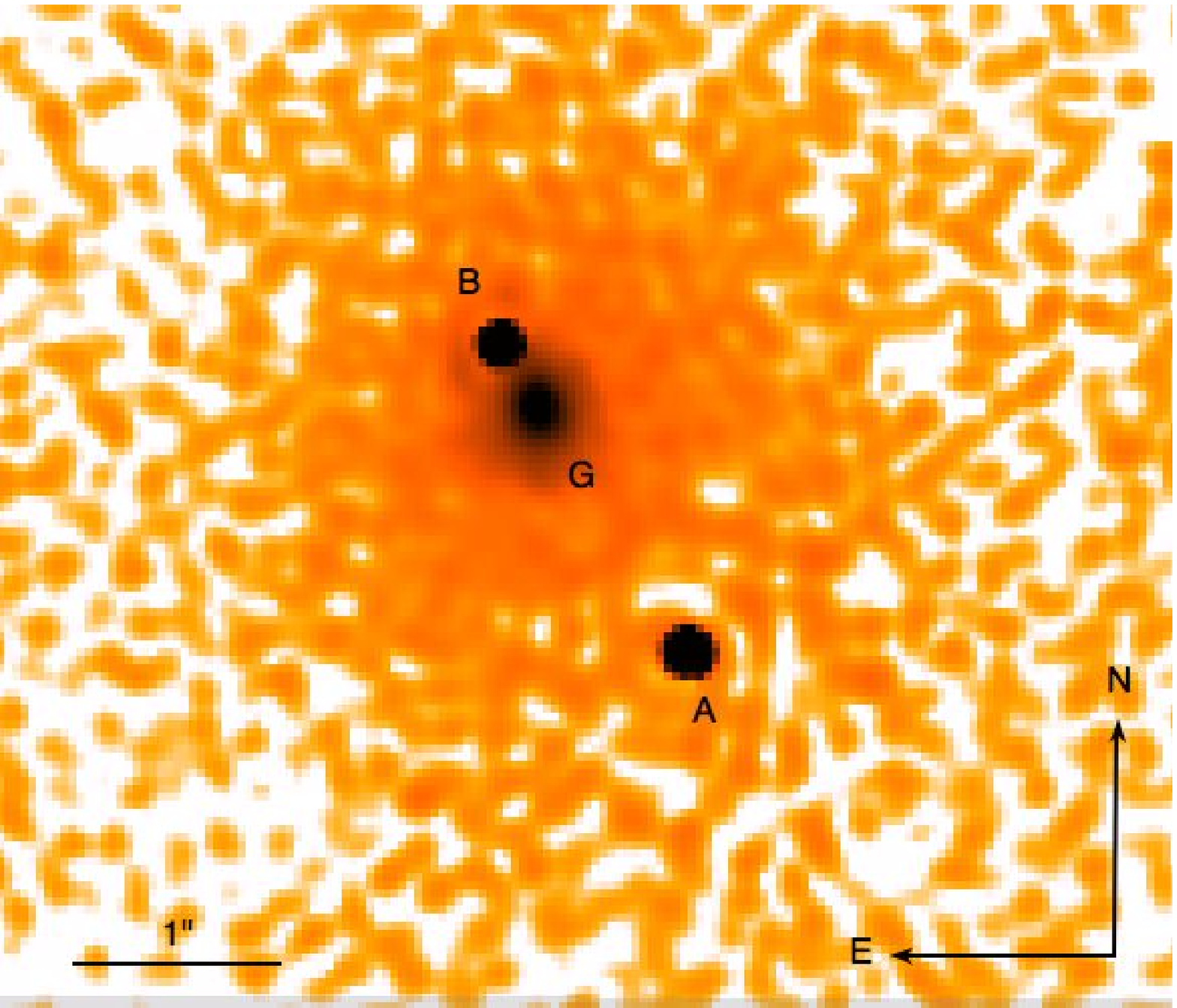}}
\setcounter{subfigure}{10}
  \subfigure[HE 2149-274]{\includegraphics[scale=0.3]{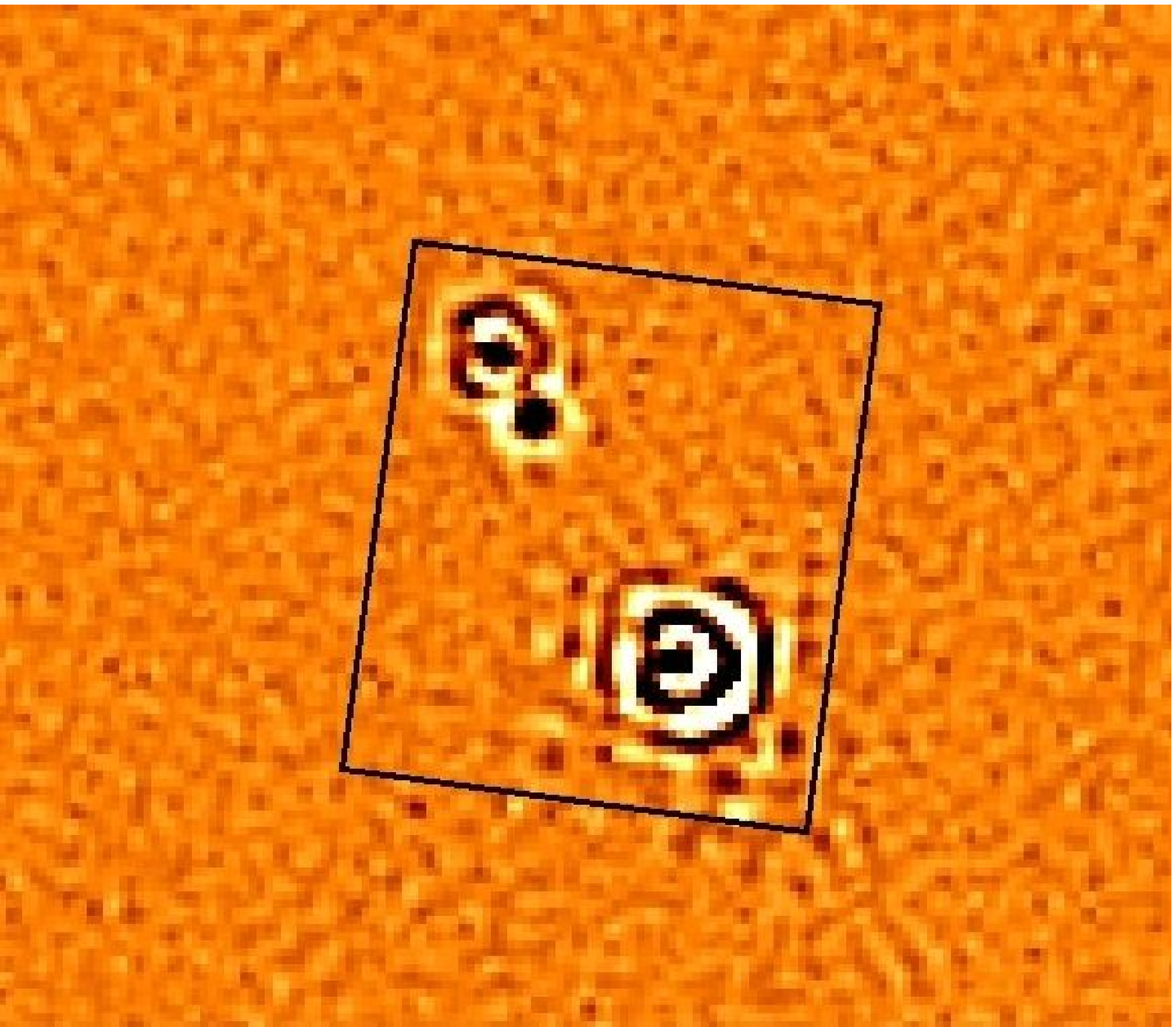}}
  \subfigure{\includegraphics[scale=0.3]{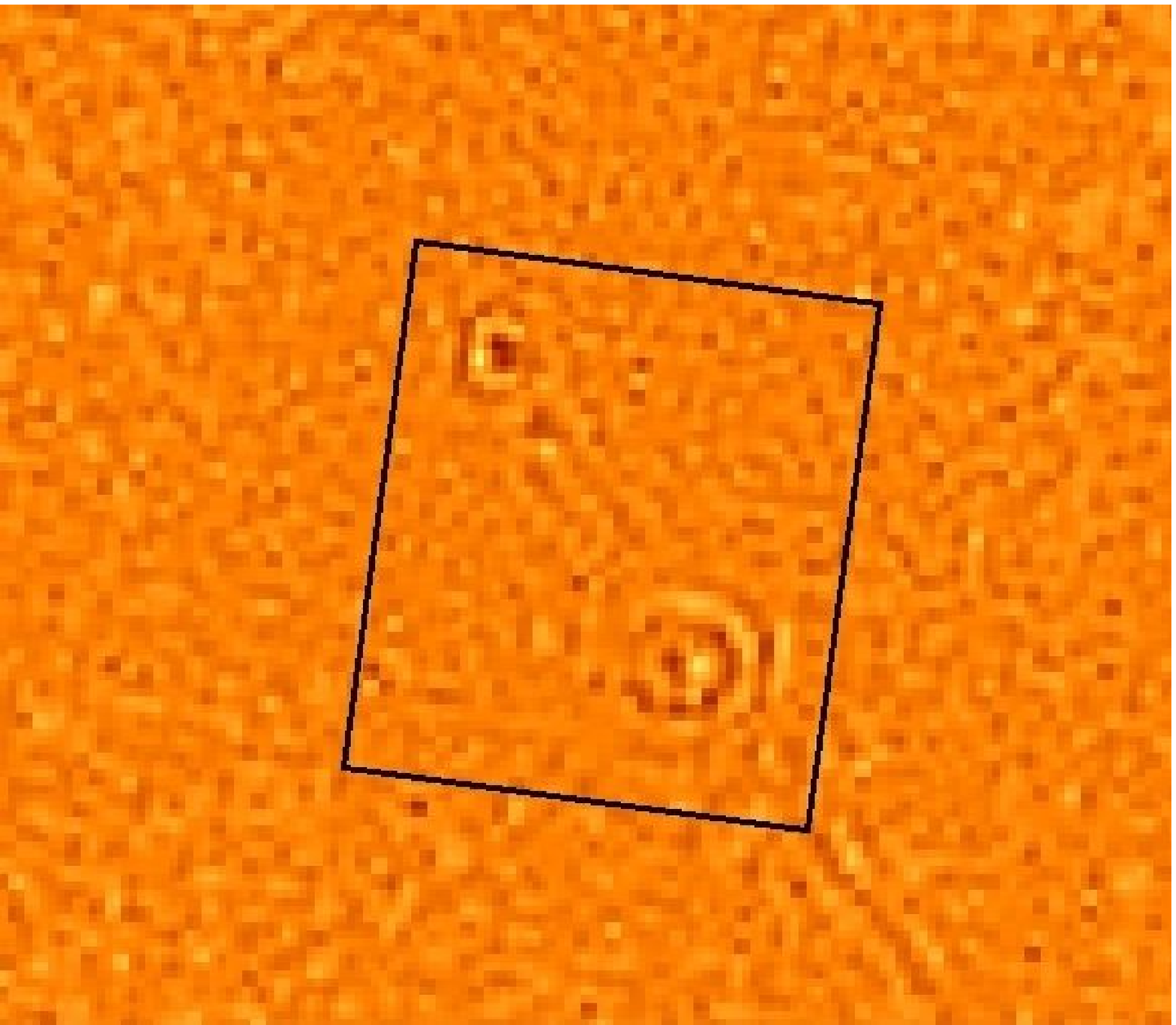}}
\setcounter{figure}{0}
\caption[]{continued.}
\end{figure*}

\begin{table*}
\centering
\begin{tabular}{lcccccc}
\hline
Object & Label & $\theta_{e}$ ($^{\circ}$) & ${e}$ & $a_{\rm eff}$ (\arcsec) & $b_{\rm eff}$ (\arcsec) & $R_{\rm eff}$ (\arcsec) \\ %& Magnitude in $r_{0}$ \\ 
\hline
\hline
(a) JVAS~B0218+357 & G & -73.60 $\pm$ 2.20 & 0.04 $\pm$ 0.007 & 0.32 $\pm$ 0.01 & 0.30 $\pm$ 0.01 & 0.31 $\pm$ 0.01 \\
(b) SBS~0909+532 & G & -48.10 $\pm$ 16.90 & 0.11 $\pm$ 0.08 & 0.57 $\pm$ 0.06 & 0.51 $\pm$ 0.04 & 0.54 $\pm$ 0.02 \\
(c) RX~J0911.4+0551 & G & -70.00 $\pm$ 4.80 & 0.11 $\pm$ 0.01 & 1.08 $\pm$ 0.01 & 0.97 $\pm$ 0.01 & 1.02 $\pm$ 0.01 \\
(d) FBQS~J0951+2635 & G & 12.80 $\pm$ 2.20 & 0.47 $\pm$ 0.03 & 1.00 $\pm$ 0.04 & 0.60 $\pm$ 0.01 & 0.78 $\pm$ 0.01  \\
% (e) QSO 0957+561 & G & 48.1 $\pm$ 0.7 & 0.31 $\pm$ 0.01 & 1.35 $\pm$ 0.01 & 0.98 $\pm$ 0.01 & 1.15 $\pm$ 0.01  \\
(e) HE~1104-1805 & G & 54.70 $\pm$ 7.60 & 0.14 $\pm$ 0.03 & 1.05 $\pm$ 0.02 & 0.91 $\pm$ 0.02 & 0.98 $\pm$ 0.01  \\
(f) PG~1115+080 & G & -67.50 $\pm$ 0.50 & 0.04 $\pm$ 0.01 & 0.94 $\pm$ 0.01 & 0.90 $\pm$ 0.01 & 0.92 $\pm$ 0.01  \\
(g) JVAS~B1422+231 & G & -58.90 $\pm$ 0.80 & 0.39 $\pm$ 0.02 & 0.51 $\pm$ 0.02 & 0.33 $\pm$ 0.02 & 0.41 $\pm$ 0.02 \\
(h) SBS~1520+530 & G & -26.50 $\pm$ 1.00 & 0.49 $\pm$ 0.03 & 0.99 $\pm$ 0.05 & 0.58 $\pm$ 0.01 & 0.76 $\pm$ 0.02  \\
(i) CLASS~B1600+434 & G & 36.90 $\pm$ 2.30 & 0.75 $\pm$ 0.08 & 0.46 $\pm$ 0.01 & 0.18 $\pm$ 0.02 & 0.29 $\pm$ 0.01  \\
(j) CLASS~B1608+656 & G$_{1}$ & 73.50 $\pm$ 0.40 & 0.45 $\pm$ 0.01 & 1.93 $\pm$ 0.01 & 1.18 $\pm$ 0.01 & 1.51 $\pm$ 0.01  \\
& G$_{2}$ & -81.10 $\pm$ 0.20 & 0.55 $\pm$ 0.01 & 1.41 $\pm$ 0.01 & 0.76 $\pm$ 0.01 & 1.04 $\pm$ 0.01  \\
(k) HE 2149-2745 & G & 9.80 $\pm$ 5.10 & 0.21 $\pm$ 0.02 & 1.22 $\pm$ 0.03 & 0.99 $\pm$ 0.01 & 1.10 $\pm$ 0.02  \\
\hline
\end{tabular}
\caption{Measured shape parameters for the lensing galaxies. De Vaucouleurs light profiles are used except for JVAS~B0218+357 and B1600+434 for which we used an exponential light profile. }
\label{gal11}
\end{table*}

We list the astrometry and photometry of the point sources and lensing galaxy in the Vega system derived using our deconvolution technique in Table~\ref{astrom11}. These quantities are derived from the deconvolution of the individual frames using the PSF retrieved with the iterative method. This technique allows us to estimate the  $\pm 1 \sigma$ error bars on the relative astrometry and photometry, to correct the relative positions from the known distortions of the NIC2 camera frame by frame, as well as from the difference of pixel scale between the x and y directions. The error bars derived this way are internal error bars inherent to the images and their processing. To estimate the total error that affects our results we assumed that the main source of systematic error comes from residual geometric distortion. We have estimated the amplitude of the residual error caused by distortion (i.e. our maximum total errors MTE) by comparing the relative astrometry obtained with various HST data of the Cloverleaf \citep{Chantry2007}. Here, we derived the MTE of a system by scaling the MTE found for the Cloverleaf with the image separation of the considered lensed system. This is a maximum error because it is estimated with the maximum spatial extension of the system on the detector, regardless of the direction of this extension. It varies from 0.27 to 2.61 milliarcseconds (mas, see fifth column of Table~\ref{astrom11}). The internal error should be used over the MTE if the first is larger than the latter, which is often the case for the lensing galaxy. Finally, note that the magnitude of each galaxy is measured in an aperture equal to the effective radius, $R_{\rm eff}$, given in Table~\ref{gal11}.

\begin{table*}
\begin{center}
\begin{tabular}{lcccccc}
\hline
Object & Label & $\Delta \rm RA$ (\arcsec) & $\Delta \rm Dec$ (\arcsec) & MTE (mas) & Magnitude & Flux ratio \\ 
\hline
\hline
(a) JVAS~B0218+357$^{\dagger}$&A &  0.        & 0.        & 0.27 & 17.46 $\pm$ 0.01  &  1.\\
&B &  0.3073 $\pm$ 0.0004 & 0.1273 $\pm$ 0.0004 & 0.27 & 16.88  $\pm$ 0.02 & 1.707 $\pm$ 0.030\\
&G &  0.1980 $\pm$ 0.0081 & 0.0941$\pm$ 0.0094 & 0.27 & 17.97 $\pm$ 0.06 &  /\\
\hline
% \multicolumn{5}{c}{(b) SBS0909+532} \\
% \hline
(b) SBS~0909+532 & A &  0.        & 0.        & 0.90 & 14.53 $\pm$ 0.01 &  1.\\
& B & 0.9868 $\pm$ 0.0004 & -0.4973 $\pm$ 0.0008 & 0.90 & 14.67 $\pm$ 0.01 & 0.885 $\pm$ 0.014 \\
& G & 0.4640 $\pm$ 0.0023 & -0.0550 $\pm$ 0.0037 & 0.90 & 19.44 $\pm$ 0.1 & / \\
\hline
(c) RX~J0911.4+0551 & A$_{1}$ &  0.        & 0.        & 2.51 & 17.52 $\pm$ 0.01 & 1. \\
& A$_{2}$ & 0.2611 $\pm$ 0.0009 & 0.4069 $\pm$ 0.0010 & 2.51 & 17.58 $\pm$ 0.01 & 0.950 $\pm$ 0.008\\
& A$_{3}$ & -0.0158 $\pm$ 0.0009 & 0.9575 $\pm$ 0.0008 & 2.51 & 18.27 $\pm$ 0.01 & 0.513 $\pm$ 0.005\\
& B & -2.9681 $\pm$ 0.0013 & 0.7924 $\pm$ 0.0006 & 2.51 & 18.60 $\pm$ 0.02 & 0.372 $\pm$ 0.004\\
& G & -0.7019 $\pm$ 0.0019 & 0.5020 $\pm$ 0.0039 & 2.51 & 18.25 $\pm$ 0.06 &  /\\
& G$_2^{\dagger \dagger}$ & -1.4601 $\pm$ 0.0360 & 1.1678 $\pm$ 0.0470 & 2.51 & / & / \\
\hline
(d) FBQS~J0951+2635 & A &  0.        & 0.        & 0.90 & 15.56 $\pm$ 0.01  &  1.\\
& B & 0.8983 $\pm$ 0.0011 & -0.6336 $\pm$ 0.0013 & 0.90 & 16.93 $\pm$ 0.02 & 0.283 $\pm$ 0.005\\
& G & 0.7521 $\pm$ 0.0017 & -0.4603 $\pm$ 0.0039 & 0.90 & 17.36 $\pm$ 0.01 &  /\\
\hline
% (e) QSO 0957+561 & A &  0.        & 0.        & 15.49 $\pm$ 0.01  &  \\
% & B & -1.2297 $\pm$ 0.0006 & -6.0476 $\pm$ 0.0003 & 15.56  $\pm$ 0.01 &  \\
% & G & -1.4072 $\pm$ 0.0011 & -5.0342 $\pm$ 0.0010 & 16.23 $\pm$ 0.01 &  \\
% \hline
(e) HE~1104-1805  & A &  0.        & 0.        & 2.61 & 15.58 $\pm$ 0.01  &  1.\\
& B & 2.9020 $\pm$ 0.0008 & -1.3362 $\pm$ 0.0028 & 2.61 & 17.05 $\pm$ 0.02 & 0.259 $\pm$ 0.005 \\
& G & 0.9731 $\pm$ 0.0022 & -0.5120 $\pm$ 0.0012 & 2.61 & 17.79 $\pm$ 0.03 &  /\\
\hline
(f) PG~1115+080  & C &  0.        & 0.        & 1.98 & 17.13 $\pm$ 0.01  & 0.255 $\pm$ 0.008\\
& A$_{1}$ & 1.3286 $\pm$ 0.0002 & -2.0338 $\pm$ 0.0015 & 1.98 & 15.64  $\pm$ 0.02 & 1.\\
& A$_{2}$ & 1.4772 $\pm$ 0.0024 & -1.5757 $\pm$ 0.0009 & 1.98 & 16.10 $\pm$ 0.01 & 0.656 $\pm$ 0.013\\
& B & -0.3405 $\pm$ 0.0005 & -1.9598 $\pm$ 0.0010 & 1.98 & 17.60 $\pm$ 0.01 & 0.164 $\pm$ 0.003 \\
& G & 0.3813 $\pm$ 0.0041 & -1.3442 $\pm$ 0.0040 & 1.98 & 16.97 $\pm$ 0.01 & /  \\
\hline
(g) JVAS~B1422+231 & A & 0.3860 $\pm$ 0.0004 & 0.3169 $\pm$ 0.0003 & 1.05 & 14.35 $\pm$ 0.01  &  1.\\
& B & 0. & 0. & 1.05 & 14.23 $\pm$ 0.01 & 1.116 $\pm$ 0.003 \\
& C & -0.3360 $\pm$ 0.0003 & -0.7516 $\pm$ 0.0005 & 1.05 & 14.92  $\pm$ 0.01 & 0.590 $\pm$ 0.002 \\
& D & 0.9470 $\pm$ 0.0006 & -0.8012 $\pm$ 0.0005 & 1.05 & 18.11 $\pm$ 0.02 & 0.032 $\pm$ 0.001 \\
& G & 0.7321 $\pm$ 0.0037 & -0.6390 $\pm$ 0.0054 & 1.05 & 18.16 $\pm$ 0.05 & / \\
\hline
(h) SBS~1520+530 & A &  0.        & 0.        & 1.28 & 17.21 $\pm$ 0.01  &  1.\\
& B & 1.4274 $\pm$ 0.0006 & -0.6525 $\pm$ 0.0004 & 1.28 & 18.03 $\pm$ 0.01 & 0.471 $\pm$ 0.002 \\
& G & 1.1395 $\pm$ 0.0016 & -0.3834 $\pm$ 0.0017 & 1.28 & 18.22 $\pm$ 0.02 & / \\
\hline
(i) CLASS~B1600+434 & A &  0.        & 0.        & 1.14 & 19.98 $\pm$ 0.03  & 1. \\
& B & 0.7300 $\pm$ 0.0024 &  -1.1891$\pm$ 0.0006 & 1.14 & 20.22 $\pm$ 0.07 & 0.897 $\pm$ 0.056\\
& G & 0.6044 $\pm$ 0.0040  & -0.8444 $\pm$ 0.0041 & 1.14 & 19.48 $\pm$ 0.14 & / \\
\hline
(j) CLASS~B1608+656 & A &  0.        & 0.        & 1.71 & 18.37 $\pm$ 0.07 & 1.086 $\pm$ 0.056\\
& B & 0.7464 $\pm$ 0.0026 & -1.9578 $\pm$ 0.0026 & 1.71 & 18.66 $\pm$ 0.09 & 1. \\
& C & 0.7483 $\pm$ 0.0038 & -0.4465 $\pm$ 0.0033 & 1.71 & 18.90 $\pm$ 0.07 & 0.855 $\pm$ 0.021 \\
& D & -1.1181 $\pm$ 0.0025 & -1.2527 $\pm$ 0.0018 & 1.71 & 19.50 $\pm$ 0.04 & 0.418 $\pm$ 0.024 \\
& G$_{1}$ & -0.4561 $\pm$ 0.0061 & -1.0647 $\pm$ 0.0037 & 1.71 & 16.85 $\pm$ 0.01 & / \\
& G$_{2}$ & 0.2821 $\pm$ 0.0015 & -0.9359 $\pm$ 0.0023 & 1.71 & 17.29 $\pm$ 0.01 & / \\
\hline
(k) HE~2149-2745 & A &  0.        & 0.        & 1.39 & 15.62 $\pm$ 0.01  & 1. \\
& B & 0.8901 $\pm$ 0.0007 & 1.4461 $\pm$ 0.0003 & 1.39 & 17.19 $\pm$ 0.01 & 0.234 $\pm$ 0.001\\
& G & 0.7198 $\pm$ 0.0036 & 1.1498 $\pm$ 0.0051 & 1.39 & 17.69 $\pm$ 0.01 & / \\
\hline
\end{tabular}
\end{center}
\vspace{0.2cm}
{\tiny{Notes: ($\dagger$) Possibly large systematic errors (see Sect \ref{Results_11}); ($\dagger \dagger$) Position measured based on Gaussian fitting.}}
\caption{Relative positions, maximum total error (\textquotedblleft MTE\textquotedblright), magnitudes and flux ratios of the lensed images and lensing galaxy (see Fig.~\ref{dec_NICMOS11} for the labels). Magnitudes are measured within 1 $R_{\rm eff}$ for the galaxy and are PSF magnitudes for the point-like sources. }
%DISTORTION COEFFICIENTS : NICMOS Data Handbook 2002 (PRE NCS)
%Mag galaxie vient de deconvo simult.
\label{astrom11}
\end{table*}

A few remarks can be made about the results:
\begin{itemize}
\item \textit{JVAS~B0218+357 (a)}. \citet{Lehar2000} derived a position for the lens by averaging the results they obtained with the same frames as ours and with archival HST/NIC1 images. Our position agrees with theirs because of the large error bars associated to their result. These authors also warn about additional systematic errors affecting their measurements. We may similarly underestimate the error on the lens galaxy position. A first indication for this is that our positions are affected by larger internal errors than for most of the other objects. In addition, our position is significantly different from the one deduced from the lens model of the radio ring \citep[][ $\Delta \rm RA=0\farcs06$, $\Delta \rm Dec=0\farcs02$]{Wucknitz2004} and from the ACS observations of \citet{york2005}. The combination of the poor NICMOS sampling with the small separation between the galaxy and image B, i.e. less than 1 NIC2 pixel $\sim $ 75\,mas, precludes us from deriving accurate positions for B and G. The ellipticity we find for the bulge of this late-type galaxy agrees with the one extracted from HST/ACS images by \citet{york2005}. However, the spiral arms are not visible on NIC2 images. Additionally, our image separation of 333 mas is larger than the one found by \citet{york2005} in the optical, i.e. 317 mas, but really close to the 334 mas found in the radio. 

\item \textit{SBS~0909+532 (b)}. This is the first published measurement of the lensing galaxy position and morphology. Contrary to the claim of \citet{Lehar2000}, we do not find that the galaxy has a low surface brightness. Moreover, its effective radius is fairly small, i.e. 0\farcs54. Regarding the residuals, we notice that at a long distance from the core, the spikes of the PSF of the bright point sources, which are not very well sampled (they have a FWHM of less than two pixels), appear in the background and on the residual map. This is because (1) these spikes are not accurately modeled by the Tiny Tim software which we use to create a first guess of the PSF, (2) the lensed images of SBS~0909+532 are brighter than in most other systems (3) we did not attempt to correct the PSF at distances which are much longer than the extent of the whole lensed system.\\

\item \textit{FBQS~0951+2635 (d)}. The ellipticity we find with ISMCS, i.e. $e=0.47 \pm 0.03$, is higher than the one found by \citet{Jakobsson2005}, i.e. $e=0.25 \pm 0.04$. The latter is likely underestimated because of a too large smoothing parameter or an oversimplified PSF because they used the MCS deconvolution algorithm but not combined with an iterative strategy.\\

\item \textit{CLASS~B1600+434 (i)}. The high blending between the lens and image B prevents us from reaching the desired level of accuracy on the point source position, and the error bars inherent to the deconvolution are larger than the estimated maximum total error (MTE) for the right ascension of image B.

\vspace{0.05cm}

\end{itemize}

Additionally, the deconvolution is unsatisfactory, although better than with a pure Tiny Tim fit, for three systems, i.e. HE~1104-1805, PG~1115+080, and CLASS~B1608+656. The source in CLASS~B1608+656 might not be exactly point-like in the near-IR, as suggested by the remnant structures in the residual map and because it is not an AGN, but the core of a post-starburst radio galaxy. Moreover, as for PG~1115+080, we highlight a partial Einstein ring, which likely contributes to the loss of accuracy for these two objects: it is very bright and can affect the PSFs. For HE~1104-1805, the angular separation is larger than in most cases and, as a consequence, the effect of the distortions must be stronger. Indeed, the distortions differ according to the position of the point sources on the detector and to the position of the object on the detector, which varies from image to image as they are dithered. This probably results in a disagreement between each frame and consequently in larger error bars because they come from the individual deconvolution of each frame. We note that the convergence criterion was respected in the three cases and that the last iteration did not improve the results. %For these systems, we recommend to use the largest internal error bar as MTE.

We also note that our results for the lensed images position are compatible with those found in the CASTLES database within their error bars (our error bars are smaller by typically a factor $>$ 2), which is not always the case for the position of the lensing galaxies. Indeed, these extended structures are more sensitive to the shape of the PSF.

\section{Lens modeling}
\label{sec:model}

We modeled the mass distribution of the lensing galaxies as in \citet{Chantry2010}, using the LENSMODEL software v1.99o \citep{Keeton2001a}. Two different mass models were systematically considered for the main lensing galaxy. First, an isothermal profile \citep{Kassiola1993, Keeton2001a} with a surface density profile of the form\\
\begin{equation}
\kappa=\frac{b}{2\,(x^2+y^2/q^2)^{1/2}}, 
\end{equation}

with $q$ being the axis ratio and $b$ a normalization factor. Second, a de Vaucouleurs profile \citep{Keeton2001a} of the form \\

\begin{equation}
\kappa = \kappa_0\,\exp\left[-k\,\left(\frac{x^2+y^2/q^2}{R_e}\right)^{1/4}\right] ,
\end{equation}
where k=7.66925 is a constant, $\kappa_0$ is the mass scaling parameter and $R_e$ is the effective radius. For the two systems JVAS~B0218+357 and B1600+434, we also used an exponential profile \citep{Keeton2001a} that has a surface density of the form\\
\begin{equation}
\kappa = \frac{\kappa_0}{q}\,\exp\left[\left(-\frac{x^2+y^2/q^2}{R_e}\right)\right] ,
\end{equation}
where the parameters are similar to those described above.
The effect of the environment on the gravitational potential is accounted for using a shear term characterized by an amplitude $\gamma$ and orientation $\theta_{\gamma}$, in our conventions, $\theta_{\gamma}$ points toward the mass producing the shear. The agreement between the model and the parameters is quantified through a $\chi^2$ term calculated in the image plane:

\begin{equation}
  \chi^2 = \sum_i {\frac{(p_{obs, i}-p_{mod, i})^2}{\sigma_i^2}},
\label{chidef}
\end{equation}

where $p_i$ are all the observable quantities that are used to constrain the model and $\sigma_i$ are the associated standard errors. The $\chi^2$ can furthermore be expressed as a sum of four contributions \citep{Keeton2001}:
\begin{equation}
\chi^2 =\chi^2_{\rm {img}}+\chi^2_{\rm {lens}}+\chi^2_{\rm{morph}}+\chi^2_{\rm{flux}}\,,    
\label{eq:chidecomp}
\end{equation}
where $\chi^2_{\rm {img}}$ encodes the contribution of the astrometry of the lensed images and $\chi^2_{\rm {lens}}$ is the contribution of the astrometry of the lensing galaxy, in both cases using MTE as error bars when they are larger than intrinsic errors. The term $\chi^2_{\rm{morph}}$ refers to the contribution of the morphological parameters of the lens, namely the ellipticity $e$, the position angle $\theta_e$ and the effective radius $R_{eff}$ (Table \ref{gal11}). This term is different from zero only when we use non-spherically symmetric mass model, i.e. singular isothermal ellipsoid (SIE), deVaucouleurs (DV), and exponential-disk (ExpD). When we modeled quads with SIE models, we used a large error for the mass ellipticity ($\sigma_e=1$) to allow it to deviate from that of the light. We used the uncertainty associated to the observed light (Tab.~\ref{gal11}) profile in all the other cases. When we modeled doubles, in addition to elliptical mass distributions, we also considered singular isothermal sphere (SIS) models. 

The last term of Eq.~\ref{eq:chidecomp}, i.e. $\chi^2_{\rm{flux}}$, has the form
\begin{equation}
\chi^2_{\rm{flux}} = \sum_j { \left( f_j - M_j f_{src} \right)^2 \over \sigma_{f,j}^2 }\ ,
\end{equation}
where the index $j$ denotes individual lensed images that have an observed flux $f_j \pm  \sigma_{f,j}$. The lens model gives the magnification $M_j = |\det(\mu_j)|$ of an image $j$, which has a magnification tensor $\mu_j$, and the intrinsic flux of the source is noted $f_{src}$. To avoid that our models for doubles became underconstrained, we used the flux constraint through the measured NIR flux ratio of the two point sources with a 1$\sigma$ error of 10\% to account for microlensing, dust extinction, and time-delay \citep{Yonehara2008}. For two systems (i.e. RX~J0911.4+0551, B1608+656), a second lensing galaxy is visible in the images and is included in the model as an SIS, assumed to be at the redshift of the main lensing galaxy. The position of this second deflector is fitted to the observed position using the uncertainties displayed in Table~\ref{astrom11}. All models were computed for a flat universe with the following cosmological parameters: $\rm H_{0}=70$ km/s/Mpc, $\Omega_{m}=0.3$, and $\Omega_{\Lambda}=0.7$. 

The strategy used to search for the best model and to estimate the uncertainties is the same as in Paper I. First local and global minima were searched for by  running the minimization from an ensemble of at least 1000 randomly distributed initial parameters. Second, we sampled the parameter space in the vicinity of the identified global minimum  using an adaptive Metropolis Hastings Monte Carlo Markov Chain (MCMC) algorithm implemented in LENSMODEL \citep{Fadely2010}. For each point of the MCMC, we calculated the relative likelihood of a parameter $p$ based on the $\chi^2$ statistics (i.e. $L(D|p) = \exp(-\chi^2/2)$), and calculated a 68\% confidence interval for each parameter.

The parameters of the best-fit models are displayed in Table \ref{lensmodel11}. The columns display the following items: the name of the object, the type of mass distribution used, the mass scale parameter (i.e. the angular Einstein radius $R_{Ein}$ in arcseconds), the mass distribution ellipticity $e$ and its orientation $\theta_{e}$ in degrees positive east of north, the effective radius $R_{eff}$ in arcseconds for a DV or an ExpD model, the intensity of the shear $\gamma$ and its orientation $\theta_{\gamma}$ in degrees (east of north), the model flux ratios, the number of degree(s) of freedom (d.o.f.), the $\chi^{2}$ of the fit, and the predicted time delays in days. For the quads and in the same column as the $\chi^{2}$, we also give the $\chi^{2}_{ima}$ (resp. $\chi^{2}_{lens}$) as defined in Eq.~\ref{eq:chidecomp}. A positive time delay, $\Delta t_{AB} >$ 0, means that the flux of A varies sooner than that of B. The median value of each parameter and its 68\% confidence level is shown in Table \ref{MCMC11}.

% Results of the parametric modeling.
% Table generated with script_table.tabbest()
% The SIE models for the models are generated in a separate table using the same procedure and included manually here
% tabbest.tex
\begin{table*}[hpt!]
\begin{center} 
\begin{tabular}{l||c|cccccccc}
\hline
Object & Model & $R_{Ein}$ & $e, \theta_{e}$ & $R_{eff}$ & $\gamma, \theta_{\gamma}$ & {\it {Flux ratios}} & d.o.f. & $\chi^{2}$ & {\it {Time delays}} \\ 
\hline
\hline

(a) JVAS~B0218+357$\dagger$& 	SIS+$\gamma$	 &  0.151	 & /	 & /	 & 0.096, -28.87	 & {\it {$f_B/f_A=$ 1.707}}	& 0 & 0.0	 & {\it {$\Delta t_{AB} = $ 4.9 }}  \\ 
& SIE+$\gamma$           & 0.152	 & (0.04, -73.56)	 & /	 & 0.093, -25.26	 &  {\it {$f_B/f_A=$ 1.702}}		& 0  & 0.0	 & {\it {$\Delta t_{AB} = $ 4.8}} \\ 
& 	DV+$\gamma^{\dagger\dagger}$	 &  0.150	 & (0.04, -73.60)	 & (0.31)	 & 0.125, -26.28	 &  {\it {$f_B/f_A=$ 1.707}}	 & 0 & 0.0	 & {\it {$\Delta t_{AB} = $ 6.0 }}  \\ 
& ExpD+$\gamma$ & 0.182 &  (0.04, -73.60)	 & (0.31)	 & 0.011, 21.86 & {\it {$f_B/f_A=$}1.707} & 0 & 0.0 & {\it {$\Delta t_{AB} = $ 1.9 }} \\
\hline
(b) SBS~0909+532& 	SIS+$\gamma$	 &  0.557	 & /	 & /	 & 0.066, 54.72	 &  {\it {$f_B/f_A=$0.885}}	 & 0 & 0.0	 & {\it {$\Delta t_{AB} = $ -29.2}}  \\ 
& SIE+$\gamma$           &  0.548	 & (0.11, -48.05)	 & /	 & 0.101, 49.96	 &  {\it {$f_B/f_A=$ 0.885}}	 & 0  & 0.0	 & {\it {$\Delta t_{AB} = $ -27.2}} \\ 
& 	DV+$\gamma$	 &  0.543	 & (0.11, -48.12)	 & (0.54)	 & 0.125, 50.18	 &  {\it {$f_B/f_A=$0.885}}	 & 0 & 0.0	 & {\it {$\Delta t_{AB} = $ -39.2 }}  \\ 
\hline
(c) RX~J0911.4+0551& 	SIE+$\gamma$	 &  1.086	 & 0.24, (-72.00)	 & /	 & 0.327, 9.39	 & {\it {$f_{A2}/f_{A_{1}}=$1.872}}	 & 1& 0.4	 & {\it {$\Delta t_{A_{1}B} = $ -148.9 }}  \\ 
   & 		 &  0.225	 & /	 & /	 & /	 & {\it {$f_{A_{3}}/f_{A_{1}}=$0.881}}	 & & $\chi^2_{\rm ima} = $ 0.0	 & {\it {$\Delta t_{A_{2}B} = $ -147.9 }}  \\ 
   & 		 & 	 & /	 & /	 & /	 & {\it {$f_{B}/f_{A_{1}}=$ 0.341}}	 & & $\chi^2_{\rm lens} = $ 0.1	 & {\it {$\Delta t_{A_{3}B} = $ -149.5 }}  \\ 
 & 	DV+$\gamma$	 &  1.085	 & (0.16, -65.47)	 & (1.03)	 & 0.380, 8.06	 & {\it {$f_{A2}/f_{A_{1}}=$1.921}}	 & 2 & 112.7	 & {\it {$\Delta t_{A_{1}B} = $ -213.2 }}  \\ 
 & 		 &  0.255	 & /	 & /	 & /	 & {\it {$f_{A_{3}}/f_{A_{1}}=$0.888}}	 & & $\chi^2_{\rm ima} = $ 56.2	 & {\it {$\Delta t_{A_{2}B} = $  -211.9 }}  \\ 
 & 		 & 	 & /	 & /	 & /	 & {\it {$f_{B}/f_{A_{1}}=$0.476}}	 & & $\chi^2_{\rm lens} = $ 28.9	 & {\it {$\Delta t_{A_{3}B} = $ -214.2  }}  \\ 
\hline
(d) FBQS J0951+2635& 	SIS+$\gamma$	 &  0.541	 & /	 & /	 & 0.098, 0.18	 & {\it {$f_B/f_A=$0.283}}	 & 0 & 0.0	 & {\it {$\Delta t_{AB} = $ 13.8 }}  \\ 
& SIE+$\gamma$           &  0.583	 & (0.47, 12.80)	 & /	 & 0.133, -66.57	 & {\it {$f_B/f_A =$0.283}}	 & 0 & 0.0	 & {\it {$\Delta t_{AB} = $ 16.5}}  \\
& 	DV+$\gamma$	 &  0.526	 & (0.47, 12.81)	 & (0.78)	 & 0.017, -33.02	 & {\it {$f_B/f_A=$0.283}}	 & 0 & 0.0	 & {\it {$\Delta t_{AB} = $ 20.8 }}  \\ 
\hline
(e) HE~1104-1805& 	SIS+$\gamma$	 &  1.382	 & /	 & /	 & 0.136, 22.44	 & {\it {$f_B/f_A=$0.259}}	 & 0 & 0.0	 & {\it {$\Delta t_{AB} = $ -185.2}}  \\ 
& SIE+$\gamma$           &  1.405	 & (0.14, 54.71)	 & /	 & 0.120, 12.96	 & {\it {$f_B/f_A=$0.259}}	 & 0 & 0.0	 & {\it {$\Delta t_{AB} = $ -189.9}} \\ 
& 	DV+$\gamma$	 &  1.359	 & (0.14, 54.70)	 & (0.98)	 & 0.201, 19.50	 & {\it {$f_B/f_A=$0.259}}	 & 0 & 0.0	 & {\it {$\Delta t_{AB} = $ -268.9}}  \\ 
\hline
(f) PG~1115+080& 	SIE+$\gamma$	 &  1.141	 & 0.14, (-67.59)	 & /	 & 0.125, 53.80	 &  {\it {$f_C/f_{A_{1}}=$0.271}}	 & 2 & 29.9	 & {\it {$\Delta t_{CA_{1}} = $ 10.4 }}  \\ 
& 		 & 	 & /	 & /	 & /	 &  {\it {$f_{A_{2}}/f_{A_{1}}=$0.926}}	 & & $\chi^2_{\rm ima} = $ 2.9	 & {\it {$\Delta t_{CA_{2}} = $ 10.5 }}  \\ 
& 		 & 	 & /	 & /	 & /	 &  {\it {$f_{B}/f_{A_{1}}=$0.218}}	 & & $\chi^2_{\rm lens} = $ 27.0	 & {\it {$\Delta t_{CB} = $ 17.0 }}  \\ 
& 	DV+$\gamma$	 &  1.137	 & (0.14, -67.59)	 & (0.92)	 & 0.178, 59.91	 & {\it {$f_C/f_{A_{1}}=$0.294}}	 & 3 & 152.7	 & {\it {$\Delta t_{CA_{1}} = $ 15.7 }}  \\ 
& 		 & 	 & /	 & /	 & /	 & {\it {$f_{A_{2}}/f_{A_{1}}=$0.908}}	 & & $\chi^2_{\rm ima} = $ 41.0	 & {\it {$\Delta t_{CA_{2}} = $ 15.9  }}  \\ 
& 		 & 	 & /	 & /	 & /	 & {\it {$f_B/f_{A_{1}}=$0.203}}	 & & $\chi^2_{\rm lens} = $ 18.4	 & {\it {$\Delta t_{CB} = $ 26.1 }}  \\ 
\hline
(g) JVAS B1422+231& 	SIE+$\gamma$	 &  0.771	 & 0.27, (-57.45)	 & /	 & 0.177, -52.25	 & {\it {$f_B/f_{A}=$1.270}}	 & 2 & 10.0	 & {\it {$\Delta t_{AC} = $ -0.7 }}  \\ 
& 		 & 	 & /	 & /	 & /	 & {\it {$f_C/f_{A}=$0.641}}	 & & $\chi^2_{\rm ima} = $ 1.6	 & {\it {$\Delta t_{BC} = $ -0.9  }}  \\ 
& 		 & 	 & /	 & /	 & /	 & {\it {$f_D/f_{A}=$0.048}}    & & $\chi^2_{\rm lens} = $ 5.1	 & {\it {$\Delta t_{DC} = $ -24.6 }}  \\ 
& 	DV+$\gamma$	 &  0.722	 & (0.47, -55.80)	 & (0.49)	 & 0.375, -53.63	 & {\it {$f_B/f_{A}=$1.045}}	 & 3 & 110.5	 & {\it {$\Delta t_{AC} = $ -1.3 }}  \\ 
& 		 & 	 & /	 & /	 & /	 & {\it {$f_C/f_{A}=$0.674}}	 & & $\chi^2_{\rm ima} = $ 37.1	 & {\it {$\Delta t_{BC} = $ -1.7 }}  \\ 
& 	         & 	 & /	 & /	 & /	 & {\it {$f_D/f_{A}=$0.029}}	 & & $\chi^2_{\rm lens} = $ 28.0	 & {\it {$\Delta t_{DC} = $ -39.7 }}  \\ 
\hline
(h) SBS 1520+530& 	SIS+$\gamma$	 &  0.745	 & /	 & /	 & 0.128, -1.56	 & {\it {$f_B/f_{A}=$0.471}}	 & 0 & 0.0	 & {\it {$\Delta t_{AB} = $ 93.6 }}  \\
& SIE+$\gamma$           &  0.732	 & (0.49, -26.50)	 & /	 & 0.144, 43.33	 & {\it {$f_B/f_A = $0.471}}	 & 0 & 0.0	 & {\it {$\Delta t_{AB} =$  95.9}} \\
& 	DV+$\gamma$	 &  0.644	 & (0.49, -26.50)	 & (0.76)	 & 0.229, 21.80	 & {\it {$f_B/f_{A}=$0.471}}	 & 0 & 0.0	 & {\it {$\Delta t_{AB} = $ 122.3 }}  \\ 
\hline
(i) CLASS~B1600+434& 	SIS+$\gamma$	 &  0.613	 & /	 & /	 & 0.143, 46.80	 & {\it {$f_B/f_{A}=$0.897}}	 & 0 & 0.0	 & {\it {$\Delta t_{AB} = $ 27.6 }}  \\
& SIE+$\gamma$           & 0.721	 & (0.75, 36.90)	 & /	 & 0.265, -54.85 & {\it {$f_B/f_A =$ 0.897}}	 & 0 & 0.0	 & {\it {$\Delta t = $ 39.5}} \\
& 	DV+$\gamma^{\dagger\dagger}$	 &  0.579	 & (0.75, 36.90)	 & (0.29)	 & 0.175, 58.36	 & {\it {$f_B/f_{A}=$0.897}}	 & 0 & 0.0	 & {\it {$\Delta t_{AB} = $ 44.6 }}  \\ 
& ExpD+$\gamma$ & 0.673 &  (0.75, 36.90)	 & (0.29)	 & 0.086, -78.89 & {\it {$f_B/f_A=$}0.897} & 0 & 0.0 & {\it {$\Delta t_{AB} = $ 44.6 }} \\
\hline
(j) CLASS B1608+656& 	SIE+$\gamma$	 &  1.124	 & 0.80, (72.89)	 & /	 & 0.195, -7.76	 &  {\it {$f_A/f_{B}=$ 2.009}}	 & 1 & 7.4	 & {\it {$\Delta t_{AB} = $ -48.1}}  \\ 
& 		 &  0.113	 & /	 & /	 & /	 &  {\it {$f_C/f_{B}=$ 0.974}}	 & & $\chi^2_{\rm ima} = $ 0.8	 & {\it {$\Delta t_{CB} = $ -53.9  }}  \\ 
& 	         & 	 & /	 & /	 & /	 &  {\it {$f_D/f_{B}=$0.190}}	 & & $\chi^2_{\rm lens} = $ 4.3	 & {\it {$\Delta t_{DB} = $ -115.8 }}  \\ 
& 	DV+$\gamma$	 &  1.072	 & (0.59, 78.73)	 & (1.51)	 & 0.102, 48.51	 &  {\it {$f_A/f_{B}=$1.768}}	 & 2 & 595.0	 & {\it {$\Delta t_{AB} = $ -27.7 }}  \\ 
& 		 &  0.357	 & /	 & /	 & /	 &  {\it {$f_C/f_{B}=$1.389}}	 & & $\chi^2_{\rm ima} = $ 125.1	 & {\it {$\Delta t_{CB} = $ -32.1 }}  \\ 
& 		 & 	 & /	 & /	 & /	 &  {\it {$f_C/f_{B}=$0.213}}	 & & $\chi^2_{\rm lens} = $ 111.2	 & {\it {$\Delta t_{DB} = $ -102.0 }}  \\ 
\hline
(k) HE2149-2745& 	SIS+$\gamma$	 &  0.858	 & /	 & /	 & 0.016, 8.30	 & {\it {$f_B/f_{A}=$0.234}}	 & 0 & 0.0	 & {\it {$\Delta t_{AB} = $ 93.7 }}  \\ 
& SIE+$\gamma$           & 0.834	 & (0.21, 9.80)	 & /	 & 0.063, -80.35	 & {\it {$f_B/f_A =$ 0.234}}	 & 0 & 0.0	 & {\it {$\Delta t_{AB} = $ 88.6}} \\ 
& 	DV+$\gamma$	 &  0.784	 & (0.21, 9.79)	 & (1.10)	 & 0.076, -69.73	 & {\it {$f_B/f_{A}=$0.234}}	 & 0 & 0.0	 & {\it {$\Delta t_{AB} = $ 116.8 }}  \\ 

\hline
\end{tabular}
\end{center}
\vspace{0.2cm}
{\tiny{Notes: ($\dagger$) The models of this lensed quasar might be affected by strong systematic errors owing to the large uncertainty on the exact location of the lens from the NICMOS data. ($\dagger\dagger$) an exponential profile better reproduces the light distribution of the spiral lens galaxy and therefore the ExpD mass model should be prefered to the DV mass model. The $R_{\rm eff}$ used for our DV mass model is the effective radius derived in Tab.~\ref{gal11} using exponential light distribution.}}
\caption{Results of the parametric modeling of the 11 newly deconvolved quads. Quantities which are results of the modeling (i.e. not explicit model parameters) are displayed in {\it {italic}} and parameters which are fitted to their observed values are displayed into parentheses. Note that flux ratios are used to constrain the lens model only for doubles. }
\label{lensmodel11}
\end{table*}

% Results of the MCMC fit 
% Table generated with script_table.taberr() -> modres.tex
\begin{table*}%[t!]
\begin{center}
\begin{tabular}{l||c|cccccc}
\hline
Object & Model & $R_{Ein}$ & $e$ & $\theta_{e}$ & $R_{eff}$ & $\gamma$ & $\theta_{\gamma}$ \\ 
\hline
\hline
(a) JVAS B0218+357$^\dagger$	 & SIS+$\gamma$	 & $ 0.151^{+0.003} _{-0.003}$ 	 & /	 & /	 & /	 & $ 0.096^{+0.019} _{-0.018}$ 	 & $ -29.24^{+3.56} _{-3.26}$ \\ 
         & SIE+$\gamma$ & $ 0.150^{+0.003} _{-0.003}$ 	 & $ 0.04^{+0.01} _{-0.00}$ 	 & $ -73.83^{+1.58} _{-1.66}$ 	 & /	 & $ 0.096^{+0.019} _{-0.017}$ 	 & $ -25.31^{+3.54} _{-3.51}$ \\ 
	 & DV+$\gamma^{\dagger\dagger}$	 & $ 0.149^{+0.003} _{-0.004}$ 	 & $ 0.04^{+0.00} _{-0.00}$ 	 & $ -73.84^{+1.60} _{-1.46}$ 	 & $ 0.31^{+0.01} _{-0.01}$ 	 & $ 0.127^{+0.027} _{-0.022}$ 	 & $ -26.74^{+3.61} _{-2.96}$ \\ 
         & ExpD$+\gamma$ & 0.182$^{+0.005}_{-0.004}$ & $0.04^{+0.00}_{-0.00}$   & $-73.76^{+1.45}_{-1.50}$  & $0.31^{+0.01}_{-0.01}$ & $0.008^{+0.005}_{-0.006}$ & $-16.99^{+25.24}_{-24.73}$ \\
\hline
(b) SBS~0909+532	 & SIS+$\gamma$	 & $ 0.556^{+0.003} _{-0.003}$ 	 & /	 & /	 & /	 & $ 0.066^{+0.003} _{-0.003}$ 	 & $ 54.47^{+1.64} _{-1.61}$ \\ 
  & SIE+$\gamma$ & $ 0.549^{+0.005} _{-0.006}$ 	 & $ 0.10^{+0.06} _{-0.05}$ 	 & $ -52.18^{+12.95} _{-12.69}$ 	 & /	 & $ 0.094^{+0.020} _{-0.016}$ 	 & $ 49.87^{+4.13} _{-4.95}$ \\ 
	 & DV+$\gamma$	 & $ 0.787^{+0.026} _{-0.067}$ 	 & $ 0.11^{+0.05} _{-0.06}$ 	 & $ -51.73^{+12.65} _{-11.93}$ 	 & $ 0.54^{+0.01} _{-0.01}$ 	 & $ 0.122^{+0.013} _{-0.014}$ 	 & $ 49.73^{+2.70} _{-2.84}$ \\ 
\hline
(c) RX J0911.4+0551	 & SIE+$\gamma$	 & $ 0.921^{+0.011} _{-0.013}$ 	 & $ 0.24^{+0.02} _{-0.02}$ 	 & $ -72.42^{+2.27} _{-2.14}$ 	 & /	 & $ 0.326^{+0.004} _{-0.004}$ 	 & $ 9.34^{+0.24} _{-0.24}$ \\ 
	 & DV+$\gamma$	 & $ 0.959^{+0.006} _{-0.006}$ 	 & $ 0.16^{+0.01} _{-0.01}$ 	 & $ -65.15^{+1.56} _{-1.50}$ 	 & $ 1.02^{+0.01} _{-0.01}$ 	 & $ 0.381^{+0.003} _{-0.004}$ 	 & $ 7.93^{+0.09} _{-0.09}$ \\ 
\hline
(d) FBQS J0951+2635	 & SIS+$\gamma$	 & $ 0.546^{+0.025} _{-0.027}$ 	 & /	 & /	 & /	 & $ 0.092^{+0.031} _{-0.029}$ 	 & $ -6.63^{+11.00} _{-13.40}$ \\ 
& SIE+$\gamma$ & $ 0.581^{+0.006} _{-0.006}$ 	 & $ 0.46^{+0.02} _{-0.02}$ 	 & $ 12.54^{+1.53} _{-1.57}$ 	 & /	 & $ 0.128^{+0.017} _{-0.016}$ 	 & $ -66.81^{+2.28} _{-2.36}$ \\
	 & DV+$\gamma$	 & $ 0.516^{+0.024} _{-0.019}$ 	 & $ 0.47^{+0.02} _{-0.02}$ 	 & $ 12.29^{+1.62} _{-1.46}$ 	 & $ 0.78^{+0.01} _{-0.01}$ 	 & $ 0.014^{+0.013} _{-0.010}$ 	 & $ -25.72^{+38.72} _{-21.60}$ \\ 
\hline
(e) HE~1104-1805	 & SIS+$\gamma$	 & $ 1.381^{+0.005} _{-0.005}$ 	 & /	 & /	 & /	 & $ 0.136^{+0.003} _{-0.003}$ 	 & $ 22.43^{+0.11} _{-0.12}$ \\ 
& SIE+$\gamma$ &  $ 1.404^{+0.009} _{-0.010}$ 	 & $ 0.14^{+0.02} _{-0.02}$ 	 & $ 54.00^{+5.58} _{-5.51}$ 	 & /	 & $ 0.119^{+0.010} _{-0.009}$ 	 & $ 12.97^{+1.70} _{-1.88}$ \\ 
	 & DV+$\gamma$	 & $ 1.358^{+0.006} _{-0.007}$ 	 & $ 0.14^{+0.02} _{-0.02}$ 	 & $ 53.63^{+6.01} _{-5.31}$ 	 & $ 0.98^{+0.01} _{-0.01}$ 	 & $ 0.200^{+0.007} _{-0.006}$ 	 & $ 19.48^{+0.57} _{-0.59}$ \\ 
\hline
(f) PG~1115+080	 & SIE+$\gamma$	 & $ 1.141^{+0.001} _{-0.001}$ 	 & $ 0.14^{+0.00} _{-0.00}$ 	 & $ -67.64^{+0.34} _{-0.35}$ 	 & /	 & $ 0.124^{+0.001} _{-0.001}$ 	 & $ 53.72^{+0.33} _{-0.32}$ \\ 
	 & DV+$\gamma$	 & $ 1.137^{+0.001} _{-0.001}$ 	 & $ 0.14^{+0.00} _{-0.00}$ 	 & $ -67.62^{+0.36} _{-0.36}$ 	 & $0.92^{+0.01}_{-0.01}$	 & $ 0.177^{+0.002} _{-0.002}$ 	 & $ 59.89^{+0.15} _{-0.16}$ \\ 
\hline
(g) B1422+231	 & SIE+$\gamma$	 & $ 0.771^{+0.001} _{-0.001}$ 	 & $ 0.27^{+0.01} _{-0.01}$ 	 & $ -57.50^{+0.29} _{-0.31}$ 	 & /	 & $ 0.176^{+0.004} _{-0.004}$ 	 & $ -52.29^{+0.20} _{-0.20}$ \\ 
	 & DV+$\gamma$	 & $ 0.722^{+0.001} _{-0.001}$ 	 & $ 0.46^{+0.01} _{-0.01}$ 	 & $ -55.85^{+0.30} _{-0.30}$ 	 & $ 0.49^{+0.01} _{-0.01}$ 	 & $ 0.374^{+0.003} _{-0.003}$ 	 & $ -53.64^{+0.10} _{-0.10}$ \\ 
\hline
(h) SBS~1520+530	 & SIS+$\gamma$	 & $ 0.744^{+0.010} _{-0.008}$ 	 & /	 & /	 & /	 & $ 0.127^{+0.006} _{-0.006}$ 	 & $ -2.03^{+1.94} _{-2.44}$ \\ 
& SIE+$\gamma$ & $ 0.732^{+0.011} _{-0.011}$ 	 & $ 0.49^{+0.02} _{-0.02}$ 	 & $ -26.58^{+0.69} _{-0.74}$ 	 & /	 & $ 0.142^{+0.012} _{-0.013}$ 	 & $ 43.14^{+2.76} _{-2.60}$ \\ 
	 & DV+$\gamma$	 & $ 0.645^{+0.010} _{-0.009}$ 	 & $ 0.49^{+0.02} _{-0.02}$ 	 & $ -26.49^{+0.70} _{-0.69}$ 	 & $ 0.76^{+0.01} _{-0.01}$ 	 & $ 0.224^{+0.015} _{-0.016}$ 	 & $ 21.64^{+0.98} _{-0.99}$ \\ 
\hline
(i) CLASS B1600+434	 & SIS+$\gamma$	 & $ 0.613^{+0.007} _{-0.006}$ 	 & /	 & /	 & /	 & $ 0.141^{+0.008} _{-0.009}$ 	 & $ 46.49^{+0.85} _{-1.02}$ \\ 
& SIE+$\gamma$ &  $ 0.720^{+0.013} _{-0.014}$ 	 & $ 0.74^{+0.06} _{-0.06}$ 	 & $ 36.51^{+1.66} _{-1.60}$ 	 & /	 & $ 0.258^{+0.059} _{-0.051}$ 	 & $ -55.41^{+2.17} _{-2.36}$ \\
	 & DV+$\gamma ^{\dagger\dagger}$	 & $ 0.580^{+0.006} _{-0.006}$ 	 & $ 0.76^{+0.06} _{-0.05}$ 	 & $ 36.73^{+1.41} _{-1.41}$ 	 & $ 0.29^{+0.01} _{-0.01}$ 	 & $ 0.169^{+0.013} _{-0.013}$ 	 & $ 58.40^{+1.08} _{-0.98}$ \\ 
         & ExpD$+\gamma$ & 0.673$^{+0.008}_{-0.007}$ & $0.74^{+0.05}_{-0.06}$   & $36.71^{+1.67}_{-1.66}$  & $0.29^{+0.01}_{-0.01}$ & $0.084^{+0.011}_{-0.013}$ & $-81.01^{+5.01}_{-4.86}$ \\
\hline
(j) CLASS B1608+656	 & SIE+$\gamma$	 & $ 1.023^{+0.007} _{-0.007}$ 	 & $ 0.80^{+0.01} _{-0.01}$ 	 & $ 72.91^{+0.25} _{-0.26}$ 	 & /	 & $ 0.193^{+0.006} _{-0.005}$ 	 & $ -7.70^{+0.64} _{-0.64}$ \\ 
	 & DV+$\gamma$	 & $ 0.844^{+0.005} _{-0.006}$ 	 & $ 0.59^{+0.01} _{-0.01}$ 	 & $ 78.69^{+0.21} _{-0.20}$ 	 & $ 1.51^{+0.01} _{-0.01}$ 	 & $ 0.101^{+0.002} _{-0.001}$ 	 & $ 48.36^{+0.33} _{-0.29}$ \\ 
\hline
(k) HE~2149-2745	 & SIS+$\gamma$	 & $ 0.855^{+0.008} _{-0.007}$ 	 & /	 & /	 & /	 & $ 0.014^{+0.007} _{-0.005}$ 	 & $ 2.89^{+11.82} _{-20.71}$ \\ 
& SIE+$\gamma$ & $ 0.835^{+0.009} _{-0.010}$ 	 & $ 0.21^{+0.01} _{-0.02}$ 	 & $ 9.18^{+3.45} _{-3.42}$ 	 & /	 & $ 0.062^{+0.009} _{-0.010}$ 	 & $ -82.39^{+7.02} _{-5.35}$ \\ 
	 & DV+$\gamma$	 & $ 0.787^{+0.010} _{-0.011}$ 	 & $ 0.21^{+0.02} _{-0.01}$ 	 & $ 9.40^{+3.85} _{-3.80}$ 	 & $ 1.10^{+0.01} _{-0.01}$ 	 & $ 0.069^{+0.016} _{-0.017}$ 	 & $ -72.51^{+4.83} _{-5.69}$ \\ 

\hline
\end{tabular}
\end{center}
{\tiny{Notes: ($\dagger$) The models of this lensed quasar might be affected by strong systematic errors owing to the large uncertainty on the exact location of the lens from the NICMOS data. ($\dagger\dagger$) an exponential profile better reproduces the light distribution of the spiral lens galaxy and therefore the ExpD mass model should be prefered to the DV mass model. The $R_{\rm eff}$ used for our DV mass model is the effective radius derived in Tab.~\ref{gal11} using exponential light distribution.}}
\vspace{0.2cm}
\caption{Median value of the model parameters and 68\% confidence interval of the 11 newly deconvolved quads.}
\label{MCMC11}
\end{table*}

\section{Doubly imaged systems}
\label{sec:doubles}

Owing to the limited number of constraints provided by doubles, their use to derive $H_0$ is still a matter of debate. On the one hand, owing to the generally large opening angle between the lens and the two images (i.e. opening angle close to 180$^{\circ}$ when the lens and the images are nearly aligned), the time delays depend little on the structure of the quadrupole term of the potential, and so on the intrinsic ellipticity of the lens \citep{Kochanek2002}. On the other hand, the convergence produced by the lens environment may lead to large systematic errors on the time delays, which may be difficult to identify and disentangle from intrinsic ellipticity because the few number of constraints provided by doubles \citep{Keeton2004}. Although it is generally true that intrinsic and extrinsic sources of shear are difficult to disentangle in those systems, this is not always the case. From the comparison of an SIS+$\gamma$ model (i.e. spherically symmetric potential for the lensing galaxy) with a DV+$\gamma$ and with an SIE+$\gamma$, which encode information about the lens morphology and allow one to probe two different steepnesses of the mass distribution, we test how the intrinsic ellipticity affects the shear estimate. We then check whether the position angle of the shear $\theta_\gamma$, derived from our models, is supported by published environmental studies. Finally, we propose a ranking of the systems based on our ability to disentangle the shear caused by intrinsic ellipticity and nearby/line-of-sight environment. For this analysis we considered six out of the seven doubles presented above, excluding B0218+357 for which possible systematic errors on the galaxy position may strongly bias the models, and the four systems HE~0047-1756, SDSS~J1155+6346, SDSS~J1226-0006, and HS~2209+1914 presented in Paper I. A table compiling all our models (including the new SIE models of the four doubles analyzed in Paper I) for doubles is presented in Appendix~\ref{appendixdouble}.

Although the DV and the SIE models encode information regarding the intrinsic structure of the lens, the shear position angle predicted by these models is in general similar (within $\sim$ 10$^{\circ}$) to the PA of the shear predicted by the SIS model. Only three out of ten doubles of our sample (SBS~1520+530, HE~2149-2745, and CLASS~B1600+434) show evidence for a different $\theta_\gamma$ once the observed ellipticity of the lens galaxy is included in the models. FBQ~J0951+2635 also shows a significant disagreement once the SIE model is considered. Assuming that ($e$, $\theta_e$) provide a good proxy on the structure of the mass profile (see discussion in Sect.~\ref{sec:quads_e}), the lack of change in $\theta_\gamma$ suggests that the internal galaxy structure is not the main source of shear in most of the systems. This is also supported by the generally small change (typically $<$~15\%) in the shear amplitude between the SIS and SIE in these systems{\footnote{HE~0047-1756, and in a less significant way SBS~0909+532, show the strongest changes in $\gamma$.}}. The difference of shear amplitude between the SIS and the DV model is well understood and explained in ~\citet{Kochanek1991}.

For the three systems where the internal structure has a significant effect on $\theta_\gamma$, we checked if the $\theta_\gamma$ provided by the elliptical models is more realistic than the one provided by the SIS+$\gamma$ model:
\begin{itemize}

\item {For  SBS~1520+530, $\theta_\gamma$ derived from the elliptical models points roughly toward the object M identified in Fig.~\ref{dec_NICMOS11}, the agreement is better for the SIE than for the DV model. This object was suggested by \citet{Faure2002} to be an important source of shear together with a galaxy cluster/group located about WSW of the lens. Therefore, it sounds plausible that these two overdensities are the main sources of external shear.  }
\item {The case of  HE~2149-2745 is more subtle as the shear associated to the SIS model is less than 2\%. This does not automatically mean that HE~2149-2745 offers an unperturbed line of sight but that either several sources of shear cancel out or that the lens lies at the center of a group/cluster, as indeed suggested by other studies. \citet{Faure2004}, based on a weak-lensing analysis of the field of view, found an overdensity in the cluster map that they associated with a cluster nearly centered on the lens. The spectroscopic study of the field of view carried out by \citet{Momcheva2006} shows that there are in fact three major galaxy groups on the l.o.s. toward HE~2149-2745. These authors found that the total shear produced by these groups at the location of the lens is ($\gamma$, $\theta_\gamma$)=(0.022, -68$^{\circ}$). The good agreement with the value of $\theta_\gamma$ derived with both the SIE and the DV models ($\theta_\gamma ({\rm DV}) \sim -72.5\pm 5$, $\theta_\gamma ({\rm SIE}) \sim -82.4\pm 7$) may not be fortuitous and indicates that for this system proper account of the environment and intrinsic ellipticity are primordial, a conclusion that was not obvious based on the results of the SIS+$\gamma$ model only.}
\item{CLASS~B1600+434 is more difficult to assess. The amount and direction of the shear derived with the ExpD, DV, and SIE models are significantly different. The environment has been studied by \citet{Williams2006}, \citet{Auger2007} and \citet{Fassnacht2008} based on imaging, spectroscopic, and X-ray data. These studies unveiled that CLASS~B1600+434 is part of a small galaxy group ($\sigma_v \sim 100$\,km/s), too faint to be detected in the X-ray and with only seven confirmed members (including the lens). The main contributor to the external shear seems to be a relatively ``bright'' SBa galaxy located about 4.35\arcsec SE of the lens. The other group members seem to produce a shear mostly in the direction $\theta \sim $ 82$^{\circ}$ \citep{Auger2007}, which deviates from any of our estimates of $\theta_\gamma$. Because the spectroscopic identification of group candidates may be incomplete \citep{Auger2007}, the global effect of the group remains uncertain. Owing to the small number of group members, it is possible that a single unidentified member may be the key to solve the discrepancy. On the other hand, because we do not know if all group members share the same halo, because the ellipticity of the companion is unknown and because the main lens is a late-type galaxy with a strickingly high ellipticity likely to differ from the halo ellipticity, a proper estimate of the external shear may be challenging in this system. }

\item {FBQS~0951+2635 shows a strong decrease in the shear amplitude when the DV model is used but an increase when an SIE model is considered. This increase in the shear amplitude also affects the shear direction, which becomes incompatible with $\theta_\gamma ({\rm {SIS}})$. The value of $\theta_\gamma ({\rm {SIE}}) \sim$ -66.8$^\circ$ deviates by about 30$^\circ$ from the main candidate galaxy group identified about 1 arcmin SE from FBQS~0951+2635 by~\citet{Williams2006} in the direction $\theta \sim $-30$^\circ$. Other groups identified by these authors might also play a role.}
\end{itemize}

In summary, heterogeneous environment information exists for six out of the ten systems. For two of them (SBS~1520+530, HE~2149-2745), our simple models seem to catch the effect of the environment and internal structure well. For two other systems (HE0047-1746 and J1226-006), a nearby galaxy produces a fair fraction of the shear needed by the model (see Paper I for discussion). Therefore, we think that these four systems are those that are better suited for time-delay studies such as COSMOGRAIL. For B1600+434, despite of the low density of the lens environment, our simple models miss to  properly identify the source of shear. Four of the remaining systems show either a very high amplitude of shear without obvious contribution in the field (J1155+6346, HE~1104-1805), a likely small contribution of the environment (HS~2209+1914) or a significant dependence of the shear with the exact mass profile of the lensing galaxy (FBQS~0951+2635). The last system (SBS~0909+532) has $\theta_e$ nearly orthogonal to $\theta_\gamma$ such that the intrinsic and extrinsic source of shear will be hard to disentangle. These last six systems, with the exception of HS~2209+1914, which needs more detailed observations, seem less well suited for deriving $H_0$ from time-delay studies.  

\section{Quadruply imaged systems}
\label{sec:quads}

The four quadruply imaged lensed quasars we modeled (Tables~\ref{lensmodel11} and ~\ref{MCMC11}) are poorly fitted  by the DV+$\gamma$ model while the SIE+$\gamma$ model provides an acceptable fit to the observed image position. Still, the observed position of the lensing galaxy in  PG~1115+080 and JVAS~B1422+231 is not well reproduced by the SIE+$\gamma$ model, which requires a lens offset by resp. 0.02$\arcsec$ and 0.01$\arcsec$ (i.e. 68 and 37 $h^{-1}$ pc) w.r.t. the observed position. As pointed out earlier in the literature, it is necessary to account explicitly for the lens galaxy environment in these systems to reproduce both the lensed images and the lens galaxy position \citep[e.g.][]{Impey1998, Kundic1997b}. The modeling of the four quads studied above illustrates the shortcuts of single mass models and advocates for mass models that account explicitly for the local environment of the lens. Therefore we decided to model most of the quads with milli-arcsec accuracy astrometry, taking into account the effect of satellites/companion galaxies detected in the images. We focused on the effect of the objects located in projection in the vicinity of the lens that contributes to the higher order terms of the potential, and are therefore more likely to influence the image positions. 

\subsection{New models and extended sample}
\label{subsec:newmodels}

We gathered a sample of 14 quads that we modeled with an SIE model for the main lensing galaxy, and a SIS model for any visible companion/satellite galaxy or nearby group. The remainder of the environmental perturbation to the gravitational potential was modeled as an external shear.  The modeling methodology is similar to the one outlined in Sect.~\ref{sec:model}, except that we left ($e$, $\theta_e$) free during the fit in order to compare the mass and the light distribution of the lens. Our sample includes all quads analyzed here and those studied in Paper I (i.e. RXS~J1131-1231, SDSS~J1138+0314, WFI~J2026-4536). We have added to this sample the seven systems B0128+437, MG0414+0534, HE0435-1223, SDSS~0924+0219, H1413+117, WFI~2033-4723, B2045+265. In three cases (B0128+437, J0924+0219, J1138+0314), the literature data did not allow us to identify any obvious secondary lens{\footnote{\cite{Faure2011} identified a bright substructure near the center of the lens J0924+0219, but they demonstrate this is very unlikely to be a nearby satellite galaxy.}} and only SIE+$\gamma$ models have been used. The data we used and the references are provided in Appendix~\ref{appendixdata}.

% Table generated with script_table.tabbestquad()
% tabbest02c.tex
\begin{table*}%[t!]
\begin{center}
\begin{tabular}{l||c|lllllll}
\hline
 \multicolumn{1}{c||}{Object} & 
 \multicolumn{1}{c|}{N$_{\rm lens}$} & 
 \multicolumn{1}{c}{$R_{\rm Ein}$} & 
\multicolumn{1}{c}{$e, \theta_{e}$} &  
\multicolumn{1}{c}{$\gamma, \theta_{\gamma}$} & 
\multicolumn{1}{c}{$\Delta G$} & 
\multicolumn{1}{c}{RMS} & 
\multicolumn{1}{c}{$\chi^{2}$ (d.o.f.)}&
\multicolumn{1}{c}{$ \chi^2_{ima}, \chi^2_{lens}$} \\ 
\hline
\hline

B0128+437& 1 	 &  0.235	 & 0.46, -27.72	 & 0.213, 41.17	 & 0.006, 38	 & 0.000	 & 0.4 (1)& 0.0, 0.4\\ 
\hline
MG0414+0534& 1 	 &  1.182	 & 0.23, -38.26	 & 0.119, 79.30	 & 0.057, 321	 & 0.000	 & 26.6 (1) & 0.0, 26.6\\ 
MG0414+0534& 2 	 &  1.100, 0.181	 & 0.22, 82.65	 & 0.099, -55.03	 & 0.000, 2	 & 0.000 	 & 0.0 (0) & 0.0, 0.0\\ 
\hline
HE0435-1223& 1 	 &  1.201	 & 0.05, 4.60	 & 0.067, 16.00	 & 0.003, 12	 & 0.000	 & 1.1 (1) & 0.2, 1.0 \\ 
HE0435-1223& 2 	 &  1.191, 0.074	 & 0.09, 11.75	 & 0.059, 18.60	 & 0.000, 0	 & 0.000 & 0.0 (0) & 0.0, 0.0 \\ 
\hline
RX J0911.4+0551& 1 	 &  1.120	 & 0.27, -36.13	 & 0.315, 12.87	 & 0.043, 227	 & 0.005	 & 138.2 (1) & 15.9, 122.3\\ 
RX J0911.4+0551& 2 	 &  0.916, 0.232	 & 0.25, -73.48	 & 0.329, 9.25	 & 0.000, 0. & 0.000	 & 0.0 (0) & 0.0, 0.0\\ 
\hline
J0924+219& 1 	 &  0.874	 & 0.11, -53.87	 & 0.068, 68.00	 & 0.009, 33	 & 0.001	 & 5.4 (1) & 0.3, 5.1\\ 
\hline
PG~1115+080& 1 	 &  1.145	 & 0.16, -84.23	 & 0.107, 51.06	 & 0.018, 58	 & 0.002	 & 22.2 (1) & 2.7, 19.5\\ 
PG~1115+080& 2$^{\dagger}$ 	 &  0.969, 4.773	 & 0.06, -73.90	 & 0.248, 61.25	 & 0.000, 1	 & 0.000	 & 0.8 (1) & 0.0, 0.7\\ 
\hline
RXS~J1131-1231& 1 	 &  1.837	 & 0.18, -60.15	 & 0.106, -83.29	 & 0.029, 89	 & 0.011	 & 183.6 (1) & 64.4, 119.2\\ 
RXS~J1131-1231& 2$^{\dagger\dagger}$ 	 &  1.837, 2.8e-6	 & 0.18, -60.32	 & 0.106, -83.27	 & 0.029, 90	 & 0.011	 & 183.6 (0) & 64.0, 119.6 \\ 
\hline
J1138+0314& 1 	 &  0.664	 & 0.04, -63.45	 & 0.106, 31.95	 & 0.010, 39	 & 0.000	 & 2.9 (1) & 0.1, 2.8\\ 
\hline
H1413+117& 1 	 &  0.596	 & 0.60, -35.88	 & 0.310, 49.98	 & 0.031, 176	 & 0.000	 & 2.4 (1) & 0.0, 2.4\\ 
H1413+117& 2 	 &  0.523, 1.078	 & 0.49, -38.89	 & 0.144, 63.29	 & 0.029, 164	 & 0.000	 & 2.1 (0) & 0.0, 2.1\\ 
\hline
B1422+231& 1 	 &  0.771	 & 0.28, -56.96	 & 0.174, -52.45	 & 0.012, 40	 & 0.000	 & 5.5 (1) & 0.2, 5.3\\ 
B1422+231& 2$^{\dagger}$ 	 &  0.740, 3.941	 & 0.28, -57.05	 & 0.132, -48.36	 & 0.012, 41	 & 0.000	 & 5.8 (1) & 0.2, 5.6\\ 
B1422+231& 2$^{\dagger \dagger\dagger}$ 	 &  0.785, 4.450	 & 0.21, -57.62	 & 0.091, 77.47	 & 0.000, 1	 & 0.000	 & 0.0 (1) & 0.0, 0.0\\ 
\hline
B1608+656& 1 	 &  1.155	 & 0.90, 69.86	 & 0.301, -15.04	 & 0.028, 136	 & 0.012	 & 96.3 (1) & 49.8, 46.5\\ 
B1608+656& 2 	 &  1.049, 0.094	 & 0.84, 71.69	 & 0.223, -10.70	 & 0.000, 0	 & 0.000	 & 0.0 (0) & 0.0, 0.0\\ 
\hline
WFI2026-4536& 1 	 &  0.654	 & 0.17, -19.03	 & 0.137, 81.80	 & 0.011, 50	 & 0.002	 & 71.2 (1) & 12.6, 58.6\\ 
WFI2026-4536& 2 	 &  0.604, 1.116	 & 0.20, -1.54	 & 0.099, 74.90	 & 0.000, 0	 & 0.000	 & 0.0 (0) & 0.0, 0.0\\ 
\hline
WFI2033-4723& 1 	 &  1.117	 & 0.35, 88.69	 & 0.236, 12.71	 & 0.032, 158	 & 0.009	 & 330.5 (1) & 75.3, 255.2\\ 
WFI2033-4723& 2 	 &  0.945, 1.255	 & 0.21, 15.45	 & 0.214, 10.11	 & 0.011, 57	 & 0.005	 & 60.9 (0) & 28.0, 32.9\\ 
\hline
B2045+265& 1 	 &  1.130	 & 0.05, 78.89	 & 0.174, -66.29	 & 0.003, 19	 & 0.002	 & 17.8 (1) & 7.8, 10.0\\ 
B2045+265& 2 	 &  1.101, 0.032	 & 0.11, 29.09	 & 0.203, -67.07	 & 0.000, 0 & 0.000	 & 0.0 (0) & 0.0, 0.0\\ 

\hline
\end{tabular}
\end{center}
{\tiny{Notes: ($\dagger$) This model uses the group position of \citet{Momcheva2006} and the velocity dispersion of \citet{Wong2011}; ($\dagger \dagger$) this model is identical to model \#1 because of the negligible contribution of the secondary lens; ($\dagger \dagger \dagger$) this model uses the X-ray centroid of the group by \citet{Grant2004}}}
\vspace{0.2cm}
\caption{Best SIE+$\gamma$ (N$_{\rm lens} = 1$) and SIE+SIS+$\gamma$ (N$_{\rm lens} = 2$) models for the sample of 14 quads (Sect.~\ref{subsec:newmodels}). The Einstein radius ($R_{\rm Ein}$) is in arcsec, the offset between the observed and model lens galaxy position $\Delta G$ is expressed in arcsec and in $h^{-1}$ pc and the RMS is in arcsec. The last column provides the contribution of the lens and image astrometry to the total $\chi^2$ (Eq.~\ref{eq:chidecomp}). Note that the flux ratios are not used to constrain the models.  } 
\label{allquad02}
\end{table*}

We always assumed the second galaxy to lie at the redshift of the lens. For PG~1115+080 and B1422+231, the second component of our mass model is a galaxy group. We used the position of the group center derived by \citet{Momcheva2006} based on imaging data and we constrained the velocity dispersion of the SIS with the velocity dispersion of the group obtained by \citet{Wong2011}. Owing to the large uncertainty on the group position, we also tested an alternative model for B1422+231 with a group located at the centroid of the X-ray emission \citep{Grant2004} but keeping the velocity dispersion of the SIS unconstrained. For PG~1115+080, we have not implemented this alternative model because of the more diffuse X-ray emission, which leads to a very uncertain group centroid \citep{Grant2004, Fassnacht2008}. 

The parameters of the best models we found are presented in Table ~\ref{allquad02}. Models with one (N$_{\rm lens} = 1$) and two (N$_{\rm lens} = 2$) mass components are presented. When N$_{\rm lens} = 1$ (resp. N$_{\rm lens} = 2$), we have 1 d.o.f. (resp. 0 d.o.f.). The quality of the fit can be evaluated by means of the $\chi^2$ as defined in Eq.~\ref{eq:chidecomp}, but with $\chi^2_{\rm morph}=\chi^2_{\rm flux} = 0$ because of our modeling strategy. We also provide the root mean square (RMS) between the modeled and observed image position and the offset $\Delta G$ (in arcsec and in $h^{-1}$ pc) between the centroid of the light and of the mass distribution. We now use these results to test the ability of our models to reproduce the quads down to milli-arcsec accuracy (Sect.~\ref{sec:quads_anomaly}) and to compare the distribution of the light and of the mass in the main lensing galaxy (Sect.~\ref{sec:quads_e}). 

\subsection{Astrometric anomalies}
\label{sec:quads_anomaly}

Because the errors on the astrometry vary from system to system, we have decided to use quantities not weighted by astrometric error bars to identify astrometric anomalies. As a rule of thumb, we considered that the model disagrees with the data when the RMS between the model and observed image position is RMS $>$ 2 mas, i.e. our largest measured uncertainty on a lensed image position. Furthermore, we considered that the galaxy is not well fitted when $\Delta G > 40\,h^{-1}$\,pc, which is about twice the typical offset between light and mass found by \citet{Yoo2006} based on the analysis of four lensed systems with Einstein rings. This offset corresponds to typically 10 mas and is on average twice larger than the uncertainty on the lens position (except for H1413+117 and B0128+437). When a single lens deflector is considered, the lensed images and galaxy position are well reproduced in 6 out of 14 systems (B0128+437, HE~0435-1223, J0924+219, J1138+0314, B1422+231 and B2045+265). For four other systems (MG~0414+0534, PG~1115+080, H1413+117, WFI~2026-4536), only the lensed image positions are well reproduced, while for the remaining systems, both image and galaxy positions are discrepant. The inclusion of a second lens allows a nearly perfect match between the model and the observations except for RXS~J1131-1231, WFI~2033-4723 and H1413+117. For RXS~J1131-1231 the visible faint companion would increase the anomaly if too massive, as already pointed out in \citet{Claeskens2006}. The nearest galaxy in projection is located 15 arcsec (75 $h^{-1}$\,kpc at $z_{\rm lens}$) away from the lens and may be unable to explain the anomaly. Current studies of the environment \citep{Momcheva2006, Wong2011}, although they identify at least a foreground galaxy group, also fail to explain the shear needed in the models. For H1413+117, the companion only marginally modifies the predicted centroid of the main lensing galaxy but leads to a significantly lower external shear. For WFI~2033-4723, the companion improves the modeling but not enough to lead to an acceptable fit. Other nearby companions visible on the images might be the cause of the discrepancy but we have not investigated this possibility because of the small number of observational constraints. 

A few additional remarks should be made regarding the modeling of the quads. First, our best models for WFI~2026-4536 and B2045+265 predict a value of $\theta_e$ that is significantly tilted compared to the observed PA. Although we cannot exclude that the gravitational potential of the main lens is highly perturbed by the environment, this may also be a good indication that additional ingredients are missing in the lens model. This can be an effect of dark matter substructure or of visible perturbers not accounted for by our models. The ellipticity of the second mass component and the slope of the radial profile might also play a role in the model. Second, even in the models with two mass components, the amplitude of external shear reaches $\gamma > 0.2$ for 6/14 systems. In three cases (RX~J0911.4+0551, PG~1115+080, B2045+265), the modeling with two lenses even leads to a stronger shear than the modeling with a single SIE+$\gamma$. This is not alarming and reflects the fact that the shear is not a scalar quantity and has to be added vectorially. The higher amplitude of the shear observed in our sample of quads compared to our sample of doubles may be caused by the increased probability of observing a quad in rich environment \citep{Holder2003, Keeton2004, Huterer2005}. Nevertheless, the selection bias is too strong to allow us to properly investigate statistical occurrence of quads and doubles. Third, the formally good fit of H1413+117 is dominated by the large uncertainty (0.02\arcsec) on the lens galaxy position. The model indeed predicts the lens to be offset by about 170 $h^{-1}$\,pc ($z_{\rm lens} = 1$) compared to the observed position. In addition, the inclusion of the closest galaxy in the environment only marginally improves the model and only once the prior on the shear \citep[i.e. $\gamma < 0.25$, similar to][]{McLeod2009} is used. Improving the accuracy on the astrometry of the lens by a factor 5 would reveal whether this system is the only cross-like quad showing significant astrometric anomaly maybe caused by the environment. Finally, we should note that two positions of the group have been tested for B1422+231 but only the group position matching the X-ray centroid identified by \citet{Grant2004} leads to a successful fit. In that case, the SIS model of the group has a velocity dispersion $\sigma_{\rm SIS} \sim $ 436\,km/s, in excellent agreement with the velocity dispersion $\sigma_v = 421^{+99}_{-82} $ km/s of the group derived spectroscopically by \citet{Wong2011}. The X-ray group centroid of \citet{Grant2004} however disagrees with the position obtained by \citet{Fassnacht2008}, which is nearly centered on the main lens.

In summary, astrometric anomalies are frequent when a single galaxy is used for the modeling. Nevertheless, in most of the cases, the inclusion of a nearby companion/group in the lens model led to a much better model. This effect of a massive companion on the image positions and the way the lensing galaxy could absorb the perturbation induced by the latter agrees with the findings of \cite{Chen2007} based on the study of mock lenses. For two systems, WFI2033-4723 and RXS~J1131-1231, the companion galaxy is not sufficient to provide a perfect fit to the astrometry. In the case of WFI~2033-4723, other galaxies projected at an angularly small distance from the lens (i.e. typically $<$ 10 arcsec) might explain the observed discrepancy. RXS~J1131-1231 remains the clearest case in our sample of an astrometric anomaly. Its amplitude is consistent with a perturbation produced by a dark matter substructure located near the Einstein radius of the lens \citep{Chen2007}. Problems encountered in reproducing the flux ratios, which are also affected by significant microlensing \citep{Sluse2006, Sluse2007}, and problems to reproduce the preliminary time delays \citep{Morgan2006, Keeton2009, Congdon2010} strongly support this interpretation. Astrometric anomalies possibly caused by substructures might also be present in other systems such as  WFI~2026-4536 and B2045+265, for which the best simple model predicts a strong tilt between the mass and the light distributions. Finally, we mention that astrometric anomalies may appear when higher astrometric accuracies are considered. Indeed, we did not find any sign of anomalies for B0128+437 but \citet{Biggs2004} have shown that such anomalies are observed when sub-mas  accuracy on the astrometry is used.

\subsection{Position angle and ellipticity}
\label{sec:quads_e}

% Figure created using script_fig_mass_obs.py and input file lensvsmodel.txt
% The python script is in the Latex directory 
\begin{figure*}[ht!]
\begin{tabular}{cc}
 {\includegraphics[scale=0.45]{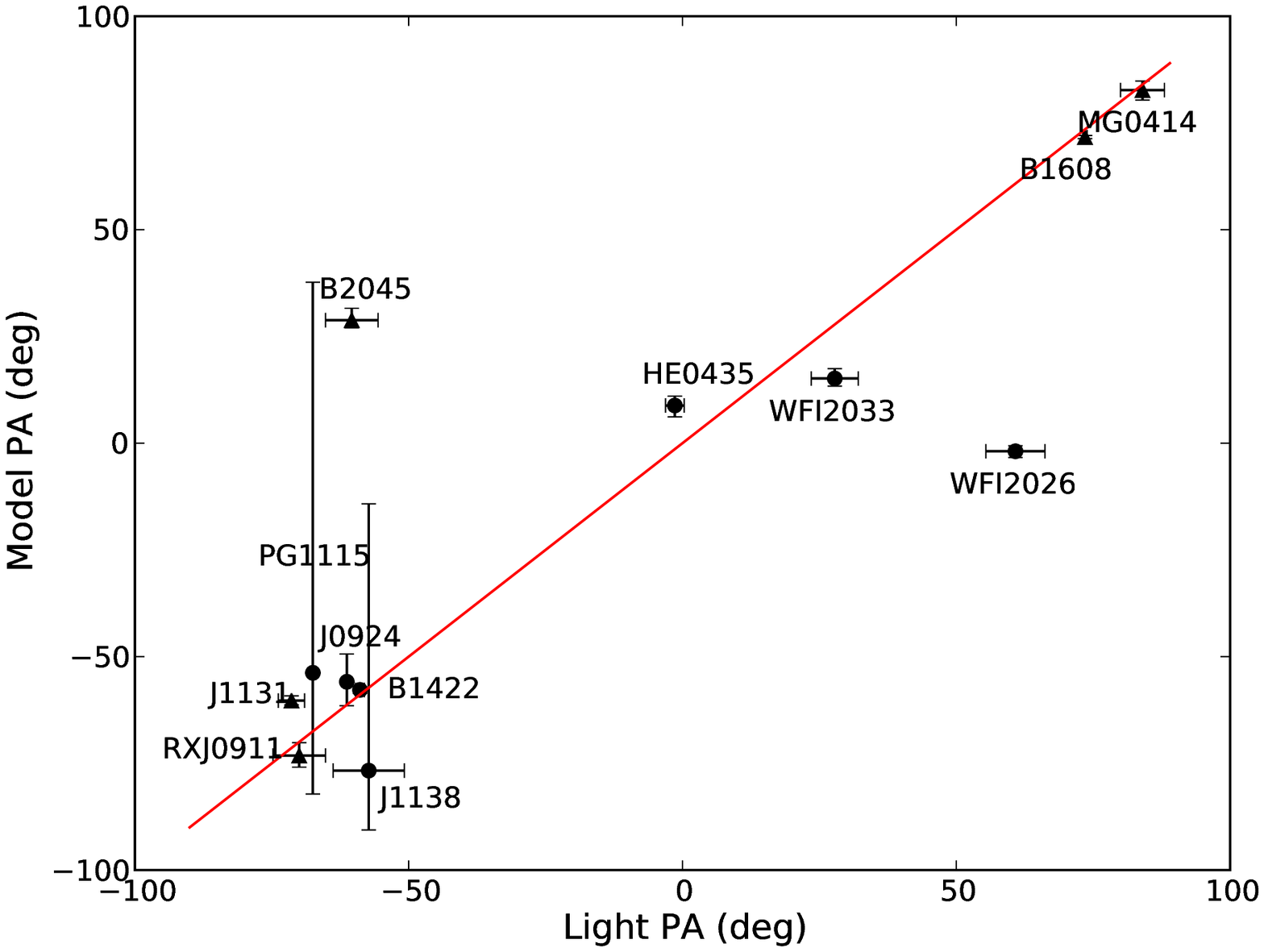}} & 
 {\includegraphics[scale=0.45]{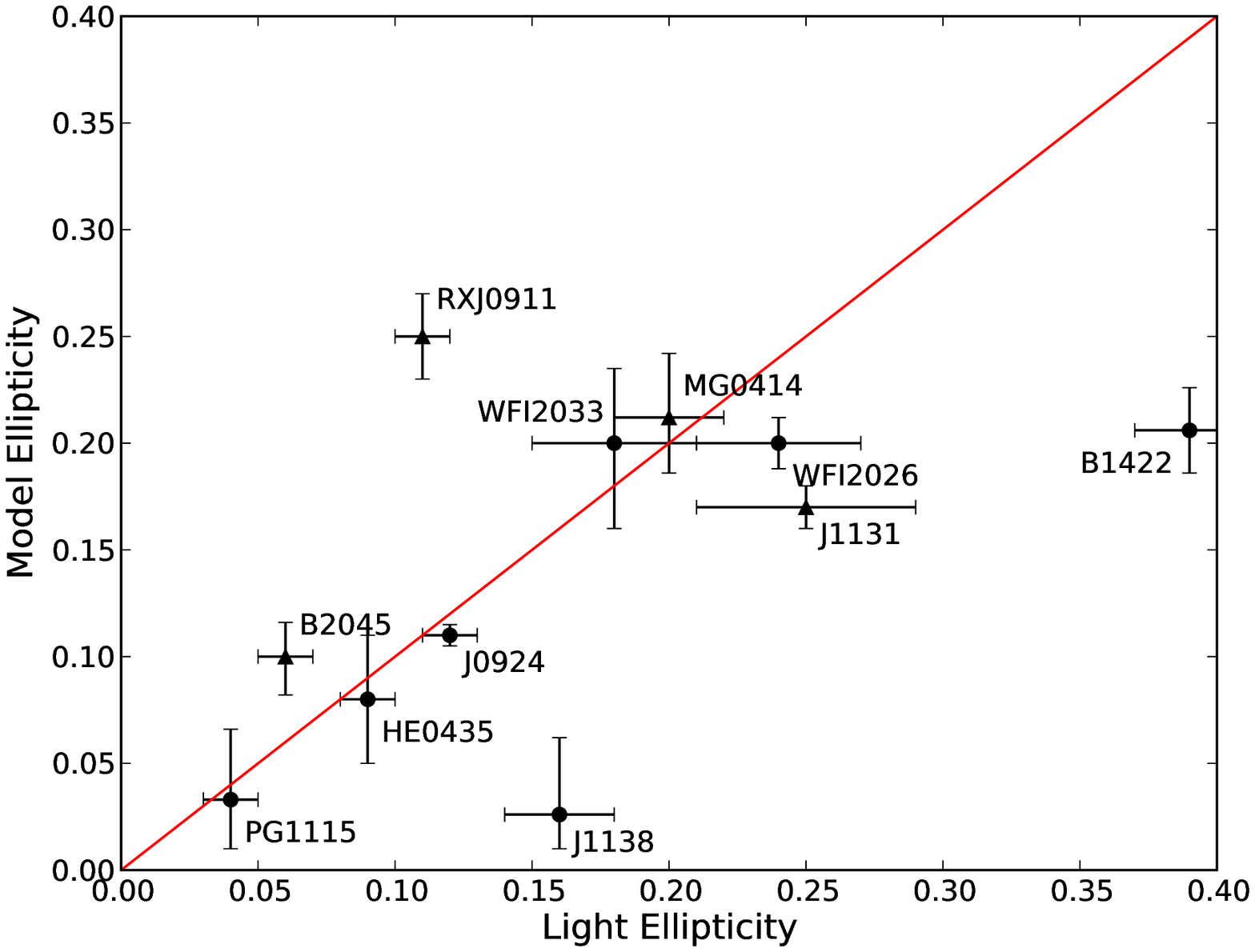}} \\
\end{tabular}
\caption{Light vs mass (model) PA ({\it{left}}) and ellipticity ({\it{right}}) for our sample of quadruply imaged quasars. The ellipticity of the main lens in B1608+656 ($e_{\rm {obs}}$=0.45$\pm$0.01, $e_{\rm{mod}}$=0.84$\pm$0.01) is not shown to ease legibility. Systems having a companion within the Eintein radius of the main lens are depicted with a filled triangle. The red solid line depicts the situation where the light perfectly traces the total mass. The error bars are 68\% CL error bars. }
\label{ellip-PA}
\end{figure*}

We now compare the ellipticity $e$ and position angle $\theta_e$ of the mass distribution of the main lens with the one derived from the light profile. The results are displayed in Fig.~\ref{ellip-PA} for the 12 quads for which measurements are available. This figure confirms the previous finding that the PA of the light and of the mass generally agree well \citep{Keeton1998, Ferreras2008, Treu2009}. A mismatch by about 10 degrees is not uncommon but it is unclear whether this is real, i.e. caused by triaxiality of the dark matter or lens environment, or caused by errors in the PA measurements. The systems B2045+265 and WFI~2026-4536 show a much larger mismatch whose origin is not yet clearly identified.

More surprisingly, Figure~\ref{ellip-PA} shows that the mismatch between the ellipticity of the mass and of the light distribution is not as large as previously claimed  \citep{Keeton1998, Ferreras2008}. For 8 out of 12 systems, the observed ellipticity of the light agrees well with that of the mass. Two of the four outliers (RX~J0911.4+0551, SDSS~J1138+0314, B1422+231, and B1608+656) show a nearby galaxy which lies within the Einstein radius of the lens, an effect which could easily bias the measurement of the light ellipticity. It is also likely that for these two systems and for B2045+265, the ellipticity of the second lens, which lies within 1 $R_E$ of the lens center, introduces internal shear not accounted for by our models or that the system is not relaxed and poorly reproduced by two separated potentials{\footnote{This might also be the case for MG~0414+0534 but the companion galaxy is possibly a foreground object \citep{Curran2011}}}. The difference between our results and those of \citet{Keeton1998} and \citet{Ferreras2008} may have several origins. First, these authors mixed doubles and quads in their analysis, while doubles do not always provide robust information on both the shear and the ellipticity (Sect.~\ref{sec:doubles}). Second, our ellipticity measurements have been derived nearly exclusively  with the same method (MCS deconvolution) applied to NICMOS data (only MG~0414+0534 and B2045+265 have morphological parameters derived with another method) while Keeton's measurements are more heterogeneous with possibly more significant biases associated to less accurate PSF models. Finally, the results of \citet{Ferreras2008} are derived with pixelated models. Despite the many tests and successes of this technique \citep[e.g.][]{Cosmograil4, Saha2006a, Read2007, Coles2008, Leier2009, Leier2011}, it is unclear whether the uniform prior on the ellipticity used to sample the parameter space is not biasing the average pixelated mass profile derived by this method. 

These results suggest that the dark matter and baryon distribution in the central regions (typically 5\,kpc) of lens galaxies do have similar shapes. This may however not be a generic result because of possible selection biases. First, we have probed systems with an observed ellipticity $e < 0.4$, such that the effect of a rounder dark matter halo on the total mass distribution may be harder to detect. Second, it may also be that there is a selection bias of the lensing galaxies towards systems where the observed correlations are more likely to take place \citep[cf.][]{Mandelbaum2009}. The alignement between the dark matter and the baryons distribution has been found in numerical simulations for spiral galaxies \citep{Gustafsson2006}, but still has to be demonstrated for massive ellipticals, where significant misalignements have been observed around one virial radius \citep{Bett2010, Skibba2011}.

\section{Conclusions}
\label{Conc11}

We applied the iterative strategy coupled with the MCS deconvolution algorithm, i.e. ISMCS, to high-resolution near-IR images of a sample of eleven lensed quasars. The ISMCS allowed us to obtain accurate astrometry, in most cases with 1 to 2.5 mas error bars, and shape parameters for the light distribution of the lensing galaxy. In three cases, the deconvolution process was not entirely satisfactory, i.e. HE~1104-1805, PG~1115+080 and CLASS~B1608+656, but still provided better results than simple PSF subtraction. For one special case, i.e. JVAS~B0218+357, the small angular separation and the poor resolution of NICMOS prevents us from reaching the desired accuracy. For one system, SBS~0909+532, we were able to detect the lensing galaxy for the first time, which is angularly small and relatively bright. Two objects reveal partial Einstein rings (PG~1115+080, CLASS~B1608+656).

We provided, for these systems, simple mass models (isothermal and de Vaucouleurs mass distributions) that give an overview of the ability of smooth mass models to reproduce the known lensed quasars. These models give some hints about the effect of the radial profile on the flux ratios and on the time delays. For the sample of ten doubly imaged quasars presented here and in Paper I  \citep[i.e.][]{Chantry2010}, we have also studied the role of intrinsic ellipticity and external shear in the models. For half of these systems{\footnote{HS~2209+1914 is more uncertain because of the lack of detailed environmental studies}} (HE0047-1746, J1226-006, SBS~1520+530, HE~2149-2745, and HS~2209+1914) one should be able to unfold intrinsic and extrinsic shear and therefore minimize systematic biases on $H_0$ based on the time-delay method. This identification of the most promising systems is partly subjective because detailed environmental studies of the whole sample are lacking. Exhaustive studies of lens environment are fortunately ongoing \citep{Williams2006, Momcheva2006, Wong2011} and should allow one to address this question more quantitatively. 

In addition to the ten doubly imaged quasars, we have performed modeling with singular isothermal ellipsoid (SIE) of a sample of 14 quadruply imaged systems for which we have relative astrometry with accuracy typically better than 2 mas. This sample allowed us to investigate two questions: 

\begin{enumerate}
\item {\it {How efficient are SIE models in reproducing the image and galaxy position of the observed quads and how does the local environment of the lens influence this result ? }} We found that the environment of most of the quads is rich and that at least the nearest galaxy or a nearby group has to be included explicitly in the model to recover the image and galaxy positions. Following this procedure leads to a formally perfect fit with an SIE model of the main lens for 11 of the 14 systems. When this perturber is not included, models generally tend to predict the position of the main lensing galaxy significantly offset with respect to the observed position, the lensed image positions being less affected. For the two systems RXS~J1131-1231 and WFI~2033-4723, we were not able to obtain an acceptable fit. The problems encountered with WFI~2033-4723 might be caused by other nearby galaxies not included explicitly in our model.  RXS~J1131-1231 remains the best candidate of our sample for which astrometric anomalies may be caused by a dark matter substructure. Two other systems (WFI~2026-4536 and B2045+265) show a significant mismatch between the $\theta_e$ of the light and of the mass. The origin of this discrepancy is unclear and might also be the signature of dark matter substructures. We mention that astrometric anomalies may become proeminent when higher accuracy images are available. This is the case for B0128+437, which is perfectly reproduced by our model but shows astrometric anomalies once sub-mas astrometry of the lensed images is considered \citep{Biggs2004}. \\
\item {\it {How do the morphology of the mass and of the light profile compare ?}} Our analysis of 12 out of 14 quads with measured structural parameters allowed us to find that the PA of the mass and of the  light distribution are aligned within typically 10 degrees, in agreement with previous studies \citep{Keeton1998, Ferreras2008, Treu2009}. We also discovered that the ellipticity of the light generally agrees well with that of the total mass distribution (8 out of 12 lenses show a good correlation). Overall, this indicates that the light provides a good proxy to the total mass distribution in these lensing galaxies{\footnote{The radial distribution of the total mass disagrees however with that of the light, as illustrated by the inability of deVaucouleurs mass profiles to model quads.}}, a result particularly interesting in the light of the rich environment of the lenses. Understanding why these correlations take place is important for both our comprehension of dark matter and galaxy formation and for time-delay studies. Therefore, an increase in the sample of lenses with shape measurements and a comparison with numerical simulations, including problems caused by selection biases \citep{Vandeven2009, Mandelbaum2009}, are necessary. This will provide important clues on the role of dark matter in shaping galaxies. \\

\end{enumerate}

\begin{acknowledgements}
We are grateful to Peter Schneider and Dandan Xu for very useful discussions and to the referee for his constructive reports. This work is supported by ESA and the Belgian Federal Science Policy (BELSPO) in the framework of the PRODEX Experiment Arrangement C-90312. DS acknowledges partial support from the Alexander von Humboldt fundation, from the German Virtual Observatory and from the Deutsche Forschungsgemeinschaft, reference SL172/1-1. COSMOGRAIL is financially supported by the Swiss National Foundation (SNSF).
\end{acknowledgements}

\bibliographystyle{aa}
\bibliography{Articles}

\newpage

\appendix

\section{Models of the 11 douby  imaged quasars}
\label{appendixdouble}

We provide in Table~\ref{MCMC11doubles} a summary of the median model parameters and 68\,\% CL for the 11 doubly imaged quasars analyzed in this paper and in Paper I. The SIE models for the lensed quasars HE~0047-1756, SDSS~1155+635, SDSS~1226-0006 and HS~2209+1914 were not presented in Paper I and are discussed in Sect.~\ref{sec:doubles}.

% table generated with the script script_table.taberr() applied to the 3 different subsets: dVC, SIS, SIE and merged into 1 table manually
\begin{table*}%[t!]
\begin{center}
\begin{tabular}{l||c|cccccc}
\hline
Object & Model & $R_{Ein}$ & $e$ & $\theta_{e}$ & $R_{eff}$ & $\gamma$ & $\theta_{\gamma}$ \\ 
\hline
\hline
HE~0047-1756	 & SIS+$\gamma$	 & $ 0.751^{+0.002} _{-0.002}$ 	 & /	 & /	 & /	 & $ 0.048^{+0.002} _{-0.002}$ 	 & $ 7.22^{+0.80} _{-0.73}$ \\ 
	 & SIE+$\gamma$	 & $ 0.760^{+0.003} _{-0.004}$ 	 & $ 0.22^{+0.01} _{-0.02}$ 	 & $ -66.80^{+3.49} _{-3.66}$ 	 & /	 & $ 0.124^{+0.006} _{-0.007}$ 	 & $ 17.67^{+2.34} _{-2.51}$ \\ 
	 & DV+$\gamma$	 & $ 0.755^{+0.003} _{-0.003}$ 	 & $ 0.22^{+0.01} _{-0.01}$ 	 & $ -66.59^{+3.90} _{-3.88}$ 	 & $ 0.91^{+0.01} _{-0.01}$ 	 & $ 0.119^{+0.005} _{-0.005}$ 	 & $ 15.78^{+2.08} _{-2.01}$ \\ 
\hline
JVAS B0218+437$^\dagger$	 & SIS+$\gamma$	 & $ 0.151^{+0.003} _{-0.003}$ 	 & /	 & /	 & /	 & $ 0.096^{+0.019} _{-0.018}$ 	 & $ -29.24^{+3.56} _{-3.26}$ \\ 
 & SIE+$\gamma$	 & $ 0.150^{+0.003} _{-0.003}$ 	 & $ 0.04^{+0.01} _{-0.00}$ 	 & $ -73.83^{+1.58} _{-1.66}$ 	 & /	 & $ 0.096^{+0.019} _{-0.017}$ 	 & $ -25.31^{+3.54} _{-3.51}$ \\ 
 & DV+$\gamma^{\dagger\dagger}$	 & $ 0.149^{+0.003} _{-0.004}$ 	 & $ 0.04^{+0.00} _{-0.00}$ 	 & $ -73.84^{+1.60} _{-1.46}$ 	 & $ 0.31^{+0.01} _{-0.01}$ 	 & $ 0.127^{+0.027} _{-0.022}$ 	 & $ -26.74^{+3.61} _{-2.96}$ \\ 
 & ExpD$+\gamma$ & $0.182^{+0.005}_{-0.004}$ & $0.04^{+0.001}_{-0.001}$   & $-73.76^{+1.45}_{-1.50}$  & $0.31^{+0.01}_{-0.01}$ & $0.008^{+0.005}_{-0.006}$ & $-16.99^{+25.24}_{-24.73}$ \\

\hline
SBS~0909+532	 & SIS+$\gamma$	 & $ 0.556^{+0.003} _{-0.003}$ 	 & /	 & /	 & /	 & $ 0.066^{+0.003} _{-0.003}$ 	 & $ 54.47^{+1.64} _{-1.61}$ \\ 
	 & SIE+$\gamma$	 & $ 0.549^{+0.005} _{-0.006}$ 	 & $ 0.10^{+0.06} _{-0.05}$ 	 & $ -52.18^{+12.95} _{-12.69}$ 	 & /	 & $ 0.094^{+0.020} _{-0.016}$ 	 & $ 49.87^{+4.13} _{-4.95}$ \\ 
	 & DV+$\gamma$	 & $ 0.787^{+0.026} _{-0.067}$ 	 & $ 0.11^{+0.05} _{-0.06}$ 	 & $ -51.73^{+12.65} _{-11.93}$ 	 & $ 0.54^{+0.01} _{-0.01}$ 	 & $ 0.122^{+0.013} _{-0.014}$ 	 & $ 49.73^{+2.70} _{-2.84}$ \\ 
\hline
FBQ J0951+2635	 & SIS+$\gamma$	 & $ 0.546^{+0.025} _{-0.027}$ 	 & /	 & /	 & /	 & $ 0.092^{+0.031} _{-0.029}$ 	 & $ -6.63^{+11.00} _{-13.40}$ \\ 
& SIE+$\gamma$	 & $ 0.581^{+0.006} _{-0.006}$ 	 & $ 0.46^{+0.02} _{-0.02}$ 	 & $ 12.54^{+1.53} _{-1.57}$ 	 & /	 & $ 0.128^{+0.017} _{-0.016}$ 	 & $ -66.81^{+2.28} _{-2.36}$ \\ 
& DV+$\gamma$	 & $ 0.516^{+0.024} _{-0.019}$ 	 & $ 0.47^{+0.02} _{-0.02}$ 	 & $ 12.29^{+1.62} _{-1.46}$ 	 & $ 0.78^{+0.01} _{-0.01}$ 	 & $ 0.014^{+0.013} _{-0.010}$ 	 & $ -25.72^{+38.72} _{-21.60}$ \\ 
\hline
HE~1104-1805	 & SIS+$\gamma$	 & $ 1.381^{+0.005} _{-0.005}$ 	 & /	 & /	 & /	 & $ 0.136^{+0.003} _{-0.003}$ 	 & $ 22.43^{+0.11} _{-0.12}$ \\ 
& SIE+$\gamma$	 & $ 1.404^{+0.009} _{-0.010}$ 	 & $ 0.14^{+0.02} _{-0.02}$ 	 & $ 54.00^{+5.58} _{-5.51}$ 	 & /	 & $ 0.119^{+0.010} _{-0.009}$ 	 & $ 12.97^{+1.70} _{-1.88}$ \\ 
& DV+$\gamma$	 & $ 1.358^{+0.006} _{-0.007}$ 	 & $ 0.14^{+0.02} _{-0.02}$ 	 & $ 53.63^{+6.01} _{-5.31}$ 	 & $ 0.98^{+0.01} _{-0.01}$ 	 & $ 0.200^{+0.007} _{-0.006}$ 	 & $ 19.48^{+0.57} _{-0.59}$ \\ 
\hline
SDSS~J1155+635	 & SIS+$\gamma$	 & $ 0.592^{+0.014} _{-0.012}$ 	 & /	 & /	 & /	 & $ 0.389^{+0.012} _{-0.015}$ 	 & $ -10.36^{+0.19} _{-0.19}$ \\ 
	 & SIE+$\gamma$	 & $ 0.617^{+0.014} _{-0.014}$ 	 & $ 0.15^{+0.01} _{-0.01}$ 	 & $ 0.40^{+2.42} _{-2.31}$ 	 & /	 & $ 0.345^{+0.015} _{-0.015}$ 	 & $ -11.38^{+0.27} _{-0.29}$ \\ 
	 & DV+$\gamma$	 & $ 0.583^{+0.012} _{-0.011}$ 	 & $ 0.15^{+0.01} _{-0.01}$ 	 & $ 0.43^{+2.33} _{-2.45}$ 	 & $ 1.14^{+0.01} _{-0.01}$ 	 & $ 0.449^{+0.013} _{-0.014}$ 	 & $ -11.02^{+0.22} _{-0.21}$ \\ 
\hline
SDSS~J1226-0006	 & SIS+$\gamma$	 & $ 0.568^{+0.003} _{-0.003}$ 	 & /	 & /	 & /	 & $ 0.100^{+0.005} _{-0.004}$ 	 & $ 7.95^{+0.38} _{-0.38}$ \\ 
	 & SIE+$\gamma$	 & $ 0.569^{+0.003} _{-0.003}$ 	 & $ 0.07^{+0.01} _{-0.01}$ 	 & $ 44.02^{+4.13} _{-4.15}$ 	 & /	 & $ 0.095^{+0.006} _{-0.006}$ 	 & $ 1.72^{+1.39} _{-1.36}$ \\ 
	 & DV+$\gamma$	 & $ 0.557^{+0.003} _{-0.003}$ 	 & $ 0.07^{+0.01} _{-0.01}$ 	 & $ 44.89^{+4.40} _{-4.42}$ 	 & $ 0.69^{+0.02} _{-0.02}$ 	 & $ 0.144^{+0.007} _{-0.007}$ 	 & $ 4.51^{+0.69} _{-0.70}$ \\ 
\hline
SBS~1520+530	 & SIS+$\gamma$	 & $ 0.744^{+0.010} _{-0.008}$ 	 & /	 & /	 & /	 & $ 0.127^{+0.006} _{-0.006}$ 	 & $ -2.03^{+1.94} _{-2.44}$ \\ 
	 & SIE+$\gamma$	 & $ 0.732^{+0.011} _{-0.011}$ 	 & $ 0.49^{+0.02} _{-0.02}$ 	 & $ -26.58^{+0.69} _{-0.74}$ 	 & /	 & $ 0.142^{+0.012} _{-0.013}$ 	 & $ 43.14^{+2.76} _{-2.60}$ \\ 
	 & DV+$\gamma$	 & $ 0.645^{+0.010} _{-0.009}$ 	 & $ 0.49^{+0.02} _{-0.02}$ 	 & $ -26.49^{+0.70} _{-0.69}$ 	 & $ 0.76^{+0.01} _{-0.01}$ 	 & $ 0.224^{+0.015} _{-0.016}$ 	 & $ 21.64^{+0.98} _{-0.99}$ \\ 
\hline
CLASS~B1600+434	 & SIS+$\gamma$	 & $ 0.613^{+0.007} _{-0.006}$ 	 & /	 & /	 & /	 & $ 0.141^{+0.008} _{-0.009}$ 	 & $ 46.49^{+0.85} _{-1.02}$ \\ 
	 & SIE+$\gamma$	 & $ 0.720^{+0.013} _{-0.014}$ 	 & $ 0.74^{+0.06} _{-0.06}$ 	 & $ 36.51^{+1.66} _{-1.60}$ 	 & /	 & $ 0.258^{+0.059} _{-0.051}$ 	 & $ -55.41^{+2.17} _{-2.36}$ \\ 
	 & DV+$\gamma^{\dagger\dagger}$	 & $ 0.580^{+0.006} _{-0.006}$ 	 & $ 0.76^{+0.06} _{-0.05}$ 	 & $ 36.73^{+1.41} _{-1.41}$ 	 & $ 0.29^{+0.01} _{-0.01}$ 	 & $ 0.169^{+0.013} _{-0.013}$ 	 & $ 58.40^{+1.08} _{-0.98}$ \\ 
         & ExpD$+\gamma$ & $0.673^{+0.008}_{-0.007} $ & $0.739^{+0.051}_{-0.055}$   & $36.71^{+1.67}_{-1.66}$  & $0.29^{+0.01}_{-0.01}$ & $0.084^{+0.011}_{-0.013}$ & $-81.01^{+5.01}_{-4.86}$ \\
\hline
HE~2149-2745	 & SIS+$\gamma$	 & $ 0.855^{+0.008} _{-0.007}$ 	 & /	 & /	 & /	 & $ 0.014^{+0.007} _{-0.005}$ 	 & $ 2.89^{+11.82} _{-20.71}$ \\ 
& SIE+$\gamma$	 & $ 0.835^{+0.009} _{-0.010}$ 	 & $ 0.21^{+0.01} _{-0.02}$ 	 & $ 9.18^{+3.45} _{-3.42}$ 	 & /	 & $ 0.062^{+0.009} _{-0.010}$ 	 & $ -82.39^{+7.02} _{-5.35}$ \\ 
& DV+$\gamma$	 & $ 0.787^{+0.010} _{-0.011}$ 	 & $ 0.21^{+0.02} _{-0.01}$ 	 & $ 9.40^{+3.85} _{-3.80}$ 	 & $ 1.10^{+0.01} _{-0.01}$ 	 & $ 0.069^{+0.016} _{-0.017}$ 	 & $ -72.51^{+4.83} _{-5.69}$ \\ 
\hline
HS~2209+1914	 & SIS+$\gamma$	 & $ 0.515^{+0.002} _{-0.002}$ 	 & /	 & /	 & /	 & $ 0.031^{+0.002} _{-0.002}$ 	 & $ -86.11^{+2.00} _{-1.79}$ \\ 
	 & SIE+$\gamma$	 & $ 0.519^{+0.002} _{-0.002}$ 	 & $ 0.05^{+0.01} _{-0.01}$ 	 & $ 62.84^{+2.31} _{-2.38}$ 	 & /	 & $ 0.027^{+0.002} _{-0.002}$ 	 & $ -70.43^{+5.61} _{-5.63}$ \\ 
	 & DV+$\gamma$	 & $ 0.515^{+0.002} _{-0.002}$ 	 & $ 0.05^{+0.01} _{-0.02}$ 	 & $ 63.05^{+2.17} _{-2.10}$ 	 & $ 0.53^{+0.01} _{-0.01}$ 	 & $ 0.041^{+0.003} _{-0.003}$ 	 & $ -83.64^{+4.07} _{-3.20}$ \\ 

\hline
\end{tabular}
\end{center}
{\tiny{Notes: ($\dagger$) The models of this lensed quasar might be affected by strong systematic errors owing to the large uncertainty on the exact location of the lens from the NICMOS data, ($\dagger\dagger$) An exponential profile better reproduces the light distribution and therefore the ExpD mass model should be prefered to the DV mass model. The $R_{\rm eff}$ used for our DV mass model is the effective radius derived in Tab.~\ref{gal11} using exponential light distribution. }}
\vspace{0.2cm}
\caption{Median value of the model parameters and 68\% confidence interval for the 11 doubly imaged quasars.}
\label{MCMC11doubles}
\end{table*}

\section{Data for the quadruply imaged quasars}
\label{appendixdata}

We provide the information associated to the 14 quadruply imaged quasars studied in Sect.~\ref{sec:quads}. Table~\ref{astroima} displays the relative RA-Dec of the individual images with their associated errors. Table~\ref{galaxinfo} provides the position, ellipticity and position angle of the lensing galaxy and the associated references.

% Table generated with topcat based on "datalens.txt"
% tab_imapos.tex

%\begin{sidewaystable}[t!]
\begin{table*}
\begin{center}

\begin{tabular}{l|crrrrrrrrrrr}
\hline
  \multicolumn{1}{c|}{Name} &
%  \multicolumn{1}{c|}{images$_{1,2,3,4}$} &
  \multicolumn{1}{c}{RA$_1$, Dec$_1$} &
%  \multicolumn{1}{c}{Dec$_1$} &
  \multicolumn{1}{c}{$\sigma_{RA_1,Dec_1}$} &
%  \multicolumn{1}{c}{Dec$_1$} &
%  \multicolumn{1}{c}{$\sigma_{Dec_1}$} &
  \multicolumn{1}{c}{RA$_2$} &
%  \multicolumn{1}{c}{$\sigma_{RA_2}$} &
  \multicolumn{1}{c}{Dec$_2$} &
  \multicolumn{1}{c}{$\sigma_{RA_2,Dec_2}$} &
  \multicolumn{1}{c}{RA$_3$} &
%  \multicolumn{1}{c}{$\sigma_{RA_3}$} &
  \multicolumn{1}{c}{Dec$_3$} &
  \multicolumn{1}{c}{$\sigma_{RA_3,Dec_3}$} &
  \multicolumn{1}{c}{RA$_4$} &
%  \multicolumn{1}{c}{$\sigma_{RA_4}$} &
  \multicolumn{1}{c}{Dec$_4$} &
  \multicolumn{1}{c}{$\sigma_{RA_4,Dec_4}$} & 
  \multicolumn{1}{c}{Ref} \\
\hline
B0128   & 0, 0& 0.0010   & 0.099    & 0.095   & 0.0010  & 0.521   & -0.17   & 0.0010  & 0.109   & -0.26   & 0.0010  & 3 \\
MG0414  & 0, 0& 8.0e-5   & -0.13396 & 0.4054  & 8.0e-5  & 0.58852 & 1.93752 & 8.0e-5  & 1.9454  & 0.29972 & 8.0e-5  & 4\\
HE0435  & 0, 0& 0.0020   & 1.4743   & 0.5518  & 0.0020  & 2.4664  & -0.6022 & 0.0020  & 0.9378  & -1.616  & 0.0020  & 5\\
RXJ0911 & 0, 0& 0.00251  & 0.2611   & 0.4069  & 0.00251 & -0.0158 & 0.9575  & 0.00251 & -2.9681 & 0.7924  & 0.00251 & 1\\
J0924   & 0, 0& 0.0010   & 0.0636   & -1.8063 & 0.0010  & -0.9648 & -0.6788 & 0.0013  & 0.5414  & -0.4296 & 0.0026  & 6 \\
PG1115  & 0, 0& 0.00198  & 1.3286   & -2.0338 & 0.00198 & 1.4772  & -1.5757 & 0.00198 & -0.3405 & -1.9598 & 0.00198 & 1\\
RXJ1131 & 0, 0& 0.00264  & 0.0347   & 1.187   & 0.00264 & -0.592  & -1.1146 & 0.00264 & -3.1154 & 0.8801  & 0.00264 & 2 \\
J1138   & 0, 0& 0.00117  & -0.1003  & 0.9777  & 0.00117 & -1.1791 & 0.8119  & 0.00117 & -0.6959 & -0.0551 & 0.00117 & 2\\
H1413   & 0, 0& 0.0010   & 0.7426   & 0.1686  & 0.0010  & -0.493  & 0.7135  & 0.0010  & 0.3526  & 1.0394  & 0.0010  & 7 \\
B1422   & 0, 0& 0.00105  & 0.386    & 0.3169  & 0.00105 & -0.336  & -0.7516 & 0.00105 & 0.947   & -0.8012 & 0.00105 & 1 \\
B1608   & 0, 0& 0.0017   & 0.7464   & -1.9578 & 0.0026  & 0.7483  & -0.4465 & 0.0038$^{\dagger}$ & -1.1181 & -1.2527 & 0.0025$^{\dagger\dagger}$  & 1 \\
WFI2026 & 0, 0& 0.00117  & 0.1613   & -1.429  & 0.00117 & 0.414   & -1.2146 & 0.00117 & -0.5721 & -1.0437 & 0.00117 & 2 \\
WFI2033 & 0, 0& 0.0020   & -2.1946  & 1.2601  & 0.0020  & -1.4809 & 1.3756  & 0.0020  & -2.1128 & -0.2778 & 0.0020  & 8\\
B2045   & 0, 0& 5.0e-4   & -0.1316  & -0.2448 & 6.0e-4  & -0.2869 & -0.7885 & 5.0e-4  & 1.6268  & -1.0064 & 0.0013  & 9\\
\hline\end{tabular}

\end{center}
{\tiny{Notes: 1: This paper, 2: \citet{Chantry2010}, 3: \citet{Biggs2004}, 4: \citet{Ros2000}, 5: \citet{Courbin2011}, 6: \citet{Keeton2006}, 7: \citet{Chantry2007}, 8: \citet{Cosmograil7}, 9: \citet{McKean2007}; ${\dagger}$: $\sigma_{RA_3}=$0.0038, $\sigma_{Dec_3}=$0.0025;  ${\dagger\dagger}$:  $\sigma_{RA_4}=$0.0018, $\sigma_{Dec_4}=$0.0018.}}
\vspace{0.2cm}
\caption{Lensed image positions (1, 2, 3, 4) of the 14 quads studied in Sect.~\ref{sec:quads}. We have shortened the names of the systems to ease legibility.}
\label{astroima}
\end{table*}

% Table generated with topcat based on "datalens.txt"
% tab_galaxylens.tex
\begin{table*}%[t!]
\begin{center}
\begin{tabular}{l|rrrrrrrrrrrrr}
\hline
  \multicolumn{1}{c|}{Name} &
  \multicolumn{1}{c}{RA$_{G1}$} &
  \multicolumn{1}{c}{$\sigma_{RA_{G1}}$} &
  \multicolumn{1}{c}{Dec$_{G1}$} &
  \multicolumn{1}{c}{$\sigma_{Dec_{G1}}$} &
  \multicolumn{1}{c}{RA$_{G2}$} &
  \multicolumn{1}{c}{$\sigma_{RA_{G2}}$} &
  \multicolumn{1}{c}{Dec$_{G2}$} &
  \multicolumn{1}{c}{$\sigma_{Dec_{G2}}$} &
  \multicolumn{1}{c}{$e$} &
  \multicolumn{1}{c}{$\sigma_e$} &
  \multicolumn{1}{c}{PA} &
  \multicolumn{1}{c}{$\sigma_{PA}$} &
  \multicolumn{1}{c}{Ref}   \\
\hline
  B0128 & 0.217 & 0.01 & -0.104 & 0.01 & 999 & 999 & 999 & 999 & 999 & 999 & 999 & 999 & 10 \\
  MG0414 & 1.072 & 0.011 & 0.665 & 0.011 & 1.457 & 0.013 & 2.122 & 0.013 & 0.2 & 0.02 & 84.0 & 4.0 & 4, 11\\
  HE0435 & 1.1706 & 0.0030 & -0.5665 & 0.0020 & 3.7558 & 0.0020 & -4.1733 & 0.0020 & 0.09 & 0.01 & -1.4 & 1.7& 5\\
  RXJ0911 & -0.7019 & 0.00251 & 0.502 & 0.0039 & -1.4601 & 0.036 & 1.1678 & 0.047 & 0.11 & 0.01 & -70.0 & 4.8 & 1\\
  J0924 & -0.1804 & 0.0039 & -0.8685 & 0.0039 & 999 & 999 & 999 & 999 & 0.12 & 0.01 & -61.3 & 0.5 & 12\\
  PG1115 & 0.3813 & 0.0041 & -1.3442 & 0.0040 & 3.0 & 11.0 & -21.7 & 9.0 & 0.04 & 0.01 & -67.5 & 0.5 & 1, 17 \\
  RXJ1131 & -2.0269 & 0.00264 & 0.6095 & 0.00264 & -1.93 & 0.0050 & 1.137 & 0.0050 & 0.25 & 0.04 & -71.4 & 2.4 & 2, 14\\
  J1138 & -0.4633 & 0.0071 & 0.534 & 0.0036 & 999 & 999 & 999 & 999 & 0.16 & 0.02 & -57.3 & 6.5 & 2\\
  H1413 & 0.1365 & 0.02 & 0.5887 & 0.02 & 1.87 & 0.1 & 4.14 & 0.1 & 999 & 999 & 999 & 999 & 7, 16\\
  B1422 & 0.7321 & 0.0037 & -0.639 & 0.0054 & 7.0 & 1.5 & -9.0 & 1.5 & 0.39 & 0.02 & -58.9 & 0.8 & 1, 13 \\
  B1608 & -0.4561 & 0.0061 & -1.0647 & 0.0037 & 0.2821 & 0.00171 & -0.9359 & 0.0023 & 0.45 & 0.01 & 73.5 & 0.4 & 1\\
  WFI2026 & -0.0479 & 0.0015 & -0.7916 & 0.0015 & -7.398 & 0.026 & 1.94 & 0.0060 & 0.24 & 0.03 & 60.8 & 5.4 & 2, 15 \\
  WFI2033 & -1.4388 & 0.0020 & 0.3113 & 0.0020 & -5.41 & 0.0020 & 0.285 & 0.0020 & 0.18 & 0.03 & 27.8 & 4.3 & 8\\
  B2045 & 1.1084 & 0.0011 & -0.8065 & 0.0011 & 0.4498 & 0.0021 & -0.6425 & 0.0021 & 0.06 & 0.01 & -60.4 & 4.8 & 9\\
\hline\end{tabular}
\end{center}
{\tiny{Notes: 1: This paper, 2: \citet{Chantry2010}, 3: \citet{Biggs2004}, 4: \citet{Ros2000}, 5: \citet{Courbin2011}, 6: \citet{Keeton2006}, 7: \citet{Chantry2007}, 8: \citet{Cosmograil7}, 9: \citet{McKean2007}, 10: \citet{Lagattuta2010}, 11: \citet{Falco1997}, 12: \citet{Cosmograil2}, 13: \citet{Grant2004} 14: \citet{Claeskens2006}, 15: \citet{Morgan2004}, 16: \citet{McLeod2009}, 17: \citet{Momcheva2006}}}
\vspace{0.2cm}
\caption{Position of the main lensing galaxy (G1) and of the nearest companion/group (G2) relative to the reference lensed image identified in Table~\ref{astroima}. The observed ellipticity and position angle (E of N) of G1 are also provided. Unknown quantities have an assigned value 999. }
\label{galaxinfo}
\end{table*}

\end{document}